\documentclass{elsarticle}
\usepackage[margin=2.5cm]{geometry}
\pdfoutput=1 
\usepackage[load=addn]{siunitx}
\usepackage{color}
\usepackage{verbatim}
\usepackage {longtable}
\usepackage {caption} 
\usepackage {subcaption} 
\usepackage {tabularx} 
\usepackage{arydshln}
\usepackage {hhline}
\usepackage{caption}
\usepackage{comment}

 \usepackage{amsthm}
 \usepackage{amsmath}

\usepackage{lineno}

\usepackage{graphicx}

\usepackage[export]{adjustbox}

\makeatletter
\def\ps@pprintTitle{%
 \let\@oddhead\@empty
 \let\@evenhead\@empty
 \def\@oddfoot{}%
 \let\@evenfoot\@oddfoot}
\makeatother

\begin{document}

\begin{frontmatter}



 \title{A multi-physics methodology for the simulation of the two-way interaction of reactive flow and elastoplastic structural response}

\author[cav]{L.˜Michael}
\ead{lm355@cam.ac.uk}

\author[cav]{N.~Nikiforakis}
\ead{nn10005@cam.ac.uk}

\address[cav]{Laboratory for Scientific Computing, Cavendish Laboratory, Department of Physics, University of Cambridge, UK}

\begin{abstract}

We propose a numerical methodology for the numerical simulation of distinct, interacting physical processes described by a combination of compressible, inert and reactive forms of the Euler equations, multiphase equations and elastoplastic equations. These systems of equations are usually solved by coupling finite element and CFD models. Here we solve them simultaneously, by recasting all the equations in the same, hyperbolic form and solving them on the same grid with the same finite-volume numerical schemes. The proposed compressible, multiphase, hydrodynamic formulation can employ a hierarchy of five reactive and non-reactive flow models, which allows simple to more involved applications to be directly described by the appropriate selection. The communication between the hydrodynamic and elastoplastic systems is facilitated by means of mixed-material Riemann solvers at the boundaries of the systems, which represent physical material boundaries. To this end we derive approximate mixed Riemann solvers for each pair of the above models based on characteristic equations. The components for reactive flow and elastoplastic solid modelling are validated separately before presenting validation for the full, coupled systems. Multi-dimensional use cases demonstrate the suitability of the reactive flow-solid interaction methodology in the context of impact-driven initiation of reactive flow and structural response due to violent reaction in automotive (e.g.\ car crash) or defence (e.g.\ explosive reactive armour) applications. {\color{black}Several types of explosives (C4, Deetasheet, nitromethane, gaseous fuel) in gaseous, liquid and solid state are considered.}

\end{abstract}


\end{frontmatter}



\section{Introduction}

The accurate numerical simulation of a wide range of industrial, automotive, aerospace and defence processes necessitates the consideration of gaseous or condensed-phase explosives initiated by the impact by and their interaction with fluid or elastoplastic solid materials. Examples include accidental fuel initiation in a car crash and fuel tank containment in the context of safety studies and explosive reactive armour (ERA).  This article is concerned with the development of numerical methods for the simultaneous solution of multiphase, reactive, inert fluid and elastoplastic solid equations suitable for the numerical modelling of such problems. {\color{black} The methodology can also be used as part of the manufacturing process for optimising car (or other device) compartments in terms of shapes and materials.}

An integrated numerical methodology for this kind of simulation has three elements; the formulation describing the elastoplastic solid (impactor and/or confiner), the formulation describing the gaseous or condensed phase fuel or explosive, including the explicit capturing of the reaction zone and the transition from reactants to products and the communication between the two systems through material coupling.  By explicit capturing of the reaction zone we mean that the reactants and products are described by distinct equations of state and transition between them occurs in a numerically mixed zone leading to the generation of the reaction front and detonation wave. This is in contrast to programmed burn approaches where assumptions are made with regards to the location of the front and the energy deposed behind it.

In this work, we present the coupling of fuel/explosive formulations with fluid and solid models suitable for a range of automotive and defence applications. We use the terms \emph{fluid} and \emph{hydrodynamic} interchangeably as well as the terms \emph{solid} and \emph{elastoplastic}. The complete explosive--inert fluid--solid system is represented in an Eulerian frame and both the hydrodynamic (for fuel/explosive and inert fluid) and elastoplastic systems of equations are solved with finite volume techniques, employing high-resolution, shock-capturing methods. The communication between the different systems is achieved by employing the Riemann ghost fluid method and the mixed-material Riemann solvers presented here. 

The mathematical description of the elastoplastic system has been traditionally done in a Lagrangian framework. The original Lagrangian form of the solid equations has been reformulated into a conservative form of equations in the Eulerian frame by Godunov and Romenskii \cite{godunov1972nonstationary}, Kondaurov \cite{kondaurov1982equations} and Plohr and Sharp \cite{plohr1988conservative}. This has the advantage of allowing the solution of the elastoplastic solid formulation in the same framework as the explosives hydrodynamic formulation, using the same (or the same family of) high-resolution, shock-capturing methods. This led to the development of high-order, shock capturing schemes for the numerical solution of such systems. For example, Miller and Colella \cite{miller2001high} and Barton et al.\ \cite{barton2009exact,barton2010plastic} have developed linearised Riemann solvers as part of a high-order numerical scheme to capture the seven waves in the (1D) solid system, while Gavrilyuk et al.\ \cite{gavrilyuk2008modelling} have presented the adaptation of the classic HLLC solver to the solid system. Centred numerical schemes and linearised Riemann solvers for the solid systems have also been presented by Titarev \cite{titarev2008musta}, while approximate and exact Riemann solvers for the conservative elastic system have been presented by Miller \cite{miller2004iterative} and Barton et al.\ \cite{barton2009exact}.

{\color{black}Inclusion of plasticity in the solid system has been presented using different approaches, as for example by Miller and Colella \cite{miller2001high}, who evolve the plastic deformation gradient ($\mathbf{F}^p$) in addition the total inverse deformation gradient ($\mathbf{G}=\mathbf{F}^{-1}$) and include an elastic predictor step followed by a `plastic' corrector step to correct any over-estimated elastic deformation that pushes the state outside the yield surface. The predictor-corrector approach allows for solving both for perfect and time-dependent plasticity models. Another approach is followed by Barton et al.\ \cite{barton2010plastic} who only evolve the elastic deformation gradient ($\mathbf{F}^e$) and include plasticity as source terms for the elastic deformation tensor equations.} 

In this work, we use the elastic deformation evolution model by Barton et al.\ \cite{barton2009exact} to describe the elastic behaviour of the solid material. Inelastic deformation is following the Miller and Colella approach of predictor-corrector method based on the principle of maximum dissipation 
and  is applied in combination with perfect plasticity and time-dependent plasticity models.

The mathematical description of the fuel/explosive could vary in complexity, depending on the physical {\color{black} degree of inhomogeneity} of the material or of the physical properties of the material that are dominant in the application of consideration. 
These models can be divided in two broad classes, depending on the {\color{black}description details of the reactive material} and hence on whether the mathematical model is based on some augmented form of the Euler equations or on a  multiphase approach.

Formulations of the augmented Euler class (e.g.\ \cite{banks2007high,banks2008study,wang2004thermodynamically,shyue1998efficient,shyue1999fluid,shyue2001fluid}) evolve the conservation equations for mass, momentum and energy and describe the distinct components of a mixture by means of one or more additional evolution equations, for example for the mass fraction of one of the constituents. Due to the nature of the limited physical information that is conveyed by these models, they assume mechanical and often thermal equilibrium between the components. This type or formulation works well for gaseous fuel materials.

Formulations of the multiphase class (e.g.\ \cite{baer1986two,saurel1999multiphase,kapila2001two,murrone2005five}) can be considered to be forms (full or reduced) of the Baer-Nunziato (BN) system, which have separate mass, momentum and energy conservation laws for each component (phase). Additional advection equations are necessary to differentiate between the components, and exchange between them takes place through source terms. Their mathematical (and hence numerical)\  complexity and the computational cost increases with the number of phases. Reduced versions of this system, which capture more physical information as compared to the augmented Euler approach, at a lower complexity and CPU cost than the full BN system have previously been proposed (e.g.\ \cite{saurel1999multiphase,kapila2001two,murrone2005five,petitpas2009modelling, schoch2013multi}). They do not assume thermal equilibrium and, depending on their assumptions, they may or may not assume mechanical equilibrium and exchanges between the different components are allowed. Obviously, any approximation of a complete mathematical description of a physical system is likely to come with its own limitations or/and undesirable side effects. This type or formulation works well for condensed-phase porous materials.

In this work we consider a formulation (MiNi16) presented by Michael and Nikiforakis \cite{michael2016hybrid}, which integrates the advantages of augmented-Euler and BN-type formulations while allowing for the interaction of an inert component with the reactant-product mixture, through a diffused interface approach. This would allow, for example, for the inclusion of an air gap between the explosive stick and the metal confiner. Reduced versions of this formulation are also presented to model cases when the inert component is not present, or when the explicit modelling of the products of reaction is not required or even when the phases are all non-reacting and could form free-surfaces.

There are various approaches for the numerical solution of coupled solid-fluid problems. {\color{black} Traditionally, Lagrangian techniques are followed. These, however, present difficulties for large deformations of the materials as these are inherently translated to large deformations in the underlying mesh. Others include Smoothed Particle Hydrodynamics (SPH) and arbitrary Lagrangian-Eulerian (ALE). Also, traditionally, the modelling of solid-fluid problems is using one-way coupled finite element codes for modelling the solid part and CFD codes to model the fluid part. As a result, the two processes are solved in a `co-simulation' environment, each on it's own grid with a distinct numerical method. This {\color{black} may lead to discretisation errors} passing from one method to the other. Even though each class has its merits and shortcomings, in this work, we retain a regular, Cartesian mesh (and data) structure, as this is relatively easy to implement in our existing structured, hierarchically adaptive mesh refinement (AMR) framework. }

Studies including solid-reactive  coupling in the Eulerian frame include work by Miller and Colella \cite{miller2002conservative} and Barton et al.\ \cite{barton2011conservative}. The solid material in these studies is described by a full elastoplastic system, while the explosive system considered incorporates either a single-phase Euler formulation or program burn. Although these are suitable for gaseous combustion, they are not adequate for more complex or condensed phase explosives. Schoch et al.\ \cite{schoch2013eulerian} couple the full elastoplastic system to a two-phase, five-equation model for condensed explosives, which is more complex and as a result more restrictive than the model used in this work. Examples of inert Euler-solid coupling can also be found, as for example in \cite{ortega2014numerical,ortega2015richtmyer}.

The coupling in multi-material simulations (including fluid-fluid, solid-solid or fluid-solid interaction) can follow inherently from the Lagrangian framework, as for example by Howell and Ball \cite{howell2002free} who apply the coupling in solid-solid applications. In the Eulerian framework, interface fitting approaches can be followed, such as Volume of Fluid (VOF) and Ghost Fluid Methods (GFM) as in the work by Barton et al.\ \cite{barton2010solidsolid} who use the modified GFM and Schoch et al.\ \cite{schoch2013eulerian} who use the `real GFM'. {\color{black} We define as \emph{multi-material} the framework where material interfaces are tracked rather than captured. This includes different sets of equations being solved on either side of the material interface and special methods used to apply `boundary conditions' across the material interface.} In this work we consider the Riemann ghost fluid method as presented by Sambasivan and Udaykumar \cite{sambasivan2009ghost1}, {\color{black}based on the pioneering work by Fedkiw et al.\ \cite{fedkiw1999non}}  and we provide the coupling by deriving mixed-material Riemann solvers to be used at material interfaces. To this end, we derive characteristic equations that lead to the formulation of a linearised  mixed Riemann solver, applied to the one side of the material interface {\color{black}for the full hydrodynamic formulation by Michael and Nikiforakis \cite{michael2016hybrid}} and its reduced models. Depending on the application or the fuel/explosive in consideration, the full or the reduced versions of the model is employed. 

In the remainder of this article, we first present the distinct mathematical formulations describing the explosive (including the reduced versions) and the elastoplastic solid. The coupling technique is then presented and derivations of the mixed Riemann solvers for fluid-fluid and fluid-solid coupling are presented for each explosives model considered. Then, we validate separately the solid and explosives components by invoking solid-only and explosive-only test problems with known solutions. The explosive-solid coupling is first tested in one dimension and then multi-material inert and reactive simulations are considered, illustrating the applicability of the coupling in impact-initiation of condensed-phase explosives in ERA examples and gaseous fuel in car-crash examples for fuel containment.

\section{Mathematical models}
\label{models}
\begin{figure}[!htb]
\centering
\includegraphics[width=\textwidth]{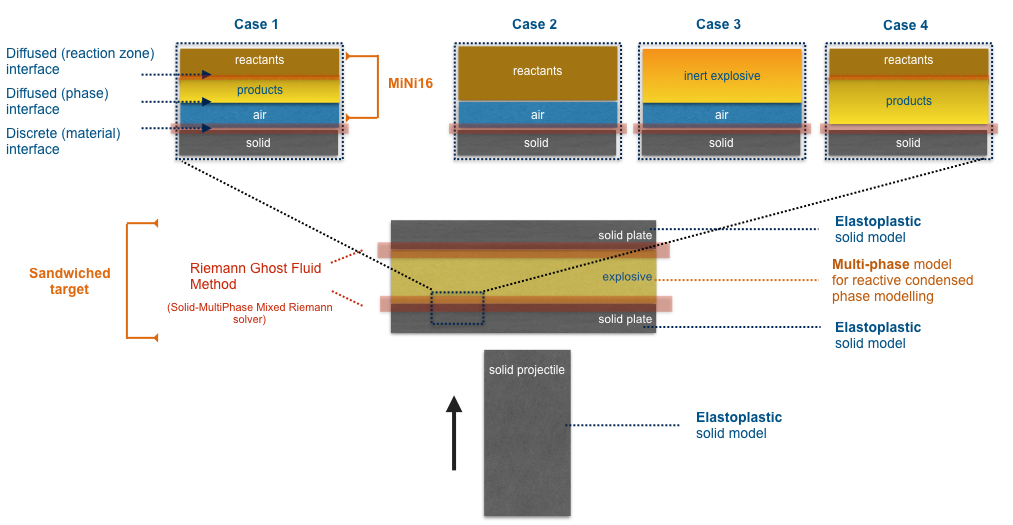}
\caption{Schematic of a sandwich-plate impact and the mathematical models used to describe each component.}
\label{Lynch}
 \end{figure}

In this section, the distinct mathematical formulations describing the materials involved in solid-fluid impact and interaction applications are presented. To aid the description of the principal components and without loss of generality, we use as a reference configuration. As the accidental impact of a fuel tank in a car crash scenario involves a complex tank geometry, we chose that for explaining the mathematical formulations and numerical methods to use the simpler sandwiched-plate impact in an ERA configuration as illustrated in Fig.\ \ref{Lynch}. An explosive (yellow) is residing between two steel plates (grey) and a steel projectile (grey) impacts this sandwiched configuration from below. The whole system could be surrounded by air or vacuum making this a solid-explosive-fluid or solid-explosive-vacuum configuration.

The choice of the explosives model depends on the physical complexity of the material (e.g.\ porous, homogeneous etc) and the dominant timescales in the mechanical and thermodynamic properties of the scenario considered. The explosives model could therefore range from the simple single-EoS augmented Euler formulation (e.g.\ suitable for gaseous combustion) to the full Baer-Nunziato formulation (e.g.\ suitable for porous materials), with the possibility of using a model of intermediate complexity like the ones proposed by Banks et al.\ \cite{banks2007high}, Kapila et al.\ \cite{kapila2001two} and Saurel et al.\ \cite{saurel1999multiphase}. In this work, the formulation proposed by Michael and Nikiforakis \cite{michael2016hybrid} (henceforth referred to as MiNi16) is used for the mathematical description of the explosive. This formulation additionally captures the air gap that can be included inbetween the explosive and the front plate (zoom in Figure \ref{Lynch}).
The steel plates and the steel projectile are described mathematically by an elastoplastic model in the Eulerian frame, as presented by Barton et al.\ \cite{barton2009exact}.
{\color{black}The plates and projectile are treated numerically as separate materials in the multi-material framework. This would allow easily for the physical material of one or both plates as well as of the projectile to be changed.} The surrounding material can be vacuum or air. If it is air, this is described by the compressible {\color{black} inert} Euler equations.

\subsection{{\color{black}The explosives model (Case 1)}}
\label{Hybrid}
In this section, we summarise the {\color{black}MiNi16 formulation} \cite{michael2016hybrid} based on the {\color{black}Case 1} example illustrated in Figure \ref{Lynch} (zoom) of a sandwiched-plate impact scenario with an air gap. In this configuration, the explosive reactant is described as \emph{phase} $\alpha$, with density, velocity vector and pressure ($\rho_{\alpha},\mathbf{u}_{\alpha},p_{\alpha}$) and the products of reaction as \emph{phase} $\beta$ with ($\rho_{\beta},\mathbf{u}_{\beta},p_{\beta}$) equivalently. These are assumed to form a homogeneous mixture, which we will hereafter call interchangeably the `explosive mixture' or the `reactant-products mixture' and denote this as \emph{phase 2} with mixture density, velocity vector and pressure ($\rho_{2},\mathbf{u}_{2},p_{2}$). We can then denote by $\lambda$ the mass fraction of the reactants, such that $\lambda=1$ denotes fully unburnt material and $\lambda=0$ denotes fully burnt material. The air gap (or any other inert material that could confine the explosive mixture) is denoted as \emph{phase 1} with density, velocity vector and pressure ($\rho_{1},\mathbf{u}_{1},p_{1}$).

We denote by $z$ a colour function, which can be considered to be the volume fraction of the air with respect to the volume of the total mixture of phases 1 and 2, with density $\rho$. Equivalently, the volume
fraction of the reactant-product mixture  with respect to the volume of the total mixture is given by  $1-z$. For convenience, we denote $z$ by $z_1$ and $1-z$ by $z_2$. Then, the closure condition $z_1+z_2=1$ holds. 

Velocity and pressure equilibrium applies between the all phases, such that $u_{\alpha}=u_{\beta}=u_1=u_2=u$
and $p_{\alpha}=p_{\beta}=p_1=p_2=p$. Temperature equilibrium is only assumed between the phases of the explosive mixture, i.e.\  $T_{\alpha}=T_{\beta}$, although other closure conditions can be found to be more suitable for other applications (see for example Stewart et al.\ \cite{stewart2002equation}). 

Then, the MiNi16 system is described as in \cite{michael2016hybrid} by:

\begin{eqnarray}
\label{Hyb:cty1}
\frac{\partial z_1 \rho_1}{\partial t}+ \nabla \cdot(z_1 \rho_1 \mathbf{u})  &=&0,\\
\label{Hyb:cty2}
\frac{\partial z_2 \rho_2}{\partial t}+\nabla \cdot(z_2 \rho_2 \mathbf{u}) &=& 0, \\
\label{Hyb:mom}
\frac{\partial}{\partial t}(\rho \mathbf{u_i})+\nabla \cdot(\rho \mathbf{u_iu})+\frac{\partial
p}{\partial{\mathbf{x_i}}}&=&0,\\
\label{Hyb:ene}
\frac{\partial}{\partial t}(\rho E)+\nabla \cdot (\rho E+p)\mathbf{u}&=&0,\\
 \label{Hyb:volfrac}
\frac{\partial z_1 }{\partial t} + \mathbf{u} \cdot \nabla z_1 &=& 0,\\
\label{Hyb:reac}
\frac{\partial z_2 \rho_2 \lambda}{\partial t}+\nabla \cdot(z_2 \rho_2 \mathbf{u}\lambda)
&=& z_2 \rho_2 K,
\end{eqnarray}
where $\mathbf{u}=(u,v,w)$ denotes the total vector velocity, $i$ denotes space dimension, $i=1,2,3$, $\rho$ the total density of the system and $E$ the specific total energy given by $E=e+\frac{1}{2}\sum_{i}u_i^2$, with $e$ the total specific internal energy of the system. 

 $K$ is a function giving the rate of conversion of reactants to products
and for numerical purposes, it is considered a source term to the hyperbolic
part of the system. In this work, depending on the reaction rate law form, term $K$  usually depends
on the temperature or pressure of the phases, as well as the total density,
the density of phase \emph{2} and $\lambda$.
The model is not restrictive with the reaction rate law, therefore other types of reaction rates can also be used.

In this work, all fluid components described by the MiNi16 model are assumed to be governed
by a Mie-Gr\"uneisen equation of state, of the form:
\begin{equation}
\label{MG}
p=p_{{ref}_i}+\rho_i\Gamma_i(e_i-e_{{ref}_i}), \; \mbox{for} \;i=1,\alpha,\beta.
\end{equation}
Material interfaces between the three phases are described by a diffused interface technique. Hence, mixture rules need to be defined for the diffusion zone, relating the thermodynamic properties of the mixture with those of the individual phases. The mixture rules for the specific internal energy, density and adiabatic index ($\gamma$) are:
\begin{equation}
\label{hybrid:mixtureRules_ene}
\rho e=z_1\rho_1e_1+z_2\rho_2e_2=z_1\rho_1e_1+z_2\rho_2(\lambda e_{\alpha}+(1-\lambda)e_{\beta}),
\end{equation}
\begin{equation}
\label{hybrid:mixtureRules_dens}
\rho = z_1\rho_1+z_2\rho_2 \;\; \mbox{with} \;\; \frac{1}{\rho_2} = \frac{\lambda}{\rho_{\alpha}} + \frac{1-\lambda}{\rho_{\beta}}.
\end{equation}
\begin{equation}
\label{hybrid:mixtureRules_gamma}
\xi=z_1\xi_1+z_2\xi_2 \;\; \mbox{with} \;\; \gamma_2=1+\frac{1}{\xi_{2}}=\frac{\lambda
\gamma_{\alpha}C_{v_{\alpha}}+(1-\lambda) \gamma_{\beta}C_{v_{\beta}}}{\lambda
C_{v_{\alpha}}+(1-\lambda) C_{v_{\beta}}},
\end{equation}

where $e_1,e_\alpha,$ and $e_\beta$ denote the specific internal energies of phases \emph{1, $\alpha$} and \emph{$\beta$},  $\xi=1/(\gamma - 1)$, and $C_{v_{\alpha}}$ and $C_{v_{\beta}}$ denote the specific heat at constant volume for phases $\alpha$ and $\beta$.

The model in this form can solve multi-component problems involving two miscible phases (\emph{phase $\alpha$} and \emph{phase $\beta$}) forming a mixture represented as \emph{phase 2} and one inert, immiscible component (\emph{phase 1}). Only one of the phases $\alpha$ and $\beta$ can be reactive.

The soundspeed also follows a mixture rule given as:
 \begin{equation}
\label{Hyb:soundspeed}
\xi c^2=\sum_i y_i\xi_ic_i^2,
\end{equation}
where $c_i$ are the individual soundspeeds of phases 1 and 2 and $c_2$ depends on averaging procedures of energy and density derivatives of phases $\alpha$ and $\beta$. For more information on this as well as for the numerical evaluation of the total equation of state the reader is referred to \cite{michael2016hybrid}.

\subsubsection{Reduced {\color{black}MiNi16} reactive model {\color{black}(Case 2)}}
\label{AllaireReactive}
Suppose that the explicit modelling of the combustion products is not necessary and the fluid system comprises of the reacting explosive and an inert phase, {\color{black} as illustrated in Case 2 of Fig.\ \ref{Lynch}}. The full model then reduces to a two-phase, reacting, five-equation interface model as given by Michael and Nikiforakis \cite{michael2015DetSymp}. In this case, \emph{phase} $\beta$ is dropped and the system comprises of \emph{phase 1} and \emph{phase 2 = phase} ${\alpha}$ (effectively quantities $(\;)_{2}=(\;)_{\alpha}$).

\subsubsection{Reduced {\color{black}MiNi16} inert model {\color{black}(Case 3)}}
\label{AllaireInert}
Suppose that \emph{phase 2} does not comprise a homogeneous mixture of reactant and products, but instead it represents a single, inert constituent (i.e.\ $\lambda=0$ or $\lambda=1$ and $K=0$ everywhere), {\color{black} as illustrated in Case 3 of Fig.\ \ref{Lynch}}. Then, the full model reduces to the  inert five-equation interface model, as given by Allaire et al.\ \cite{allaire2002five} (effectively quantities $(\;)_{2}=(\;)_{\alpha}=(\;)_{\beta}$ and equation (\ref{Hyb:reac}) becomes inactive):  

\begin{eqnarray}
\label{Allaire:cty1}
\frac{\partial z \rho_1}{\partial t}+ \nabla \cdot(z \rho_1 \mathbf{u})  &=& 0 \\
\label{Allaire:cty2}
\frac{\partial (1-z) \rho_2}{\partial t}+\nabla \cdot((1-z) \rho_2 \mathbf{u}) &=& 0, \\
\label{Allaire:mom}
\frac{\partial}{\partial t}(\rho \mathbf{u_i})+\nabla \cdot(\rho \mathbf{u_iu})+\frac{\partial
p}{\partial{\mathbf{x_i}}}&=&0, \\ 
\label{Allaire:ene}
\frac{\partial}{\partial t}(\rho E)+\nabla \cdot (\rho E+p)\mathbf{u}&=&0, \\ 
\label{Allaire:z}
\frac{\partial z }{\partial t} + \mathbf{u} \cdot \nabla z &=& 0.
\end{eqnarray}\\

The mixture rules of the full model also reduce to the ones given by Allaire et al.\ \cite{allaire2002five}.

\subsubsection{Augmented reactive Euler model {\color{black}(Case 4)}}
\label{Banks}
Suppose that the reactant-product mixture is not confined by an inert material (in the limit of $z_1\to0$) and it is considered its own entity, {\color{black} as illustrated in Case 4 of Fig.\ \ref{Lynch}}. The mixture rules then reduce to the mixture rules found in \cite{banks2007high} and the system reduces to the two-component fluid-mixture model (or two-phase, in the context of this work), as described in the same work:

\begin{eqnarray}
\label{Banks:cty}
\frac{\partial \rho}{\partial t}+ \nabla \cdot(\rho \mathbf{u})  &=&0,\\
\label{Banks:mom}
\frac{\partial}{\partial t}(\rho \mathbf{u_i})+\nabla \cdot(\rho \mathbf{u_iu})+\frac{\partial
p}{\partial{\mathbf{x_i}}}&=&0,\\
\label{Banks:ene}
\frac{\partial}{\partial t}(\rho E)+\nabla \cdot (\rho E+p)\mathbf{u}&=&0,\\
\label{Banks:reac}
\frac{\partial (\rho \lambda)}{\partial t}+\nabla \cdot(\rho \mathbf{u}\lambda) &=& K.
\end{eqnarray}

\subsection{The elastoplastic model}

In this work, we use the elastic solid model described by Barton et al. \cite{barton2009exact} based on the formulation by Godunov and Romenskii \cite{godunov2013elements}. The plasticity is included following the work of Miller and Colella \cite{miller2002conservative}.

Consider the steel plate of Fig.\ \ref{Lynch} in isolation. {\color{black}In an Eulerian frame, which we employ here, there is no mesh distortion that can be used to describe the solid material deformation. Thus the material distortion needs to be accounted for in a different way. Here, this is done by defining the elastic deformation gradient as:} 
\begin{equation}
\label{Fe}
F^e_{ij} = \frac{\partial x_i}{\partial X_j},
\end{equation}
which maps the coordinate $\mathbf{X}$ in the initial configuration to the coordinate $\mathbf{x}$ in the deformed configuration.

The state of the solid is characterised by the elastic deformation gradient, velocity $u_i$ and entropy $S$. Following the work by Barton et al.\ \cite{barton2009exact}, the complete three-dimensional system forms a hyperbolic system of conservation laws for momentum, strain and energy:

\begin{eqnarray} \label{cty}
  \frac{\partial{\rho u_i}}{\partial t} + \frac{\partial(\rho u_i u_m - \sigma_{im})}{\partial x_m} &=& 0, \\
  \frac{\partial{\rho E}}{\partial t} + \frac{\partial(\rho u_m E - u_i \sigma_{im})}{\partial x_m} &=& 0, \\
\frac{\partial{\rho F^e_{ij}}}{\partial t} + \frac{\partial(\rho F^e_{ij}u_m-\rho F^e_{m j}u_i)}{\partial x_m} &=& {\color{black} -u_i\frac{\partial \rho F^{e}_{mj}}{\partial x_m} + P_{ij}}, \label{F}\\
  \frac{\partial{\rho \kappa}}{\partial t} + \frac{\partial(\rho u_m
    \kappa)}{\partial x_m} &=& \rho \dot{\kappa},
\end{eqnarray}
with the vector components $\cdot_i$ and tensor components
$\cdot_{ij}$. The first two equations along with the density-deformation gradient relation:
\begin{equation}
\rho = \rho_0 / \text{det}\mathbf{F}^e,
\end{equation}
where $\rho_0$ is the density of the initial unstressed medium,  essentially evolve the solid material hydrodynamically. Here, $\sigma$ is the stress, $E$ the total energy such that $E=\frac{1}{2}|u|^2+e$, with $e$ the specific internal energy and $\kappa$ the scalar material history that tracks the work hardening of the material through plastic deformation. {\color{black} We denote the source terms associated with the plastic update as $P_{ij}$.} 

The system is closed by an analytic constitutive model relating the specific internal energy to the deformation gradient, entropy and material history parameter (if applicable):
\begin{equation}
e = e(\mathbf{F}^e,S,\kappa).
\end{equation}
The stress tensor is given by:
\begin{equation}
\sigma_{ij} = \rho F^e_{im}\frac{\partial e}{\partial F^e_{jm}}. 
\end{equation}
For $\mathbf{F}^e$ to represent a physical deformations, the equations for deformation gradient satisfy three compatibility constraints:
\begin{equation}
\frac{\partial \rho F^e_{kj}}{\partial x_k} = 0, \quad j=1,2,3,  
\end{equation}
which hold true for $t>0$ if true for initial data. This is based on the fact that $\mathbf{F}^e$ is defined as a gradient. 

The deformation is purely elastic until the physical state is evolved beyond the yield surface ($f>0$), which in this work is taken to be:

\begin{equation}
\label{yieldSurface}
f(\boldsymbol{\sigma} )=||\text{dev}\boldsymbol{\sigma}||-\sqrt{\frac{2}{3}}\sigma_Y = 0, \; \text{with} \;\; \text{dev} \boldsymbol{\sigma} =  \boldsymbol{\sigma}-\frac{1}{3}(\text{tr}\boldsymbol{\sigma})I,
\end{equation}
where $\sigma_Y$ is the yield stress and the matrix norm $||.||$ the Shur norm ($||\boldsymbol{\sigma}||^2=\text{tr}(\boldsymbol{\sigma}^T\boldsymbol{\sigma}$)).

As this identifies the maximum yield allowed to be reached by an elastic-only step, {\color{black}a predictor-corrector method is followed to re-map the solid state onto the yield surface.} Assuming that the simulation timestep is small, this is taken to be a straight line, using the associative flow rate 
($\dot{\epsilon}^p=\eta\frac{\partial F}{\partial \sigma}$), satisfying the maximum plastic dissipation principle (i.e.\ the steepest path). In general, this is re-mapping procedure is governed by the dissipation law: 
\begin{equation}
  \psi_{plast} =\boldsymbol{\Sigma} \colon ((\mathbf{F}^p)^{-1}\dot{\mathbf{F}^p}),
\end{equation}
where $\boldsymbol{\Sigma}=\boldsymbol{G}\boldsymbol{\sigma}\boldsymbol{F}$ and $\colon$ is the double contraction of tensors (e.g. $ \boldsymbol{\sigma}:\boldsymbol{\sigma}=\text{tr}(\boldsymbol{\sigma}^T\boldsymbol{\sigma})$).
{\color{black}The initial prediction is $\mathbf{F} = \mathbf{F}^e$ and  $\mathbf{F}^p = \mathbf{I}$, where $\mathbf{F}$ is the specific total deformation tensor and $\mathbf{F}^p$ the plastic deformation tensor that contains the contribution from plastic deformation. This is then relaxed to the yield surface according to the procedure of Miller and Colella  \cite{miller2002conservative}.}  

The explosive and solid mathematical formulations described in this section are solved numerically using high-resolution, shock-capturing, Riemann-problem based methods and structured, hierarchical adaptive mesh refinement, as described in previous work \cite{schoch2013eulerian,schoch2013multi,michael2016hybrid,michael2015DetSymp,michael2010numerical,schoch2013propagation}.

\section{The multi-material approach}

Ghost fluid methods, in combination with level set methods, provide a robust and efficient technique for tracking interfaces and boundaries, such as the material interfaces between solid and fluid materials. In this work, we use level set methods to track the {\color{black}solid-explosive\footnote{Note that by explosive we {\color{black}refer to} any hydrodynamic system modelled by MiNi16 or its reduced systems, including the simultaneous modelling of the reactant, the products and the air in the air-gap scenario} interface. Each component, e.g.\ the solid or the explosive is called a \emph{material} in this framework.} Such methods only give the location of the interface; they do not affect the evolution of the material components. The behaviour of the material components at the interface is modelled by the implementation of dynamical boundary conditions with the aid of the Riemann ghost fluid method and the devise of mixed-material Riemann solvers to solve the interfacial Riemann problems between materials.


A signed distance function $\phi(x,y)$, called the level set function is used, with the zero contour given by $\Phi = {(x,y)|\phi(x,y)=0}$. The sign of $\phi_i$, where $i$ is the cell index, determines which material is present in that cell. The evolution of $\phi(x,y)$, assuming no mass transfer and continuous velocity across the interface, is given by the advection equation:

\begin{equation}
\frac{\partial \phi}{\partial t} + \mathbf{u} \cdot \nabla \phi=0.
\end{equation}

The Riemann ghost fluid method by Sambasivan and Udaykumar \cite{sambasivan2009ghost1,sambasivan2009ghost2}  is utilised to model the behaviour of the material component at the interface. This method, in contrast to the original ghost fluid method \cite{fedkiw1999non}  and the `modified' ghost fluid method \cite{liu2003ghost} uses a Riemann solver to predict ghost-cell states adjacent to the interface. For every cell $i$ adjacent to the interface 
the following procedure is used:

\begin{enumerate}
\item locate the interface within the cell at the point $P=i+\phi\nabla \phi$
\item project two probes into the adjacent materials, reaching the points $P_1=P+\mathbf{n}\cdot\Delta x$ and $P_2=P-\mathbf{n}\cdot\Delta x$
\item interpolate states at each point using information from the surrounding cells 
\item solve a mixed Riemann problem (as described in Sec.\ \ref{Sec:MRS}) between the two states to extract the state of the real-material cells, adjacent to the interface (left star state $\mathbf{W^*_L}$, in Fig.\ \ref{Fig:OSRPallaire}) 
\item replace the state in cell $i$ by the computed star state.
\end{enumerate}

After the above procedure is followed for each material, a fast-marching method is used to fill in the ghost cells for each material.

\subsection{Mixed Riemann solvers}
\label{Sec:MRS}

In this section we describe how the mixed-material Riemann problem at material interfaces is solved (step 4 in the procedure described above). {\color {black} As we are considering five hydrodynamic models and one elastoplastic model, there are a lot of ways of combining them. In these section, we derive mixed Riemann solvers for each pair that can be encountered, although some extensions of the different combinations are straightforward.} The Riemann solver at the material interface takes as input two states from the two different materials that are modelled by different mathematical models, providing one-sided estimations of the interface-adjacent (star) states. These estimations are based on the characteristic equations deduced from the mathematical system describing the left material and by invoking appropriate `boundary conditions' between the two materials at the interface. {\color{black}That is why for each model described in Sec.\ \ref{models} to be coupled with another fluid or solid model, a new mixed Riemann solver has to be derived. In this section, we first describe how the full MiNi16 model and the reduced versions of it are coupled with a simple Euler system and then with a full elastoplastic solid system (including how the elastoplastic system is coupled with the Euler equations). The remainder combinations should be directly deductible from these.}

\subsubsection{Reactive MiNi16 model coupled with the Euler model}
\label{Sec:HybridMRSeuler}

Consider a cell which contains a material boundary. Without loss of generality, we assume that on the left side of the interface lies the material governed by the MiNi16 model of Sec.\ \ref{Hybrid} (hereby called the MiNi16 material) and on the right side of the interface lies a material governed by the Euler equations. Hereforth we assume that we are currently solving for the MiNi16 material (in GFM terminology, the `real' material). At the material boundary a Riemann problem is solved between the left MiNi16 and the right Euler system to provide the star state for the real material. We use a Riemann solver that takes into account the two different materials and all the wave patterns in the MiNi16 system, as described in this section.

\begin{figure}[!t]
\center
\includegraphics[width=0.7\textwidth]{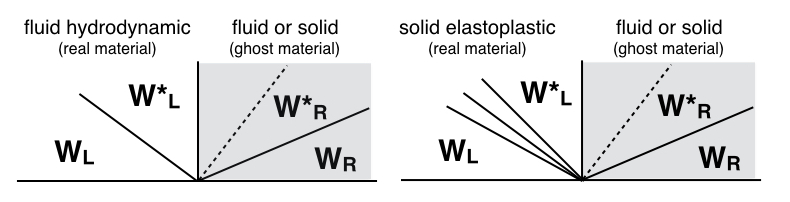}
\caption{{\color{black}The Riemann problem at the material interface between any combination of fluid hydrodynamic and solid elastoplastic materials. The shaded region represents the initial ghost region and the white region the initial 'real' material region in GFM terminology. We are solving for the material in the white, left region. If this is a fluid, the one-sided solution consists one or more degenerate and hence overlapping waves (left) and if it is a solid, the solution can consist with more than one non-overlapping waves (right).}}
\label{Fig:OSRPallaire}
\end{figure}

We write the MiNi16 model described in Sec.\ \ref{Hybrid} by equations (\ref{Hyb:cty1})--(\ref{Hyb:reac}) in primitive form as:
\begin{equation}
\label{HybridPrimitive}
\mathbf{W}_t+\mathbf{A}(\mathbf{W})\mathbf{W}_x =\mathbf{0},
\end{equation}
where
\begin{equation}
\label{HybridJacobian}
\mathbf{W} = \left[
\begin{array}{c}
\rho  \\
Y \\
u \\
w \\
p \\
z \\
\lambda
\end{array} \right], \quad\mathbf{A}(\mathbf{W}) = \left(
      \begin{array}{ccccccc}
        u & 0 & \rho & 0 & 0 & 0 & 0\\
        0 & u & 0 & 0 & 0 & 0 & 0\\
        0 & 0 & u & 0 & 1/\rho & 0 & 0\\
        0 & 0 & 0 & u & 0 & 0 & 0\\
        0 & 0 & \rho c^2 & 0 & u & 0 & 0\\
        0 & 0 & 0 & 0 & 0 & u & 0\\
        0 & 0 & 0 & 0 & 0 & 0 & u\\
      \end{array} \right).
\end{equation}\\
\\
Recall that $\rho = \rho_1z_1+\rho_2z_2$, $z_1+z_2=1$ and $Y_k=\frac{\rho_kz_k}{\rho}$ is the mass fraction\footnote{Note that $Y_k$ is not the same as $\lambda$, which is mass fraction of material $\alpha$ with respect to \emph{phase 2}.} of material $k$, $k=1,2$, with respect to the mixture of phases 1 and 2 (i.e.\ the total fluid). 
\\
\\
The Jacobian matrix $\mathbf{A}(\mathbf{W}) $ has eigenvalues $\mu_1 =\mu_2=\mu_3=\mu_4=\mu_5=u$,
$\mu_6=u-c$ and $\mu_7=u+c$.
\\
\\
The right eigenvectors are
\begin{equation}
\label{HybridEvecsR}
\mathbf{r}_1 = \left(
0 , 0, 0, 0 , 0 , 0, 1 \right)^T, \;
\mathbf{r}_2 = \left(
0 , 0, 0, 0 , 0 , 1, 0 \right)^T, \;
\mathbf{r}_3 = \left(
1  ,0 , 0 , 0 , 0 , 0, 0
 \right)^T,\;
\mathbf{r}_4 = \left(
0  ,0 , 0 , 1 , 0 , 0, 0
 \right)^T,
 \nonumber \end{equation}
 \begin{equation}
\mathbf{r}_5 = \left(
0  ,1 , 0 , 0 , 0 , 0, 0
 \right)^T, \quad
\mathbf{r}_6 = \left(
\rho  ,0 , -c, 0, \rho c^2 , 0, 0  \right)^T,\quad
\mathbf{r}_7 = \left(
\rho  ,0 , c, 0, \rho c^2 , 0, 0 
 \right)^T,
\end{equation}
and the left eigenvectors are
\begin{equation}
\label{HybridEvecsL}
\mathbf{l}_1 = \left(
0 , 0, 0 , 0 , 0 , 0, 1 \right), \;
\mathbf{l}_2 = \left(
0 , 0, 0 , 0 , 0 , 1, 0 \right), \;
\mathbf{l}_3 = \left(
-c^2  ,0 , 0 , 0 , 1 , 0, 0
 \right),\; 
\mathbf{l}_4 = \left(
0  ,0 , 0 , 1 , 0 , 0, 0
 \right),
\nonumber \end{equation}
 \begin{equation}
\mathbf{l}_5 = \left(
0  ,1 , 0 , 0 , 0 , 0, 0
 \right),\quad
\mathbf{l}_6 = \left(
0  ,0 , -\rho c , 0 , 1 , 0, 0
 \right),\quad
\mathbf{l}_7 = \left(
0  ,0 , \rho c , 0 , 1 , 0, 0
 \right).
\end{equation}
\vspace{0.001cm}
As we are solving for the MiNi16 model, we only look for the left star state, $\mathbf{W}^*_L$ (see Fig.\ \ref{Fig:OSRPallaire}). To obtain star values on the left of the interface, we use characteristic equations. 

\vspace{0.001cm}
Characteristics define directions $\frac{dx}{dt}=\mu_j$, in which
\begin{equation}
\mathbf{l}^{(i)}\cdot d\mathbf{W} = 0, \quad \text{where} \quad 
d\mathbf{W}= \left(d\rho , dY , du , dw , dp, dz, d\lambda \right)^T.
\end{equation}
\vspace{0.001cm}
So, along  $\frac{dx}{dt}=\mu_1=u$:
\begin{equation}
 \beta_1\left(
0 , 0, 0 , 0 , 0 , 1 \right)\cdot\left( d\rho , dY , du , dw , dp , dz \right)^T=0, 
\end{equation}
giving,
\begin{equation}
\label{HybridCharacteristics1}
\qquad \; \; d\lambda=0
\end{equation}
Applying the same along  $\frac{dx}{dt}=\mu_j=u$ for $j=2,3,4,5$ we obtain, respectively:
\begin{eqnarray}
\label{HybridCharacteristics2} 
dp-\color{black}c^2\color{black}d\rho&=&0,\\
\label{HybridCharacteristics2b}
dw&=&0\\
\label{HybridCharacteristics2c}
dY&=&0,\\
\label{HybridCharacteristics2d}
dz&=&0.
\end{eqnarray}
Finally, along  $\frac{dx}{dt}=\mu_6=u-c$ and $\frac{dx}{dt}=\mu_7=u+c$, we obtain:
\begin{eqnarray}
\label{HybridCharacteristics3}
dp-\color{black}\rho c\,\color{black}du&=&0 \\
\label{HybridCharacteristics3b}
dp+\color{black}\rho c\,\color{black}du&=&0
\end{eqnarray}

For a single phase described by the ideal gas or stiffened gas equation of state, the characteristic equations can be integrated directly, as the expressions for the sound speed are simple. However, for more complex equations of state or MiNi16 formulations where the sound speed is the `mixture' of the individual phase sound speeds (see equation (\ref{Hyb:soundspeed})) this might not be possible. In such case, one can obtain an approximate mixed Riemann solver by replacing the differentials with the difference of the initial and the final state without integrating (i.e.\ across the characteristics), as presented below.
\\
\\
Using $dp +  \rho c\, du=0$, we connect the states $\mathbf{W}^*_L$ and $\mathbf{W}_L$ to obtain:
\begin{equation}
\label{eqLhybrid}
p^*_L-p_L = - \rho_L c_L(u^*_L - u_L).
\end{equation}
Using $dp -  \rho c\, du=0$, we connect the states $\mathbf{W}^*_R$ and $\mathbf{W}_R$ to obtain:
\begin{equation}
\label{eqRhybrid}
p^*_R-p_R = \rho_R c_R(u^*_R - u_R).
\end{equation}
Pressure and velocity don't change across the material interface, hence $p^*_L=p^*_R=p^*$
and $u^*_L=u^*_R=u^*$. Applying these conditions to (\ref{eqLhybrid}) and (\ref{eqRhybrid})
we obtain two expressions for $p^*$ and $u^*$:
\begin{equation}
\label{ustarpstarHybrid}
p^* = p_L - \color{black}\rho_L c_L\color{black}(u^*-u_L) \quad \text{and} \quad 
u^* = \frac{p^*-p_R}{\color{black}\rho_Rc_R} + u_R.
\end{equation}
Solving the above two simultaneously, we obtain an expression for the pressure in the star region:
\begin{equation}
\label{pStarHybrid}
{\color{black}p^*} = \frac{\color{black}C_R{\color{black}p_L} - \color{black}C_L{\color{black}p_R}-\color{black}C_RC_L{\color{black}(u_R-u_L)}}{\color{black}C_R+C_L},
\end{equation}
where $\color{black}C_L = \rho_Lc_L$ and $\color{black}C_R=\rho_Rc_R$.

To calculate the left MiNi16 fluid state, connect states $\mathbf{W}^*_L$ and $\mathbf{W}_L$, using equation (\ref{eqLhybrid}) and $dp-c^2d\rho=0$ to obtain:
\begin{eqnarray}
\label{ustarHybrid}
\color{black}u^* &=& \frac{p_L-p^*}{\color{black}\rho_Lc_L} + u_L\color{black} \quad \text{and}\\
\label{rhostarHybrid}
 \color{black}\rho^*_L &=& \frac{p^*-p_L}{\color{black}c_L^2} + \rho_L\color{black}.
\end{eqnarray}
Using the remaining characteristic equations we obtain:
\begin{eqnarray}
\label{otherStarHybrid}
w^*&=&w_L,\\
\label{otherStarHybridb}
\lambda^*&=&\lambda_L,\\
\label{otherStarHybridc}
Y_1^*&=&Y_{1_L}\\
\label{otherStarHybridd}
z^* &=& z_L.
\end{eqnarray} 
Using the above and the definition for $Y_1=\frac{z_1\rho_1}{\rho}$ we get $Y^*=\frac{z^*\rho_{1_L}^*}{\rho^*_L}$ and so:
\begin{eqnarray}
\label{rho1StarHybrid}
 \rho^*_{1_L}&=&\frac{Y_{1_L} \rho^*_L}{z_L}\\
\label{rho2StarHybrid}
\rho^*_{2_L}&=&\frac{Y_{2_L}\rho^*_L}{1-z_L}.
\end{eqnarray}
Equations (\ref{pStarHybrid})--(\ref{rho2StarHybrid}) give the full state in the left star region.  The values of $C_L,C_R$ and $c^2_L$ in equations (\ref{pStarHybrid})--(\ref{rhostarHybrid}) 
are constant approximations of the $(\;)_L$ value. Alternatively, an iterative method as described in Sec.\ \ref{Sec:iterative} can be used to compute $p^*$ and extract the parameters for the remaining variables thereafter, using equations  (\ref{pStarHybrid})--(\ref{rho2StarHybrid}). 

If the real fluid is described by the reduced two-phase reactive system given in Sec.\ \ref{AllaireReactive} then the same procedure as here is used, as the same governing equations (and hence characteristic equations) hold.

\subsubsection{Iterative extension}

\label{Sec:iterative}
To improve accuracy of all of the explosive-solid mixed Riemann solvers, in the expressions for $p^*,u^*$ and $\rho^*$, the expressions for $c_L^2$ and $\rho_Lc_L$ (denoted now as $\overline{c_L}^2$ and $\overline{\rho_Lc_L}$) can approximated not just by the $(\;)_L$ state but by an average between the original and predicted star states, i.e.\  $\overline{c_L}^2=\frac{1}{2}(c^2_L+{c^*}^2_L)$ and  $\overline{\rho_Lc_L}=\frac{1}{2}(\rho_Lc_L+\rho^*_Lc^*_L)$. This can be considered as an average between the material interior state and the interface state. For other options for constructing the $\overline{(\; )}$ state see Schoch et al.\ \cite{schoch2013eulerian}.

This, however generates an implicit problem; the unknown interface state now depends on itself, i.e.\ $\overline{\mathbf{W}}=\overline{\mathbf{W}}({\mathbf{W_L},\mathbf{{W^*}_L}})$. A predictor-corrector method is used to iterate through and repeatedly update the star states until convergence (based on $p^*$). The initial guesses for the iteration process are the primitive states of the two materials.

\subsubsection{Reduced MiNi16 inert model coupled with the Euler model}
\label{Sec:AllaireMRSeuler}

Suppose now that the material on the left side of the interface is governed by the MiNi16 inert model of Sec.\ \ref{AllaireInert} (the `real material), using two phases only for demonstration purposes, while the material on the right side is governed by the Euler equations. 

We write the two-phase, inert fluid model in primitive form as given by equation (\ref{HybridPrimitive}), with
\begin{equation}
\label{AllaireJacobian}
\mathbf{W} = \left[
\begin{array}{c}
\rho  \\
Y \\
u \\
w \\
p \\
z 
\end{array} \right], \quad \mathbf{A}(\mathbf{W}) = \left(
      \begin{array}{cccccc}
        u & 0 & \rho & 0 & 0 & 0\\
        0 & u & 0 & 0 & 0 & 0 \\
        0 & 0 & u & 0 & 1/\rho & 0 \\
        0 & 0 & 0 & u & 0 & 0 \\
        0 & 0 & \rho c^2 & 0 & u & 0\\
        0 & 0 & 0 & 0 & 0 & u
      \end{array} \right).
\end{equation}\\
\\

The Jacobian matrix $\mathbf{A}(\mathbf{W}) $ has eigenvalues $\mu_1 =\mu_2=\mu_3=\mu_4=u$,
$\mu_5=u-c$ and $\mu_6=u+c$.
\\
\\
The right eigenvectors are
\begin{equation}
\label{AllaireEvecsR}
\mathbf{r}_1 = \left(
0 , 0, 0, 0 , 0 , 1 \right)^T, \quad
\mathbf{r}_2 = \left(
1  ,0 , 0 , 0 , 0 , 0
 \right)^T,
\mathbf{r}_3 = \left(
0  ,0 , 0 , 1 , 0 , 0
 \right)^T,\quad 
\mathbf{r}_4 = \left(
0  ,1 , 0 , 0 , 0 , 0
 \right)^T, \quad
 \nonumber \end{equation}
 \begin{equation}
\mathbf{r}_5 = \left(
\rho  ,0 , -c, 0, \rho c^2 , 0  \right)^T,\quad
\mathbf{r}_6 = \left(
\rho  ,0 , c, 0, \rho c^2 , 0 
 \right)^T,
\end{equation}
and the left eigenvectors are
\begin{equation}
\label{AllaireEvecsL}
\mathbf{l}_1 = \left(
0 , 0, 0 , 0 , 0 , 1 \right), \quad
\mathbf{l}_2 = \left(
-c^2  ,0 , 0 , 0 , 1 , 0
 \right),\quad 
\mathbf{l}_3 = \left(
0  ,0 , 0 , 1 , 0 , 0
 \right), \quad
\mathbf{l}_4 = \left(
0  ,1 , 0 , 0 , 0 , 0
 \right),\nonumber \end{equation}
 \begin{equation}
\mathbf{l}_5 = \left(
0  ,0 , -\rho c , 0 , 1 , 0
 \right),\quad
\mathbf{l}_6 = \left(
0  ,0 , \rho c , 0 , 1 , 0
 \right).
\end{equation}
\vspace{0.001cm}

Thus, the new system is directly reduced from the full MiNi16 system and the same procedure can be used to derive characteristic equations and connect states across characteristics to obtain left star values. As the reduced system does not contain a $\lambda-$equation, the characteristic equations are given by equations (\ref{HybridCharacteristics2})--(\ref{HybridCharacteristics3b}) and the left star states by equations (\ref{pStarHybrid})--(\ref{otherStarHybrid}) and (\ref{otherStarHybridc})--(\ref{rho2StarHybrid}).
The iterative method can be used as before for better approximations of $C_L,C_R$ and $c^2_L$.

\subsubsection{Augmented Euler reactive model coupled with the Euler model}
\label{Sec:BanksMRSeuler}

Suppose now that the material on the left side of the interface is governed by the augmented Euler reactive model of Sec.\ \ref{Banks}, while the material on the right side is governed by the Euler equations. 

We write the reactive fluid model in primitive form as given by equation (\ref{HybridPrimitive})
with
\begin{equation}
\label{BanksJacobian}
\mathbf{W} = \left[
\begin{array}{c}
\rho  \\
u \\
w \\
p \\
\lambda 
\end{array} \right], \quad \mathbf{A}(\mathbf{W}) = \left(
      \begin{array}{cccccc}
        u & \rho & 0 & 0 & 0\\
        0 & u & 0 & 1/\rho & 0 \\
        0 &  0 & u & 0 & 0 \\
        0 & \rho c^2 & 0 & u & 0\\
        0 & 0 & 0 & 0 & u
      \end{array} \right).
\end{equation}\\
\\
The Jacobian matrix $\mathbf{A}(\mathbf{W}) $ has eigenvalues $\mu_1 =\mu_2=\mu_3=u$,
$\mu_4=u-c$ and $\mu_5=u+c$.
\\
\\
The right eigenvectors are
\begin{equation}
\label{BanksEvecsR}
\mathbf{r}_1 = \left(
1  ,0 , 0 , 0 , 0
 \right)^T, \quad
\mathbf{r}_2 = \left(
0  ,0 , 1 , 0 , 0
 \right)^T,\quad 
\mathbf{r}_3 = \left(
0 , 0, 0, 0 , 1 \right)^T, \quad
 \nonumber \end{equation}
 \begin{equation}
\mathbf{r}_4 = \left(
1 , -c, 0, \rho c^2 , 0  \right)^T,\quad
\mathbf{r}_5 = \left(
1 , c, 0, \rho c^2 , 0 
 \right)^T,
\end{equation}
and the left eigenvectors are
\begin{equation}
\label{BanksEvecsL}
\mathbf{l}_1 = \left(
-c^2  ,0 , 0 , 1 , 0
 \right),\quad 
\mathbf{l}_2 = \left(
0 , 0 , 1 , 0 , 0
 \right), \quad
\mathbf{l}_3 = \left(
0, 0 , 0 , 0 , 1 \right),\nonumber \end{equation}
 \begin{equation}
\mathbf{l}_5 = \left(
0  , -\rho c , 0 , 1 , 0
 \right),\quad
\mathbf{l}_6 = \left(
0  ,\rho c , 0 , 1 , 0
 \right).
\end{equation}
\vspace{0.001cm}
Thus, the new system is directly reduced from the full MiNi16 system and the same procedure can be used to derive characteristic equations and connect states across characteristics to obtain left star values. As the reduced system does not contain a $z-$equation, the characteristic equations are given by equations (\ref{HybridCharacteristics1})--(\ref{HybridCharacteristics2b}) and (\ref{HybridCharacteristics3})--(\ref{HybridCharacteristics3b}) and the left star states by equations (\ref{pStarHybrid})--(\ref{otherStarHybridb}) and (\ref{rho1StarHybrid})--(\ref{rho2StarHybrid}).

\subsubsection{Solving for the Euler system}
When we are solving for the Euler system (i.e.\ the Euler fluid is the `real' fluid), irrespective of which system is on the right side of the interface, we repeat the above process and only compute the quantities ($p^*,u^*,w^*$ and $\rho^*_L$).

\subsubsection{MiNi16 model coupled with the elastoplastic solid model}
\label{Sec:HybridMRSsolid}
Consider a cell which contains a material boundary. Without loss of generality, we assume that on the left side of the interface lies the material governed by the MiNi16 equations (the muti-phase material) and on the right side of the interface lies a material governed by the elastoplastic solid equations (the solid material). We develop a Riemann solver that takes into account the two different materials to determine the star state in the fluid material.

We follow a similar procedure to that in Sec.\ \ref{Sec:HybridMRSeuler}. Referring to Fig.\ \ref{Fig:OSRPallaire}, $\mathbf{W}_L$ corresponds to the original MiNi16 state, $\mathbf{W}_R$ to the original elastoplastic state and $\mathbf{W}^*_L$ to the MiNi16 star state that we are looking to compute in this Riemann solver. Since we are solving for the fluid as the real material, the Riemann problem still has three types of waves (two non-linear and four overlapping linear). The same characteristic relations (\ref{HybridCharacteristics1})--(\ref{HybridCharacteristics3b}) are defined as before and we use the approach of representing the differentials with the state difference.
 Connecting fluid states  $\mathbf{W}^*_L$ and $\mathbf{W}_L$ using  $dp +  \rho c\, du=0$ and solid states $\mathbf{W}^*_R$ and $\mathbf{W}_R$ using  $dp -  \rho c\, du=0$ we obtain a mixed-material expression for $p^*$:
\begin{equation}
\label{pstarHybridSolid}
p^*=\frac{{u_S-u_F + \frac{1}{\rho_S}(\mathbf{Q}^{-1}\mathcal{D}^{-1}\mathbf{Q})^S_{11}\sigma^S_{11}+\frac{p_F}{\rho_Fc_F}}}{\frac{1}{\rho_Fc_F}-\frac{1}{\rho_S}(\mathbf{Q}^{-1}\mathcal{D}^{-1}\mathbf{Q})^S_{11}\sigma^S_{11}},
\end{equation}
where $\mathbf{Q}$ is an orthogonal matrix and $\mathcal{D}$ is the diagonal matrix of positive eigenvalues for the solid system, such that the acoustic tensor is defined by:
\begin{equation}
\Omega_{ij}=\frac{1}{\rho}\frac{\partial \sigma_{1i}}{\partial F_{jk}}{\mathbf{F}}_{1k}=\mathbf{Q}^{-1}\mathcal{D}^{-1}\mathbf{Q}.
\end{equation}
Considering $p_R=\sigma^S_{11},(\mathbf{Q}^{-1}\mathcal{D}^{-1}\mathbf{Q})^S_{11}=1/c_R$ and $C_R = \rho_Rc_R$ and $C_L=\rho_Lc_L$ we obtain equation (\ref{pStarHybrid}).

Then, using $dp+c^2d\rho=0$ to  connect states $\mathbf{W}^*_L$ and $\mathbf{W}_L$ we obtain equations (\ref{ustarHybrid})--(\ref{rhostarHybrid}) and values for $u^*_L$ and $\rho^*_L$, using the remaining characteristic equations we obtain equations (\ref{otherStarHybrid})--(\ref{otherStarHybridd}) and values for $w^*,\lambda^*, Y_k*$ and $z^*$ and using the definition of $Y_k$ we obtain $\rho^*_k$ via equations (\ref{rho1StarHybrid})--(\ref{rho2StarHybrid}). At the interface, we apply conditions
\begin{equation}
\label{slipFluid}
p^*_L = \sigma^*_{R,11}, \quad u^*_L=u^*_R.
\end{equation}
To improve accuracy, an iterative approach as described in Sec.\ \ref{Sec:iterative} is used.

If the real fluid is described by the reduced two-phase reactive system given in Sec.\ \ref{AllaireReactive} then the same procedure as here is used, as the same governing equations (and hence characteristic equations) hold.

\subsubsection{Reduced MiNi16 inert model coupled with the elastoplastic solid model}

Suppose now that on the left side of the interface lies the material governed by the MiNi16 inert equations of Sec.\ \ref{AllaireInert} (here we use two phases only here for convenience) and on the right side of the interface still lies a material governed by the elastoplastic solid equations. We follow a similar procedure to that in Sec.\ \ref{Sec:HybridMRSsolid} to determine the star state ($\mathbf{W}^*_L$) in the fluid material.

 The characteristic relations for this system are defined by equations (\ref{HybridCharacteristics2})--(\ref{HybridCharacteristics3b}). By connecting the appropriate states we obtain the mixed material expression for $p^*$ given by equation (\ref{pstarHybridSolid}) and the remaining star states by equations (\ref{ustarHybrid})--(\ref{otherStarHybrid}) and (\ref{otherStarHybridc})--(\ref{rho2StarHybrid}). The difference now is that the $(\;)_L$ quantities come from the two-phase inert model rather than the full MiNi16 model. At the interface, we apply the conditions given by (\ref{slipFluid}) as before.

As with the MiNi16-elastoplastic coupling, the expressions for $\overline{c_F}^2$ and $\overline{\rho_Fc_F}$ can be taken to be averages of the original and predicted star states and hence a predictor-corrector method is required to update the star states until convergence (based on $p^*$).

\subsubsection{Augmented reactive Euler model coupled with the elastoplastic solid model}

If now that on the left side of the interface lies the material governed by the augmented reactive Euler equations of Sec.\ \ref{Banks} and on the right side of the interface still lies a material governed by the elastoplastic solid equations. We follow a similar procedure to that in Sec.\ \ref{Sec:HybridMRSsolid} to determine the star state ($\mathbf{W}^*_L$) in the fluid material.

 The characteristic relations for this system are defined by equations (\ref{HybridCharacteristics1})--(\ref{HybridCharacteristics2b}) and (\ref{HybridCharacteristics3})--(\ref{HybridCharacteristics3b}). By connecting the appropriate states we obtain the mixed material expression for $p^*$ given by equation (\ref{pstarHybridSolid}) and the remaining star states by equations (\ref{ustarHybrid})--(\ref{otherStarHybridb}) and (\ref{rho1StarHybrid})--(\ref{rho2StarHybrid}). The difference now is that the $(\;)_L$ quantities come from the reactive augmented Euler model rather than the full MiNi16 model. At the interface, we apply the conditions given by (\ref{slipFluid}) as before.

As with the MiNi16-elastoplastic coupling, the expressions for $\overline{c_F}^2$ and $\overline{\rho_Fc_F}$ can be taken to be averages of the original and predicted star states and hence a predictor-corrector method is required to update the star states until convergence (based on $p^*$).

\subsubsection{Solving for the elastoplastic solid model}

Suppose we again have a cell interface which is a material boundary but on the left side of the interface lies an elastoplastic solid material. On the right side of the interface lies a fluid material which can be a MiNi16 material or any of its reduced versions (including the unaugmented Euler equations). We follow a similar procedure as with the explosive-fluid mixed Riemann solvers. We give a brief outline on the derivation of the solid approach and refer the reader to Barton et al.\ \cite{barton2011conservative} for more details. 

The aim here is to determine the star state for the solid cell adjacent to the solid-fluid interface. The elastoplastic solid model can be written in primitive, quasilinear form for primitive variables $\mathbf{W}=(\mathbf{u},\mathbf{F}^{{e}^T},S)$:

\begin{equation}
\label{SolidPrimitive}
\mathbf{W}_t+\mathbf{A}(\mathbf{W})\mathbf{W}_x =\mathbf{0},
\end{equation}
where
\begin{equation}
\label{SolidJacobian}
\mathbf{W} = \left[
\begin{array}{c}
u  \\
\mathbf{F}^T \mathbf{e}_1\\
\mathbf{F}^T \mathbf{e}_2 \\
\mathbf{F}^T \mathbf{e}_3 \\
S
\end{array} \right], \quad \mathbf{A}(\mathbf{W}) = \left(
      \begin{array}{ccccc}
        u_1I & -A^{\alpha 1} & -A^{\alpha 2} & -A^{\alpha 3} & -B^{\alpha} \\
        -\mathbf{F}^{{e}^T}E_{11} & u_{\alpha}I & 0 & 0 & 0\\
        -\mathbf{F}^{{e}^T}E_{12} & 0 & u_{\alpha}I & 0 & 0\\
        -\mathbf{F}^{{e}^T}E_{13} & 0 & 0 & u_{\alpha}I & 0\\
         0 & 0 & 0 & 0 & u_{\alpha}I\\
      \end{array} \right),
\end{equation}\\
where $E_{ij}$ represents the unit dyads $E_{ij}=\mathbf{e}_i\times\mathbf{e}_j^T$, $I$ is the identity matrix and 
\begin{equation}
A^{\alpha\beta}_{ij}=\frac{1}{\rho}\frac{\partial \sigma_{\alpha i}}{\partial F^e_{\beta j}}, B^{\alpha}_i=\frac{1}{\rho}\frac{\partial \sigma_{\alpha i}}{\partial S}.
\end{equation}
The eigenvalues and corresponding right and left eigenvectors are computed and used to obtain characteristic relations, allowing for the computation of the primitive elastic star state (the plastic deformation gradient $\mathbf{F}^p$ is 
not computed at this point):
\begin{equation}
\label{starSolid}
\mathbf{W}^*_L = \mathbf{W}_L+\frac{1}{\rho_L}\sum_{\mu_k>u_L}(\sigma^*_{L,1i}-\sigma_{L,1i})\hat{\mathbf{r}}_L\mathbf{e}_k,
\end{equation}
 where $\hat{\mathbf{r}}$ are the right eigenvectors of the elastic system. 

 At the material interface, slip conditions are used which specify the continuity of the normal components of velocity and traction and zero tangential stresses in the solid:
\begin{equation}
\label{slip}
\sigma^*_{L,11} = -p^*_R, \quad u^*_L=u^*_R, \quad \sigma^*_{L,12}=\sigma^*_{L,13}=0.
\end{equation}

A more accurate star state can be determined if an iterative procedure is used. Starting with initial guesses for the left and right star states the primitive solid and the primitive fluid states, the star state computation described above is iterated until convergence of $p^*$.

\section{Validation of the {\color{black}elastoplastic} component}

In this section, the implementation of the solid-solid formulation is validated in isolation, without the influence of the fluid formulation or solid-fluid mixed Riemann solvers.

\subsection{Aluminium plate impacting an aluminium target}
\label{Sec:HowellBallTest}

In this test, a projectile block of aluminium  moving at 400m/s impacts a stationary target block of aluminium. This was originally presented by Howell and Ball \cite{howell2002free}. A domain of size $[-0.6,2.4]\SI{}{\cm}\times[-2,2]\SI{}{\cm}$ is considered. The projectile initially spans $(x,y)\in [-0.5,0.0]\SI{}{\cm}\times[-0.6,0.6]\SI{}{\cm}$ and the target plate  $(x,y)\in [0,2.2]\SI{}{\cm}\times[-1.7,1.7]\SI{}{\cm}$. The remainder of the domain contains vacuum to allow for unrestricted free surface movement. Both plates are made of the same type of aluminium, modelled by the {\color{black}Romenskii hyperelastic equation of state \cite{titarev2008musta}, consisting of a two-term hydrodynamic component and a one-term shear deformation component. The full equation is given by:
\begin{equation}
\epsilon(I_1,I_2,I_3,S) = \frac{K_0}{2\alpha^2}\left(I_3^{\alpha/2}-1\right)^2+ c_v T_0 I_3^{\gamma/2}\left(e^{S/c_v}-1\right) +\frac{B_0}{2}I_3^{\beta/2}\left(\frac{I_1^2}{3}-I_2\right),
\label{RomenskiiElastic}
\end{equation}
where $I_1,I_2,I_3$ are the invariants of the Finger tensor, $K_0=c_0^2 - \frac{4}{3}b_0^2$ is the squared bulk speed of sound, $B_0$ is the reference shear wave speed, $c_v$ is the specific heat capacity at constant volume, $T_0$ is the reference temperature, $\alpha, \gamma$ are exponents determining the the non-linear dependence of the sound speed and temperature on density respectively and $\beta$ an exponent determining the non-linear dependence of this shear wave speed on density.}

The constitutive model parameters for the aluminium considered here are given in Table \ref{eosTable}. Perfect plasticity is assumed, with a yield stress of $\SI{0.4}{\giga \pascal}$. Between the two aluminium plates, a slip condition is assumed. The simulation is performed at an effective resolution of $\Delta x = \Delta y = \SI{50}{\micro \meter}$ 
up to a final time of $\SI{5}{\micro \second}$. 

\begin{table}[!h]
\centering
 \begin{tabular}{c c c c c c c c c}\hline 
\textbf{Hyperelastic and} & $\rho_0$   & $c_v$   & $T_0$  & $\alpha$ & $\Gamma_0$ & $b_0$  & $c_0$  & $\beta$ \\ 
\textbf{shear parameters} & [\SI{}{\kilogram \per \meter \tothe{3}}] &  [\SI{}{\joule \per \kilogram \per \kelvin}] &  [\SI{}{\kelvin}] & - & - &  $[\SI{}{\meter \per \second}]$ & $[\SI{}{\meter \per \second}]$ & - \\ \hline 
Aluminium & 2710 & 900 & 300 &  1 &   2.088 &  3160 &  6220 &  3.577  \\
Copper & 8930 & 390 & 300 &  1 &   2 &  2100 &  4600 &  3  \\\hline 
\end{tabular}
\caption {Romenskii equation of state parameters the elastoplastic solid materials used in this work.}
\label{eosTable}
\end{table}

\begin{figure}[!htb]
\begin{minipage}{\columnwidth}
        \centering
        \begin{subfigure}[b]{0.5\textwidth}
                \includegraphics[width=\textwidth]{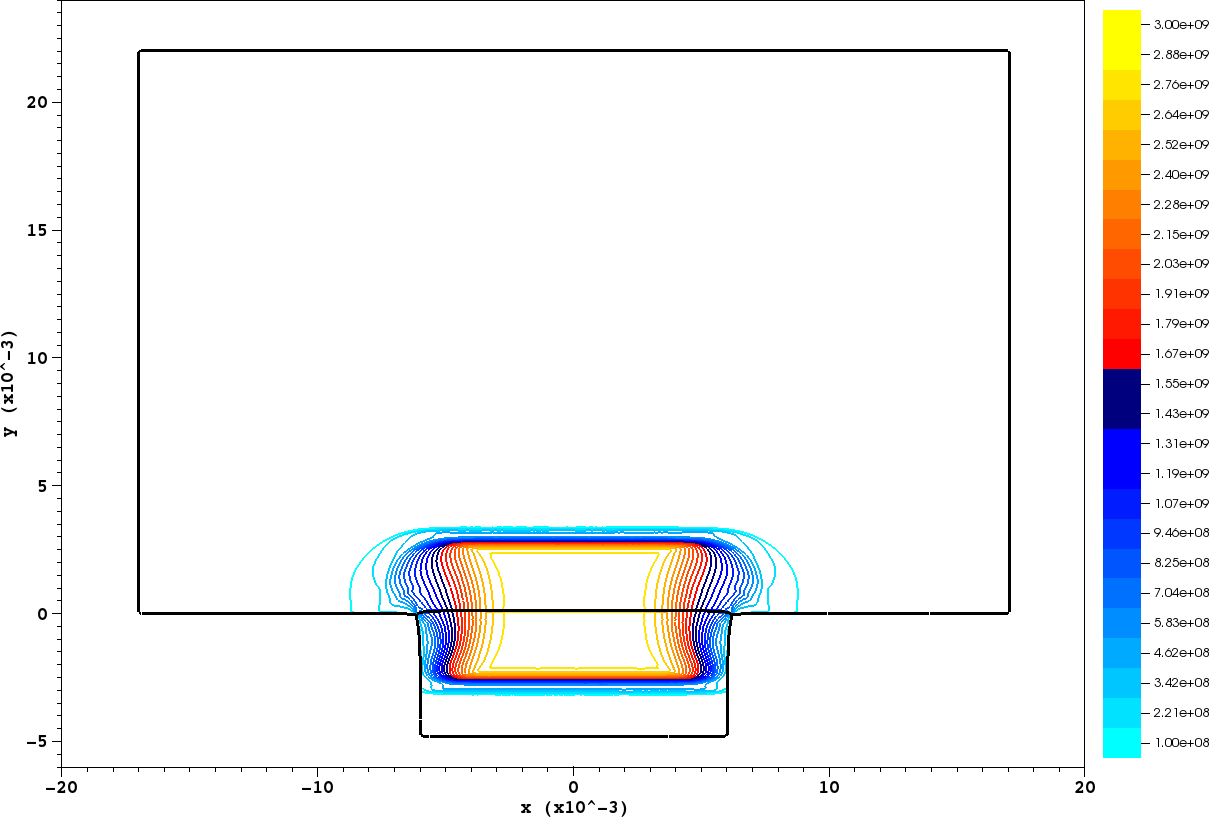}
                \label{ALonALnum}
        \end{subfigure}%
        \begin{subfigure}[b]{0.5\textwidth}
                \includegraphics[width=\textwidth]{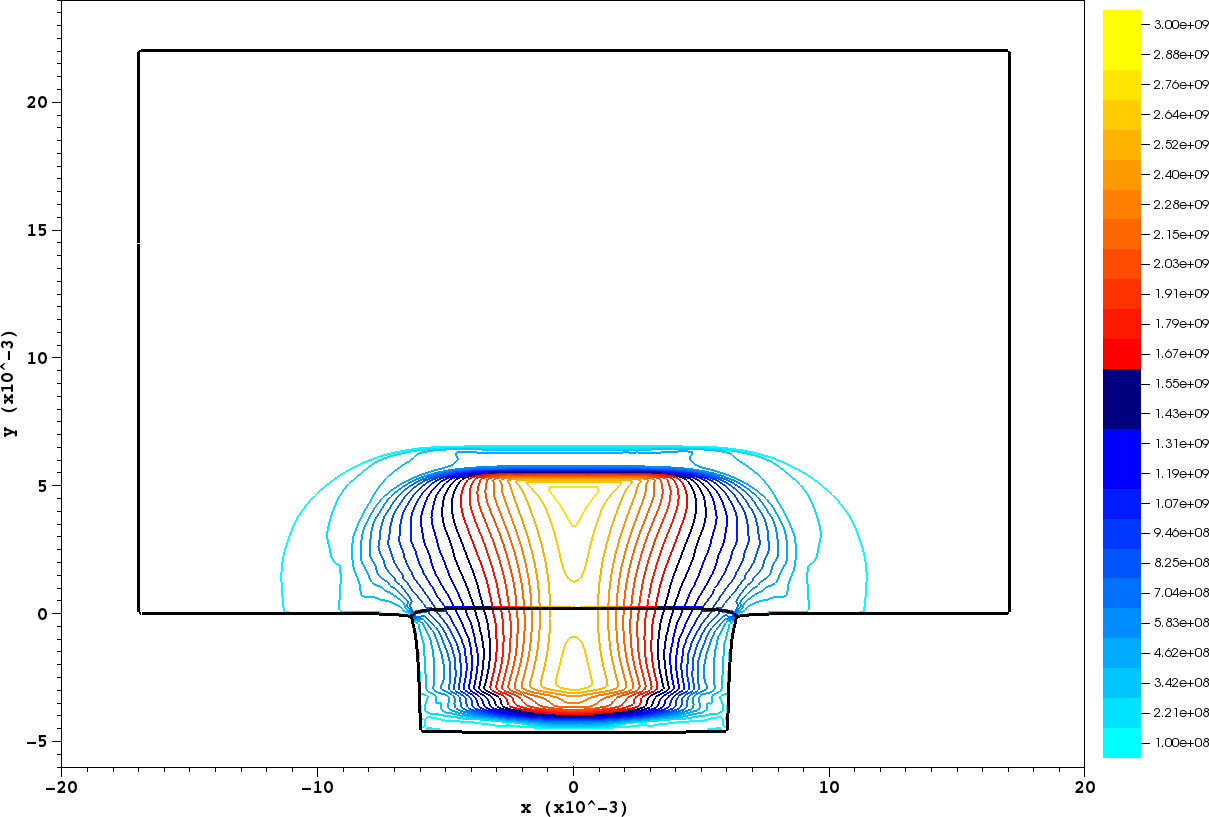}
                \label{ALonALexp}
        \end{subfigure}
\end{minipage}
\begin{minipage}{\columnwidth}
        \centering
        \begin{subfigure}[b]{0.5\textwidth}
                 \includegraphics[width=\textwidth]{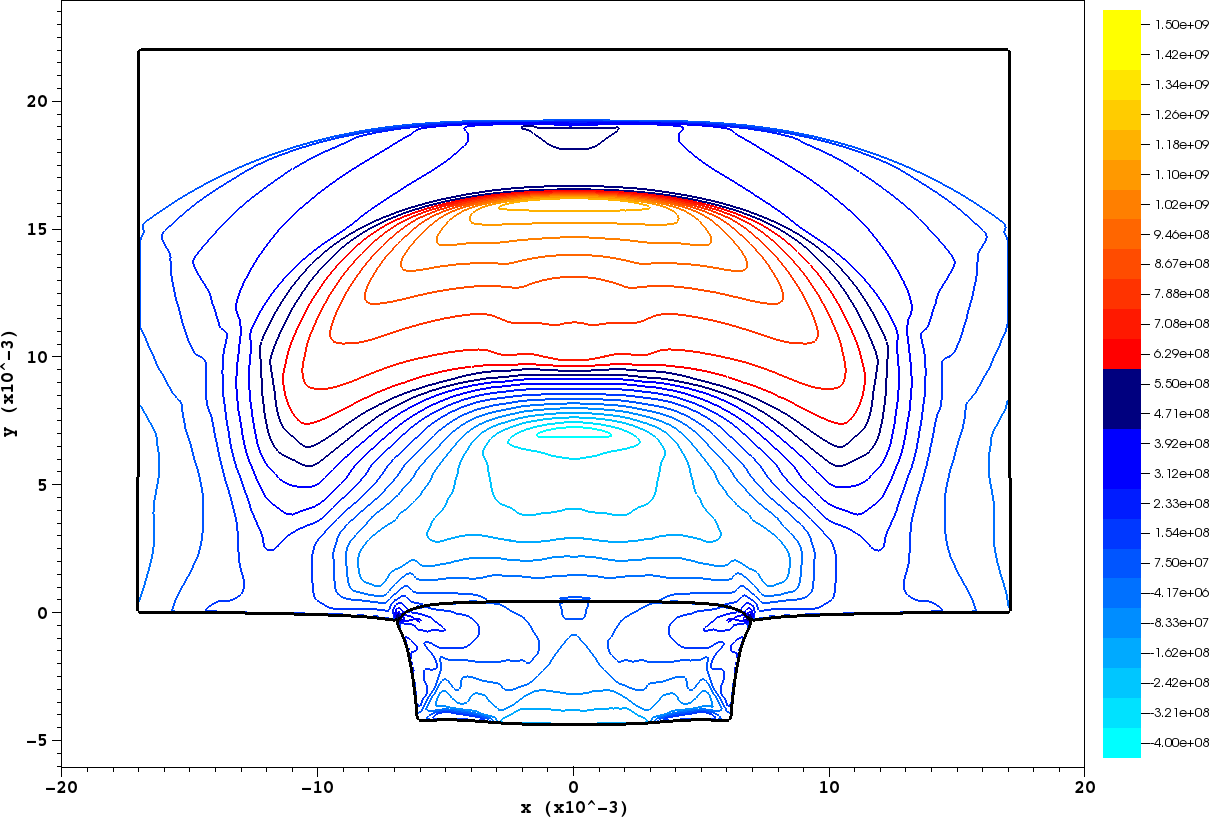}
        \end{subfigure}%
        \begin{subfigure}[b]{0.5\textwidth}
                \includegraphics[width=\textwidth]{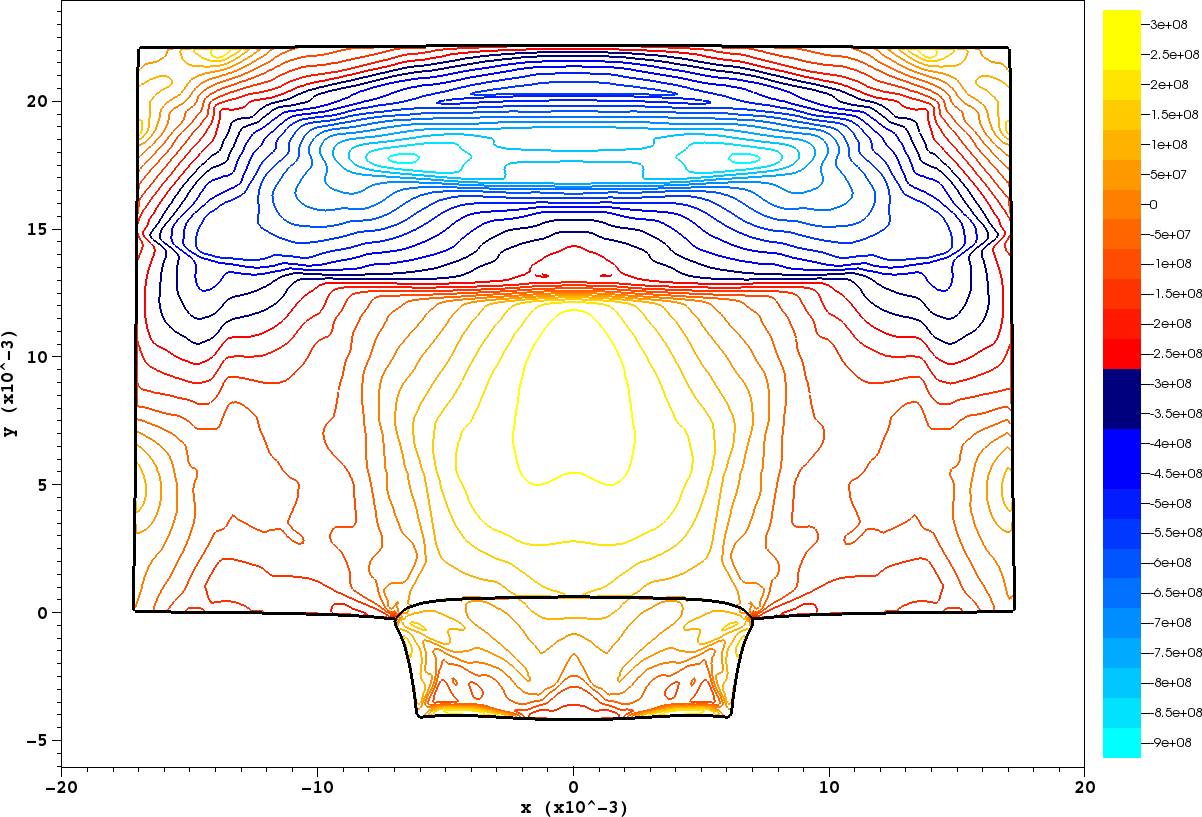}
        \end{subfigure}
\end{minipage}
\caption{Pressure contours in the aluminium projectile and aluminium target at times $t=0.5,1,3$ and $\SI{5}{\micro \second}$, for the test problem described in Sec.\ \ref{Sec:HowellBallTest}.}
\label{HowellBall} 
\end{figure}

Fig.\ \ref{HowellBall} illustrates the computed wave structure in the two plates at times $t=0.5,1,3$ and $\SI{5}{\micro \second}$. Upon impact, shock waves are generated that travel {\color{black}upwards} into the projectile and downwards into the target, as seen in Fig.\ \ref{HowellBall}(a). {\color{black}These waves are of the same strength} since the impact is symmetric (projectile and target plate are made of the same material). In Fig.\ \ref{HowellBall}(b), the shock wave travelling in the target plate is seen to have split into an elastic precursor and a trailing plastic wave. {\color{black}The shock wave travelling into the projectile reaches the rear end of the plate, where it interacts with the solid/vacuum interface. This interaction generates a release wave travelling backwards into the projectile. It then crosses into the target plate generating a region of high tensile stress in the $x$-direction.} This  weakens the plastic wave traversing the target (Fig.\ \ref{HowellBall}(c)). By $t=\SI{4}{\micro \second}$, the elastic wave reaches the rear of the target plate, generating a downwards-moving release wave. Subsequently, this release wave interacts with the still rightward-travelling plastic wave, producing a new set of waves that travel upwards and downwards and continue reflecting on the rear end of the target plate. The deformation of the two plates is also apparent in the sequence of Fig.\ \ref{HowellBall}. 

\begin{figure}[!t]
\begin{minipage}{\columnwidth}
        \centering
        \begin{subfigure}[b]{0.5\textwidth}
              \includegraphics[width=\textwidth]{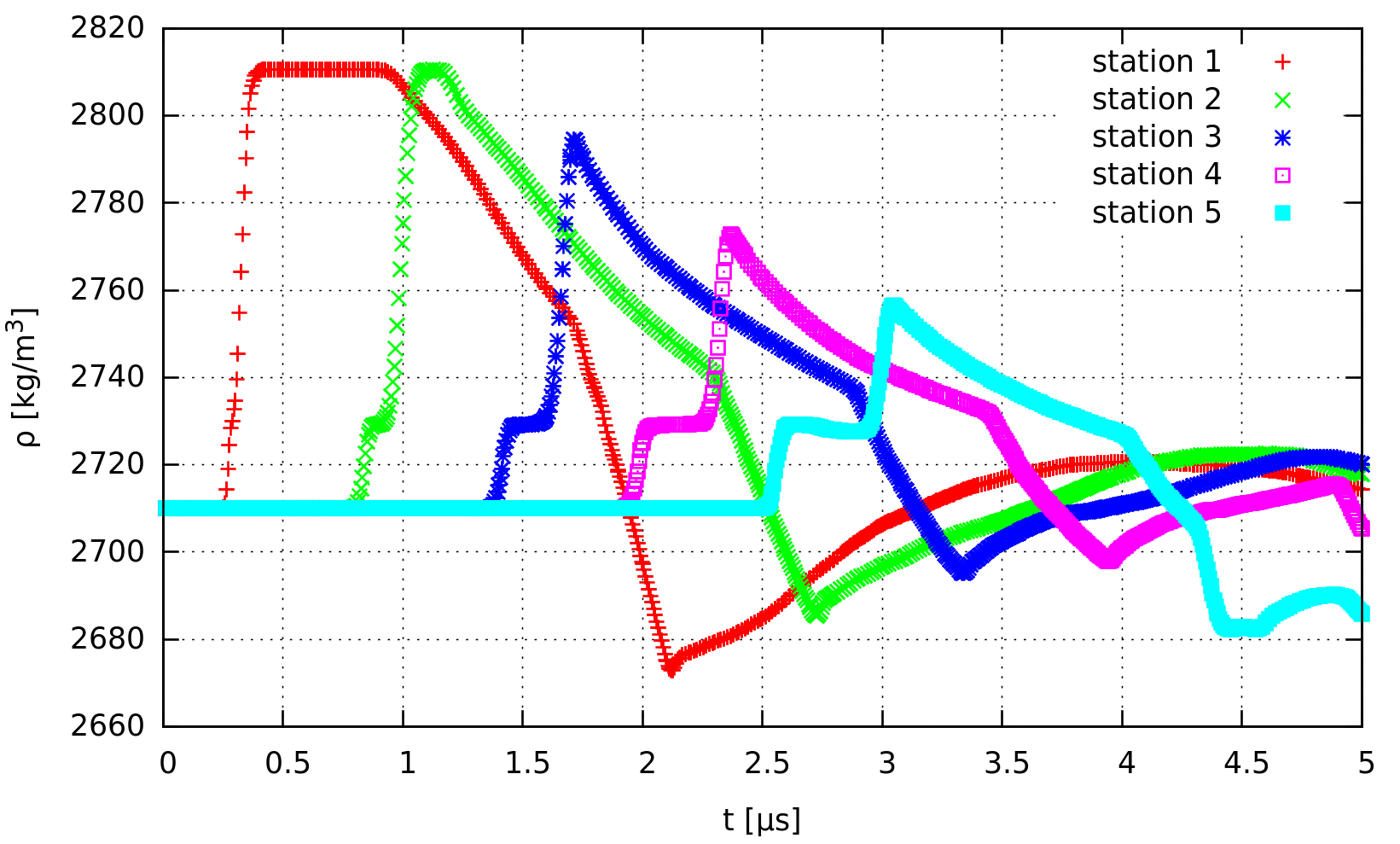}
                \caption{Density}
                \label{ALonALnum}
        \end{subfigure}%
        \begin{subfigure}[b]{0.5\textwidth}
                \includegraphics[width=\textwidth]{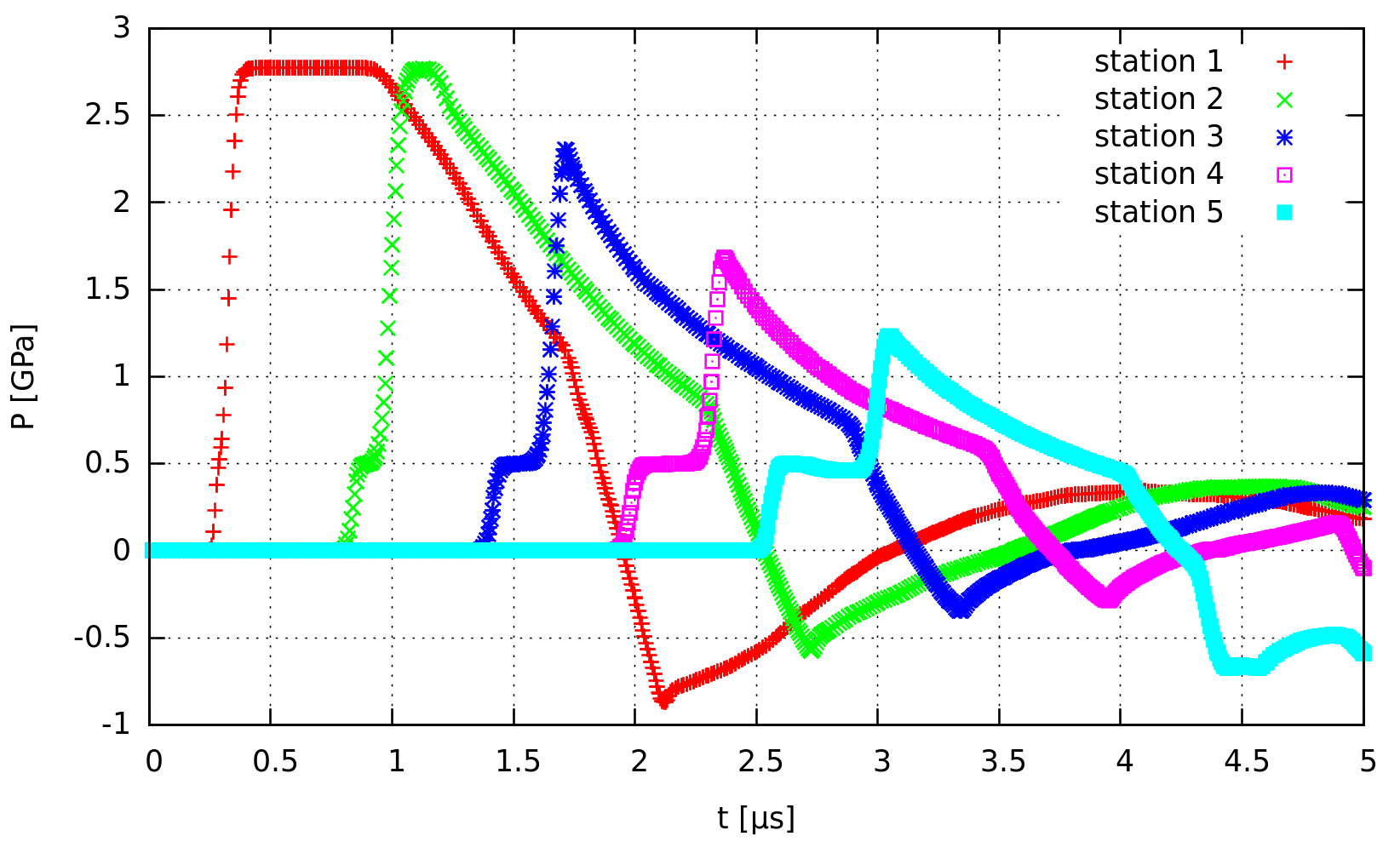}
                \caption{Pressure}
                \label{ALonALexp}
        \end{subfigure}
\end{minipage}
\caption{Time histories recorded at the five gauges along the centreline of the target plate, for the test problem described in Sec.\ \ref{Sec:HowellBallTest}. {\color{black}A good agreement is observed when comparing our results to the ones by Howell and Ball \cite{howell2002free} in terms of arrival times and amplitude of the waves.}}
\label{HowellBall1d}
\end{figure}

An excellent match is seen between our results and the computation by Howell and Ball \cite{howell2002free}. To compare further with their results we consider {\color{black}gauges} embedded in the originally stationary target block to detect strain, velocity, pressure, and density. These gauges are allowed to move with the flow as the target block deforms. Five equally spaced gauges are placed along the centreline of the target plate. The first one is placed at $\SI{1.8125}{\mm}$ from the original impact position and a distance of $\SI{3.625}{\mm}$ is allowed between consecutive gauges. The time histories for each gauge for the $x$-wise velocity, pressure, density and $x$-wise stress are seen in Fig.\ \ref{HowellBall1d}. The arrival times of the waves and their amplitude compare well with the results of Howell and Ball \cite{howell2002free}, for all gauges. The split of the shock wave generated at impact in traversing the target is clearly seen and the two waves appear to be steeper 
in our results. We choose to run the simulation longer than Howell and Ball, to capture the reflection of the elastic wave at the end of the target plate and the interaction of the reflected wave with the travelling elastic wave. These phenomena are also seen in the time-histories.

\section{Validation of the {\color{black}reactive, hydrodynamic} component}
\label{explosiveOnly}

In this section, the implementation of the MiNi16 formulation is validated in isolation, without the influence of the {\color{black} elastoplastic formulation or elastoplastic-hydrodynamic (or simply also referred to as solid-fluid) mixed Riemann solvers.} The tests are based on C4, as this is the explosive used in the later solid-fluid configurations.

\subsection{Equation of state and reaction rate}
\label{Sec:Hugoniots}
In this work, C4 and its products are modelled by the JWL equation of state, with parameters as given in Table \ref{C4eos}. The JWL equation of state
is a Mie-Gr\"uneisen equation of state, with general form given by
\begin{equation}
p(\rho,e)=p_{\text{ref}}(\rho)+\rho\Gamma(\rho)[e-e_{\text{ref}}(\rho)],
\label{MG}
\end{equation}
with reference pressure and energy curves given by:
\begin{equation}
\label{JWLeos1}
p_{ref}(\rho)=\mathcal{A}exp\Big{(}\frac{-\mathcal{R}_1\rho_0}{\rho}{\Big{)}+\mathcal{B}exp\Big{(}\frac{-\mathcal{R}_2\rho_0}{\rho}{\Big{)}}},\end{equation}
\begin{equation}
\label{JWLeos2}
e_{ref}(\rho)=\frac{\mathcal{A}}{\rho_0\mathcal{R}_1}exp\Big{(}\frac{-\mathcal{R}_1\rho_0}{\rho}\Big{)}+\frac{\mathcal{B}}{\rho_0\mathcal{R}_2}exp\Big{(}\frac{-\mathcal{R}_2\rho_0}{\rho}\Big{)}
\end{equation}
and Gr\"uneisen coefficient given as:
\begin{equation}
\label{JWLeos3}
\Gamma(\rho)=\Gamma_0.
\end{equation}
The JWL equation of state is usually used to model reaction products but it has been extensively used to model the unreacted phase of explosives as well \cite{tarver2005ignition,tarver2010theory}.

\begin{table}[!t]
\centering
\begin{tabular}{lccclcclcc}\hline 
JWL parameter  & C4 reactant & C4 products & & I\&G parameter  & & &  \\\hhline{---~-----}
$\Gamma$ & 0.8938  & 0.25 & & $\mathcal{I}[s^{-1}]$ & 4$\times 10^6$ & & $g$ & 0.667 \\
 $\mathcal{A}$ [$\SI{}{\mega\BAR}$]  & 778.1 & 6.0977 & & $\mathcal{G}_1[(10^{11}\mbox{\SI{}{\pascal}})^{-y}\mbox{\SI{}{\second}}^{-1}]$
& 149.97 & &$x$ & 7.0\\
$\mathcal{B}$ [$\SI{}{\mega\BAR}$] & -0.05031 & 0.1295 & & $\mathcal{G}_2[(10^{11}\mbox{\SI{}{\pascal}})^{-z}\mbox{\SI{}{\second}}^{-1}]$
  & 0  & & $y$ & 2.0\\
$\mathcal{R}_1$ &  11.3 & 4.5 & & $a$ & 0.0367 & & $z$ & 3.0 \\
$\mathcal{R}_2$ & 1.13 &  1.4 & & $b$ & 0.667 &  & $\phi_{IGmax}$ & 0.022 \\
$\rho_0$ [$\SI{}{\kilogram \per \cubic \meter}$] & 1601  & - & & $c$ & 0.667 &  & $\phi_{G1max}$ & 1.0 \\
$c_v  [10^5\SI{}{\mega\BAR}$$\SI{}{\per \kelvin}]$&  2.487 & 1.0 & & $d$ & 0.33 & & $\phi_{G2max}$ & 0.0 \\
Q [$\SI{}{\mega\BAR}$] & - & 0.09  & & $e$ & 0.667 \\\hline
\end{tabular}
\caption {Scaled JWL parameters for C4 and its detonation products (left) and scaled Ignition and Growth parameters for C4 (right) \cite{urtiew2006aip}.}
\label{C4eos}
\end{table}

\begin{figure}[!t]
\centering
\includegraphics[width=0.6\textwidth]{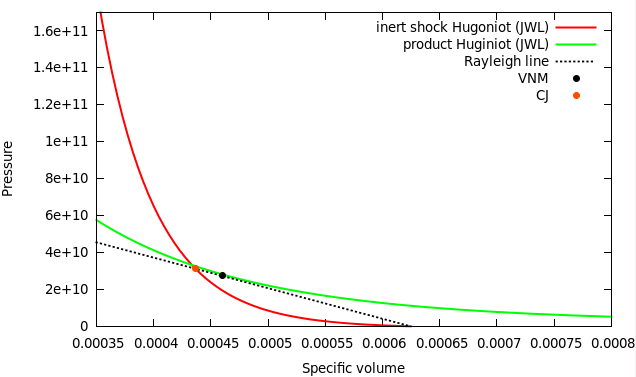}
\caption{Inert shock Hugoniot of C4 using the JWL equation of state (red), reactive Hugoniot curve (green) for the same material, and Rayleigh line, as described in Sec.\ \ref{Sec:Hugoniots}.}
\label{C4hugoniot}
 \end{figure}

Fig. \ref{C4hugoniot} illustrates the inert shock Hugoniot of the reactant in red and the fully reacted Hugoniot of the products in green. The dotted line represents the Rayleigh line. The intersection of the Rayleigh line with the inert shock Hugoniot gives the value of the von Neumann pressure predicted by the equation of state of the material. Similarly, the predicted CJ point is given by the point at which the Rayleigh line becomes tangent to the product Hugoniot. The value of the CJ pressure corresponding to the equations of state for C4 and its products used in this work is $\SI{27.5}{\giga \pascal}$ and the von Neumann pressure is $\SI{31.2}{\giga \pascal}$.

Examples of reported values for the CJ pressure of C4 at densities $1480-\SI{1600}{\kilogram \per \cubic \meter}$  in the literature \cite{andreas2005explosively} are: $\SI{27.5}{\giga \pascal}$, $\SI{22.36}{\giga \pascal}$, $\SI{24.91}{\giga \pascal}$, $\SI{22.55}{\giga \pascal}$, $\SI{25.09}{\giga \pascal}$ and $\SI{22.36}{\giga \pascal}$. 
It should be noted that in general the values for the von Neumann pressure are published considerably less frequently than the CJ pressures. For this specific explosive we could not find data in the literature reporting the von Neumann pressure. 

To describe the conversion of reactants to products we use the the Ignition and Growth model developed by Lee and Tarver \cite{lee1980phenomenological} and given by:
\begin{eqnarray}
\label{IG}
\frac{d\phi}{dt}=-K&=&I(1-\phi)^b(\rho -1 -\alpha)^xH(\phi_{IGmax}-\phi)\\&+&G_1(1-\phi)^c\phi^dp^yH(\phi_{G1max}-\phi)\nonumber\\
&+&G_2(1-\phi)^e\phi^gp^zH(\phi-\phi_{G2max}),\nonumber
\end{eqnarray}
where $\phi=1-\lambda$ is the mass fraction of the products, $H$ is the Heaviside
function and $I,G_1,G_2,a,b$, $c,d,e,g,x,y$ and $z$ are constants  chosen for
a particular explosive and a specific regime of the detonation process (i.e.\
 initiation or propagation of established detonation). The constants 
$\phi_{IGmax},\phi_{G1max}$ and $\phi_{G2max}$  determine for how long each of the
three terms is dominant. 

The form of this reaction rate was constructed so that each term  represents
one of the three stages of reaction observed during the shock initiation
and detonation of pressed solid explosives. The three terms can be interpreted
differently depending which regime of the detonation process is studied.
For shock initiation modelling, the first term represents the amount of reaction
due to the formation and ignition of hot-spots which are generated by several
mechanisms in heterogeneous explosives but mainly by void compression. The
second term then describes the reaction due to the growth of the hot-spots
and the third term the completion of reaction and transition to detonation.
For detonation modelling, the first term still describes the amount of reaction
at the initiation stage, due to the generation of hot spots.  The second
term describes the fast growth of the reaction as the reactant is converted
to products. The third term describes the relatively slow diffusion-limited
process of carbon formation \cite{tarver2005ignition}. The reaction rate law parameters for C4 are given in Table \ref{C4eos} which have been rendered non-dimensional using the CJ state for the explosive.

\subsection{C4 ZND}
\label{Sec:ZND}

Consider a one-dimensional slab of C4 at ambient conditions, initiated by a booster with pressure of $\SI{30}{\giga \pascal}$. We model this computationally in a domain that spans [0,6.432]\SI{}{\centi \meter} and the booster resides in the region [0,0.0402]\SI{}{\centi \meter}. An effective resolution of $\Delta x = \SI{33.5}{\micro \meter}$ is used. 
The initial data in the ambient region are:
\begin{equation*}
(\rho_{\text{reactant}} , \rho_{\text{product}}, u, p, \lambda) = (1590\SI{}{\kilogram \per \cubic \meter}, 1590\SI{}{\kilogram \per \cubic \meter}, 0\SI{}{\meter\per\second},1\times 10^5\SI{}{\pascal}, 1 )
\end{equation*}
and in the booster region are:
\begin{equation*}
(\rho_{\text{reactant}} , \rho_{\text{product}}, u, p, \lambda) = (1590\SI{}{\kilogram \per \cubic \meter}, 1590\SI{}{\kilogram \per \cubic \meter}, 0\SI{}{\meter\per\second},30\times 10^9\SI{}{\pascal}, 0 )
\end{equation*}

\begin{figure}[!t]
\begin{minipage}{\columnwidth}
        \centering
        \begin{subfigure}[b]{0.45\textwidth}
\includegraphics[width=\textwidth]{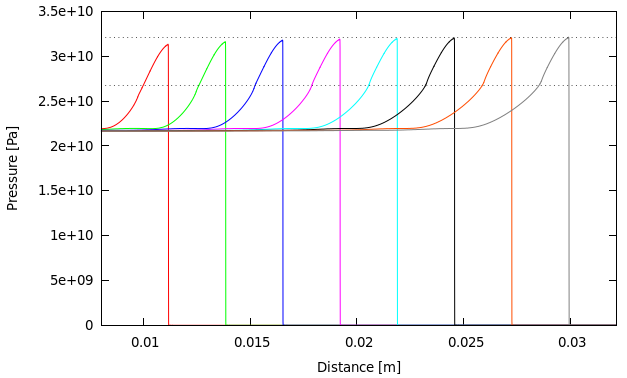}
\caption{ZND structure}
\label{zndC4}
\end{subfigure}%
        \begin{subfigure}[b]{0.49\textwidth}
\includegraphics[width=\textwidth]{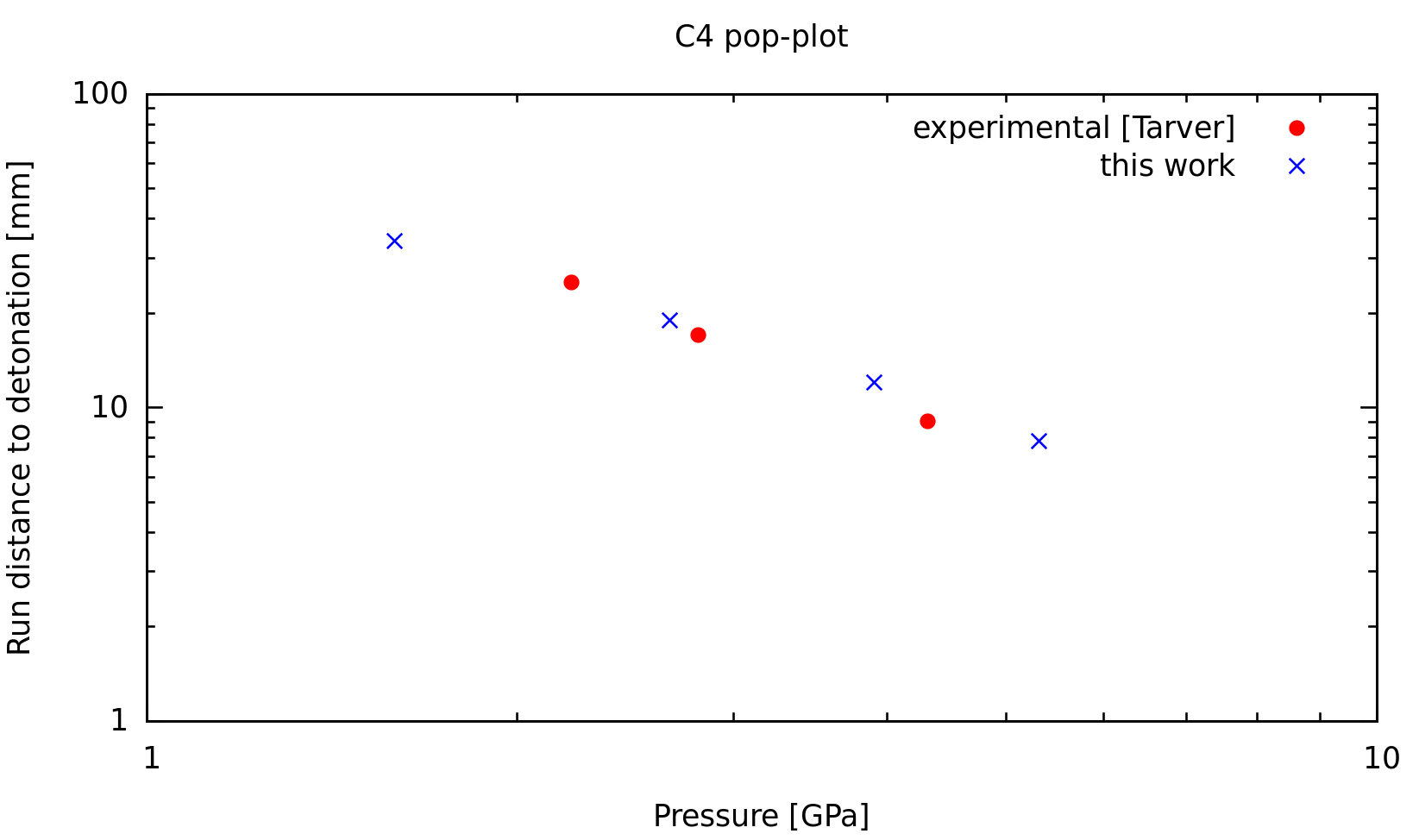}
\caption{Pop-plot}
\label{popC4}
        \end{subfigure}
\end{minipage}
\caption{Validation of C4 material model: (a) Pressure profiles of a detonation in C4, reaching steady state, as described in Sec.\ \ref{Sec:ZND}. Dashed
  lines denote the CJ and von Neumann pressure (b) Run distance to detonation \emph{vs} input pressure Pop-plot, as described in Sec.\ \ref{Sec:pop}. The filled circles represent experimental data and
the crosses numerically calculated data.}
\end{figure}

The explosive transits very quickly to detonation, which in-turn settles down to steady state. The steady detonation structure is shown in Fig.\ \ref{zndC4}.

The computed CJ and von Neumann pressure values agree with the values predicted in the previous section and fall within the range of values found in the literature.

\subsection{C4 Pop-plot}
\label{Sec:pop}

In this section we validate the the explosives model against Pop-plot data of run distance to detonation versus input pressure. This is equivalent to overtake time versus input pressure that is usually used as, for Pop-plot purposes, detonation is defined to be the overtake of precursor shock wave by the generated reactive wave. 

In Fig. \ref{popC4} we compare our numerical results for run distance to detonation to experimental data by Urtiew et al.\ \cite{urtiew2006aip} for different input pressures, where a very good match is observed.

\section{Validation and evaluation of the {\color{black}hydrodynamic-elastoplastic} coupling}
The final step towards simulating the non-linear, two-way interaction of explosives and elastoplastic materials is the validation of the solid-explosive coupling for one-dimensional and cylidrically symmetric test cases. 
Thereafter, the coupled system is used to demonstrate example applications.

\subsection{Stressed copper impacting quiescent PBX 9404}
\label{bartonTest2}

This test considers a stressed copper component for $x \in [0,0.005]\SI{}{\meter}$, impacting an initially quiescent, reacted PBX9404 gas for $x \in [0.005,1.0]\SI{}{\meter}$, as described in \cite{barton2011conservative}. The  PBX9404 is modelled by the ideal gas equation of state, with $\gamma = 2.83$   
and the copper by the elastic Romenskii equation of state (\ref{RomenskiiElastic}) with parameters as given in Table \ref{eosTable}. A domain spanning $[0,0.01]\SI{}{\meter}$ is considered and an effective resolution of  $\Delta x = \SI{20}{\micro \meter}$ 
is used. 

The initial conditions for this test are:
\begin{equation*}
\text{L:} \; \rho = \SI{1840}{\kilogram \per \cubic \meter}, \; u =  \left( \begin{array}{c}
2000 \\
0 \\
100 \end{array} \right) \SI{}{\meter \per \second}, \; S=\SI{0}{\joule \per \kilogram \per \kelvin}, \textbf{\text{F}}= \left( \begin{array}{ccc}
1 & 0 & 0 \\
-0.01 & 0.95 & 0.02 \\
-0.015 & 0 & 0.9 \end{array} \right)
\end{equation*}

\begin{equation*}
\text{R:} \; \rho = \SI{1840}{\kilogram \per \cubic \meter}, \; u = \SI{0}{\meter \per \second}, \; p=10^5\SI{}{\pascal}.
\end{equation*}

\begin{figure}[!t]
\centering
\includegraphics[width=\textwidth]{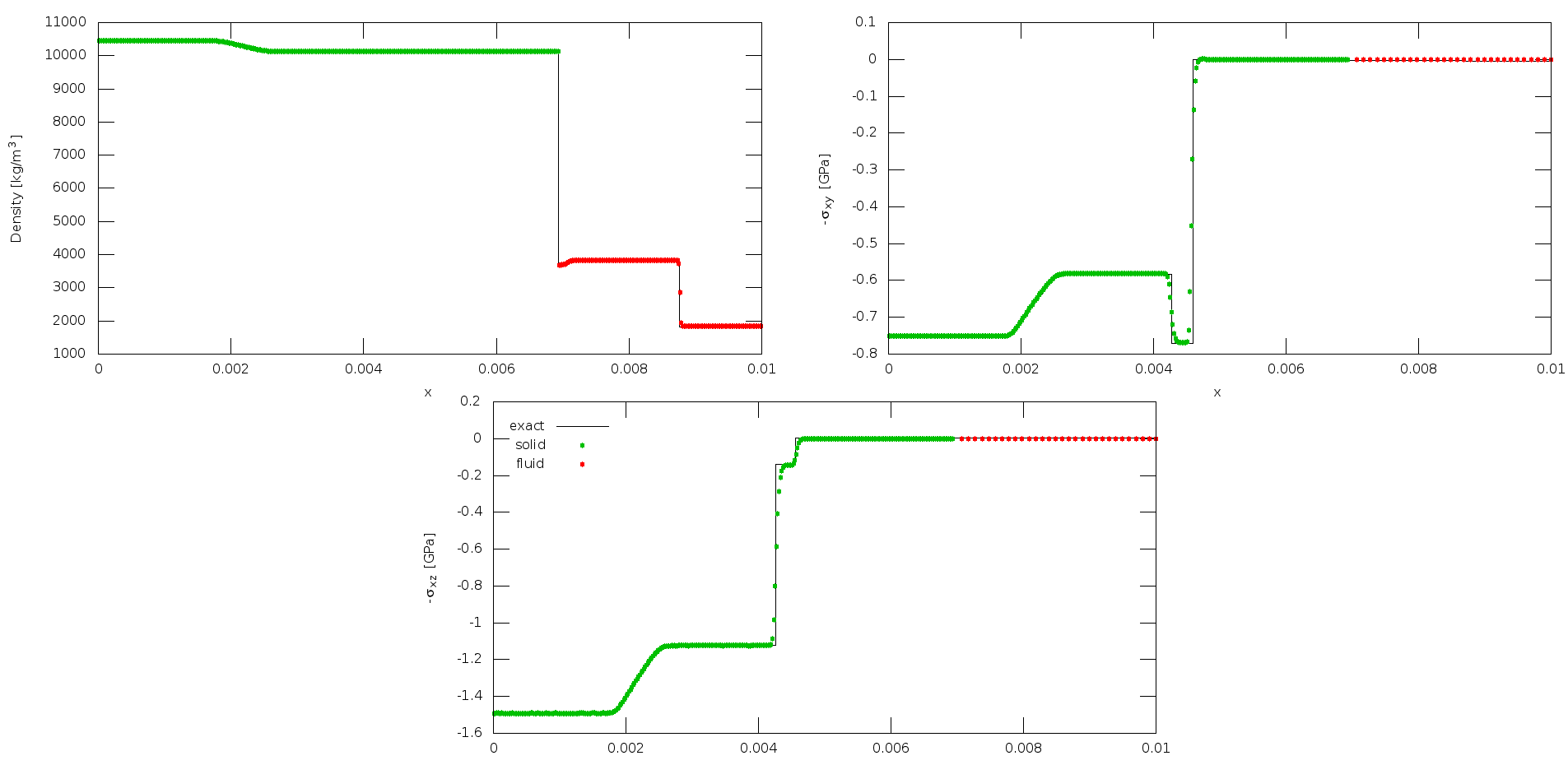}
\caption{ Comparison of our numerical solutions for total density and tangential $xy$ and $xz$ stress against the exact solution for the solid-fluid test problem described in Sec. \ref{bartonTest2}, with an initially moving and stressed copper (green) and quiescent PBX9404 at ambient pressure (red). {\color{black} The material interface between the solid and fluid component is captured as a discontinuity, low numerical diffusion is generally observed and the solution is free from any spurious oscillations.}}
\label{barton2}
 \end{figure}

The numerical and the exact solutions for density and tangential stress components $\sigma_{xy}$ and $\sigma_{xz}$ at $t=\SI{0.9}{\micro \second}$ are given in Fig.\ \ref{barton2}, where good agreement for all quantities is demonstrated. {\color {black} A small density error at the interface can be seen, which is, however, visibly smaller than the error in \cite{barton2011conservative}. } 

\subsection{Sandwich-plate impact: inert Detasheet confined by steel, impacted by steel}
\label{Sec:Detasheet}

\begin{figure}[!t]
\centering
 \includegraphics[width=0.5\textwidth]{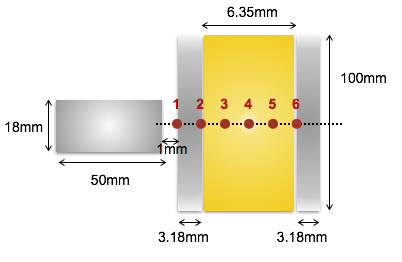}
 \caption{Schematic of the sandwich-plate impact setup described in Sec.\ \ref{Sec:Detasheet}.}
 \label{schematic}
\end{figure}

{\color{black}The next validation is in cylindrical symmetry for an elastoplastic-elastoplastic-hydrodynamic configuration and we compare our results against existing numerical solutions.} The test considers a cylindrical flyer plate impact and the subsequent response of the explosive residing behind the target plate. Specifically, a steel projectile of $\SI{18}{\mm}$ diameter and $\SI{50}{\mm}$ length impacts a steel target plate of thickness $\SI{3.18}{\mm}$ and diameter $\SI{100}{\mm}$. Behind the target plate sits a block of Detasheet explosive with the same diameter as the target plate. Two explosive thicknesses are considered; $\SI{6.35}{\mm}$ and $\SI{3.18}{\mm}$. Another steel plate is considered to sit behind the explosive, with the same diameter and thickness as the front plate. The simulation is done in axial symmetry. Initially the projectile and target plate are separated by $\SI{1}{\mm}$. A schematic of the setup is shown in Fig.\ \ref{schematic}, residing in vacuum.

The Detasheet is modelled by the linear Hugoniot equation of state \cite{lynch2008influence}, which is of Mie-Gr\"uneisen form (\ref{MG}), with reference functions given by:
\begin{equation}
p_{ref}(\rho)=p_0+\frac{\rho_0\rho c_0^2(\rho-\rho_0)}{\rho-s(\rho-\rho_0)},
\end{equation}
\begin{equation}
e_{ref}(\rho)=e_0+[p_{ref}(\rho)+p_0]\frac{\rho-\rho_0}{2\rho\rho_0}
\end{equation}
and Gr\"uneisen coefficient given as:
\begin{equation}
\Gamma(\rho)=\Gamma_0\Big{(}\frac{\rho_0}{\rho}\Big{)}^a.
\end{equation}
The equation of state parameters for this explosive are: 
\begin{equation}
(\Gamma, c_0 [\SI{}{\meter \per \second}], s, \rho_0 [\SI{}{\kilogram \per \meter \tothe{3}}] )=(0.3, 1850, 2.32, 1480).
\end{equation}

The hydrodynamic, elastic and plastic behaviour of the steel is described by  a linear Hugoniot equation as well, in combination with a constant shear model with $G=95.4\times 10^9$$\SI{}{\pascal}$ and the Johnson Cook plasticity model \cite{johnson1983constitutive}. The Hugoniot parameters for steel are:

\begin{equation}
(c_0 [\SI{}{\meter \per \second}], s, \rho_0 [\SI{}{\kilogram \per \meter \tothe{3}}], T_0 [\SI{}{\kelvin}] )=( 4610, 1.275, 7860, 298).
\end{equation}

The Johnson-Cook plasticity model assumes yield stress given by:
\begin{equation}
  \sigma_Y = (A + B \varepsilon^n) \left(1 + C \ln{\frac{\dot{\varepsilon}}{\dot{\varepsilon}_0}} \right) \left( 1 - \left( \frac{T-T_0}{T_m - T_0} \right)^m \right),
\end{equation}
with $ A$ the base yield stress, $B$ the hardening constant multiplier, $C$ the strain rate dependence constant multiplier, $n$ the hardening exponent, $m$ the temperature dependence exponent, $\dot{\varepsilon}_0$ the reference strain rate, $T_0$ the reference temperature and $T_m$ the reference melt temperature.

For this test, the Johnson-Cook parameters for steel are:
{
\begin{equation}
A=0.53\times 10^9\SI{}{\pascal}, B=0.229\times 10^9\SI{}{\pascal}, C=0.027, n=0.302, m=1.0, \dot{\varepsilon}_0=1.0\SI{}{\per \second}
\end{equation}
\begin{equation*}
T_0=298\SI{}{\kelvin}, T_m=1836\SI{}{\kelvin}.
\end{equation*}}
 A domain  $(x,y)\in [-50,50]\SI{}{\mm}\times[0,65]\SI{}{\mm}$ is considered, with effective resolution $\Delta x = \Delta y = \SI{0.5}{\mm}$. 
The region outside the plates and explosive contains vacuum. To compare directly with the results by Lynch \cite{lynch2008influence}, the Detasheet explosive is taken to be inert and an impact velocity of $u_z= \SI{1800}{\meter \per \second}$ is considered.

The initial data for this test are: 
\begin{equation*}
\text{Steel Projectile:} \; \rho = \SI{7860}{\kilogram \per \cubic \meter}, \; u_r = \SI{0}{\meter \per \second},  u_z= \SI{1800}{\meter \per \second}, \; S=\SI{0}{\joule \per \kilogram \per \kelvin}, \; \textbf{\text{F}}=\textbf{\text{I}}. 
\end{equation*}
\begin{equation*}
\text{Steel Plates:} \; \rho = \SI{7860}{\kilogram \per \cubic \meter}, \; u_r = u_z= \SI{0}{\meter \per \second}, \; S=\SI{0}{\joule \per \kilogram \per \kelvin}, \; \textbf{\text{F}}=\textbf{\text{I}}. 
\end{equation*}
\begin{equation*}
\text{Detasheet:} \; \rho = \SI{1400}{\kilogram \per \cubic \meter}, \; u_r = u_z= \SI{0}{\meter \per \second}, \; p=10^5\SI{}{\pascal},
\end{equation*}
\begin{figure}[!t]
\centering
 \includegraphics[width=0.5\textwidth]{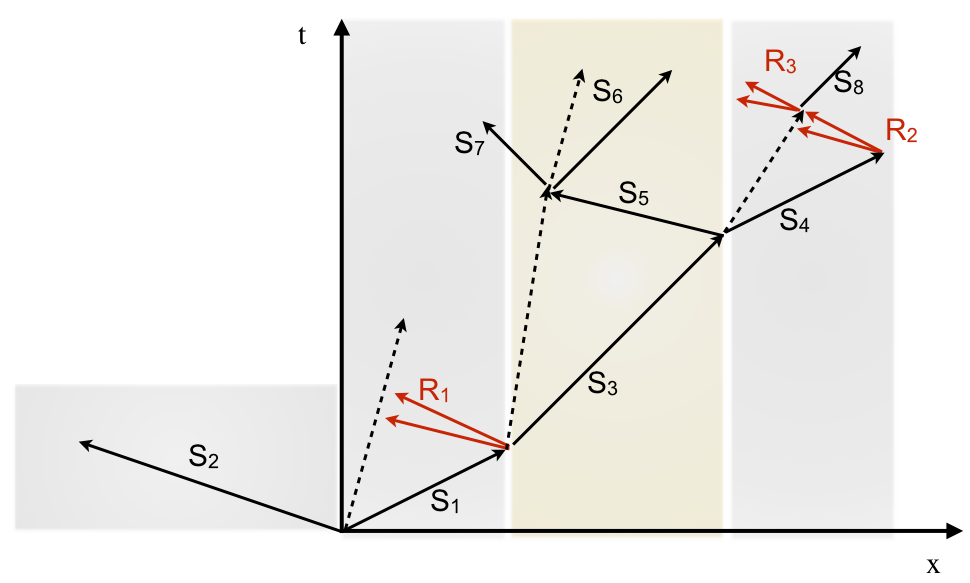}
 \caption{Wave diagram observed in the sandwich-plate impact application of Sec.\ \ref{Sec:Detasheet}. Solid black lines show the paths of shock waves, solid red lines show the paths of rarefaction waves, and dashed lines show the motion of material interfaces.}
 \label{wavediagram}
\end{figure}
The wave pattern in this scenario is complex so a wave diagram is included as Fig.\ \ref{wavediagram} to keep track of the waves generated. The solid black lines denoted by `S' represent the shock waves, the pairs of red lines denoted by `R' the rarefaction fans and the black dotted lines the impulsively accelerated material interfaces.  
Upon impact, two shocks are generated; one travelling in the target plate ($S_1$) and one travelling back into the projectile ($S_2$). As a result of the impulse, the face of the plate is accelerated. The shock $S_1$ reaches the front plate/explosive interface, a high impedance/low impedance interface (HI-LI), where it is transmitted in the explosive as a shock ($S_3$) and reflected back into the front plate as a release wave ($R_1$). Again, the material interface is accelerated. The shock $S_3$ reaches the end of the explosive and encounters a low impedance-high impedance interface (LI-HI). Hence, it is transmitted into the rear plate as shock $S_4$ and reflected into the explosive as shock $S_5$. It should be noted that if there was no rear plate, the wave diagram would stop after the generation of $S_3$ and no other waves would have affected the explosive. The shock $S_5$ reaches the rear end of the explosive, now a LI-HI interface, where two shocks are generated; $S_6$, which is reflected back into the explosive and $S_7$, which is transmitted into the front target plate. In the meantime, shock $S_4$ reaches the end of the rear plate and, since there is vacuum on the other side of the plate, the shock is reflected as a rarefaction wave ($R_2$) only, into the plate. The release wave $R_2$ reaches the rear plate/explosive interface which is now a HI-LI interface where it is reflected as a shock ($S_8$) back into the plate and transmitted as a rarefaction wave ($R_3$) into the explosive. The release wave ($R_3$) interacts with the shock ($S_6$) within the explosive, leading to its weakening and making the wave pattern much more complex from here on. It is worth noting that if the rear plate was infinitely thick  there would be no release waves coming into the explosive and a series of shock reflections would occur (like for $S_5$ and $S_6$), continuously increasing the pressure in the explosive. This is the general one-dimensional behaviour along the centreline of the experiment. Additional effects are generated due to the deformation of the solid materials and the weld assumption between the solid components.

\begin{figure}[!t]
\begin{minipage}{\columnwidth}
        \centering
        \begin{subfigure}[b]{0.5\textwidth}
                \includegraphics[width=\textwidth]{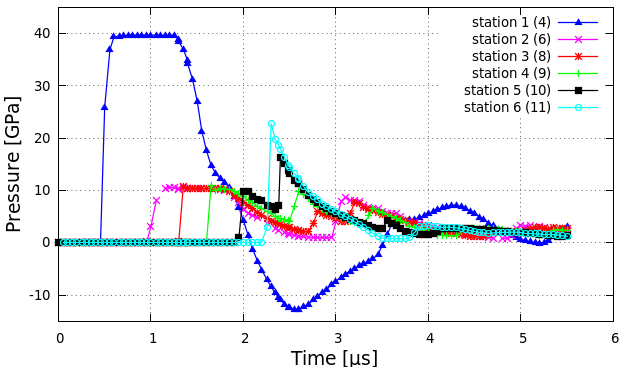}
                \caption{Thickness 6.35mm}
                \label{}
        \end{subfigure}%
        \begin{subfigure}[b]{0.5\textwidth}
          \includegraphics[width=\textwidth]{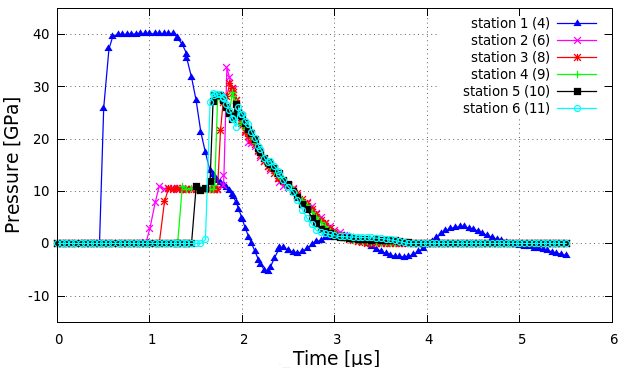}
                \caption{Thickness 3.18mm}
                \label{}
        \end{subfigure}
\end{minipage}
\caption{Pressure time-histories for explosives of thickness $\SI{6.35}{\mm}$ and $\SI{3.18}{\mm}$ for the test problem of Sec.\ \ref{Sec:Detasheet}. These match well with the results by Lynch \cite{lynch2008influence}. The station numbers in the parentheses correspond to the station numbers in the reference. }
\label{deetasheet} 
\end{figure}

Pressure gauges are placed at the point of impact (station 1), at the front plate/explosive interface (station 2), at equally spaced points in the body of the explosive (stations 3-5) and at the explosive/rear plate interface (station 6). The pressure at each of these gauges over time is seen in Fig.\ \ref{deetasheet} for both explosive thicknesses and we compare our results to the results by Lynch \cite{lynch2008influence}. The station 1 curve (station 4 in \cite{lynch2008influence}) shows the shock ($S_1$) as it is generated at the front plate/explosive interface, increasing the pressure at $\SI{40}{\giga \pascal}$. 
The wave $S_3$ is seen traversing the explosive in the stations 2-5 (stations 6-10 in \cite{lynch2008influence}), followed by {\color{black}release waves}. 
In the meantime, the release wave $R_1$ in the front plate lowers considerably the pressure in the steel and the weld condition allows the pressure (or rather, stress) to go as low as $\SI{-10}{\giga \pascal}$. The wave $S_5$ is seen in stations 5 and 6. The release wave $R_3$ travels behind the shock $S_5$ within the explosive leading to its weakening. 

A similar behaviour is seen for the explosive of thickness $\SI{3.18}{\mm}$, though the several features are seen to happen a lot faster (of course the gauges have also been moved). Moreover, the strength of the shock $S_5$ is greater than before and also the minimum stress achieved in the front plate is higher than before. 

For both thicknesses our results agree well with the results by Lynch  \cite{lynch2008influence}. We note that in the results by Lynch \cite{lynch2008influence} some oscillations are seen that are attributed to the numerical scheme used therein.

\subsection{Sandwich-plate impact: reactive C4 confined by steel, impacted by steel}
\label{sandwiched_orig}

The previous test is repeated with a different explosive, namely C4, which is now chemically active. The explosive has been tested and validated individually in  Sec.\ \ref{explosiveOnly}. 
We use the JWL equation of state as given by equations (\ref{MG})--(\ref{JWLeos3}) and the ignition and growth reaction rate model (\ref{IG}) to represent the explosive. The parameters for these are found in Table \ref{C4eos}.

The initial data for steel are as in the previous section, with the exception of the impact velocity which here is taken to be  $u_z= \SI{700}{\meter \per \second}$. The initial data for the explosive are:

\begin{equation}
\text{C4:} \; \rho = \SI{1590}{\kilogram \per \cubic \meter}, \; u_r = u_z= \SI{0}{\meter \per \second}, \; p=10^5\SI{}{\pascal},\; \lambda=1.0.
\end{equation}

\begin{figure}[!t]

\begin{minipage}[t]{\columnwidth}
\centering
        \begin{subfigure}[t]{0.4\textwidth}
                \includegraphics[width=\textwidth]{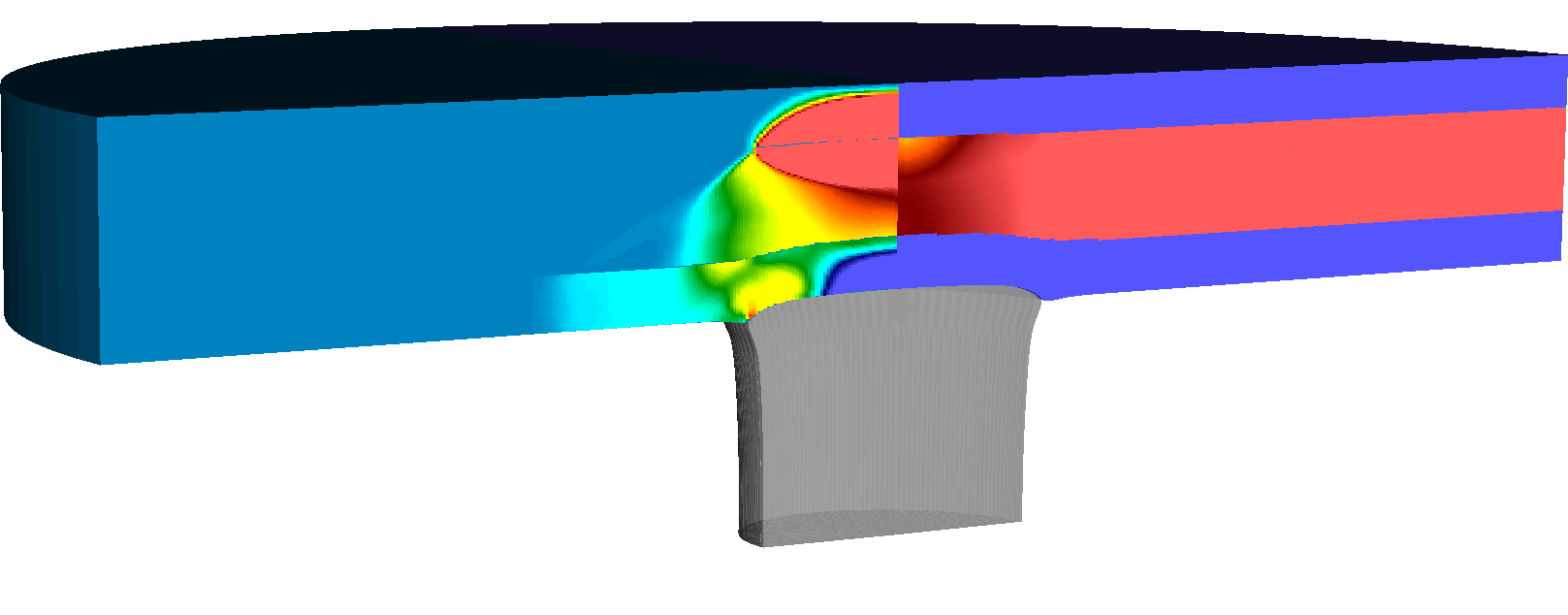}
        \end{subfigure}%
\quad
        \begin{subfigure}[t]{0.4\textwidth}
                \includegraphics[width=\textwidth]{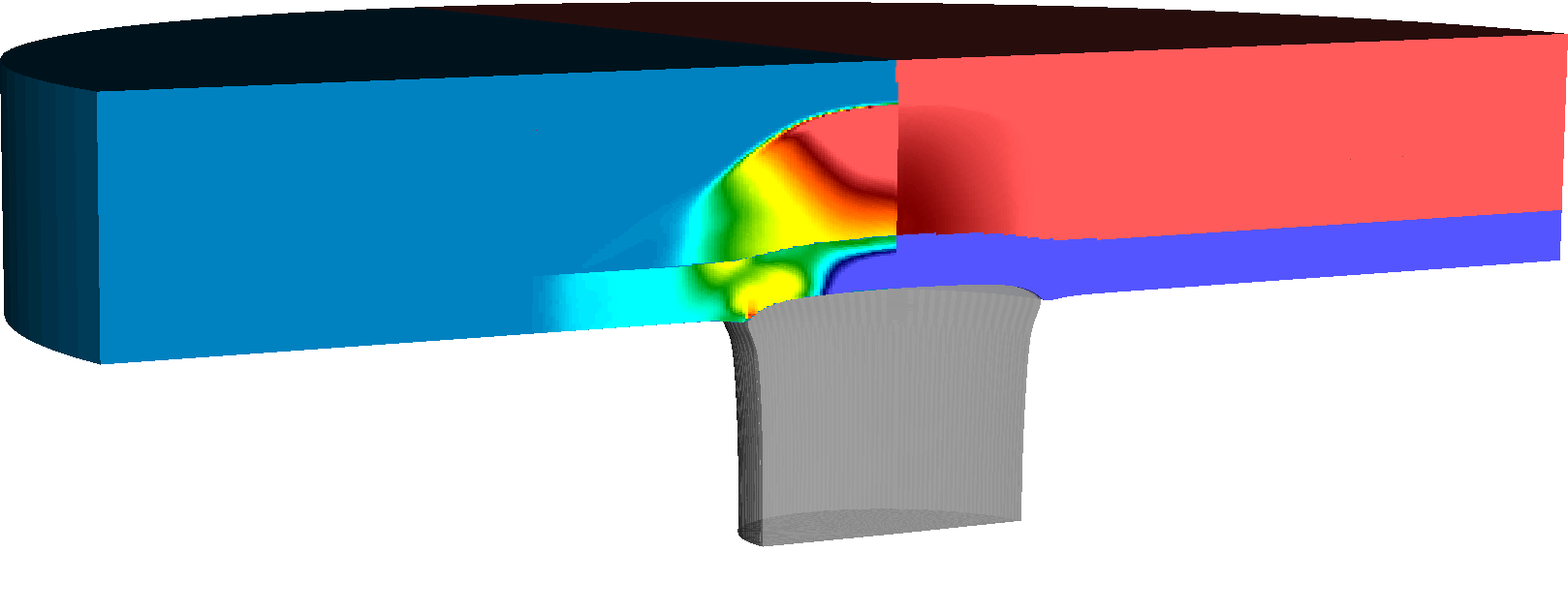}
        \end{subfigure}
\subcaption{$t=\SI{2.4}{\micro \second}$}
\end{minipage}
\begin{minipage}[t]{\columnwidth}
\centering
        \begin{subfigure}[t]{0.4\textwidth}
                \includegraphics[width=\textwidth]{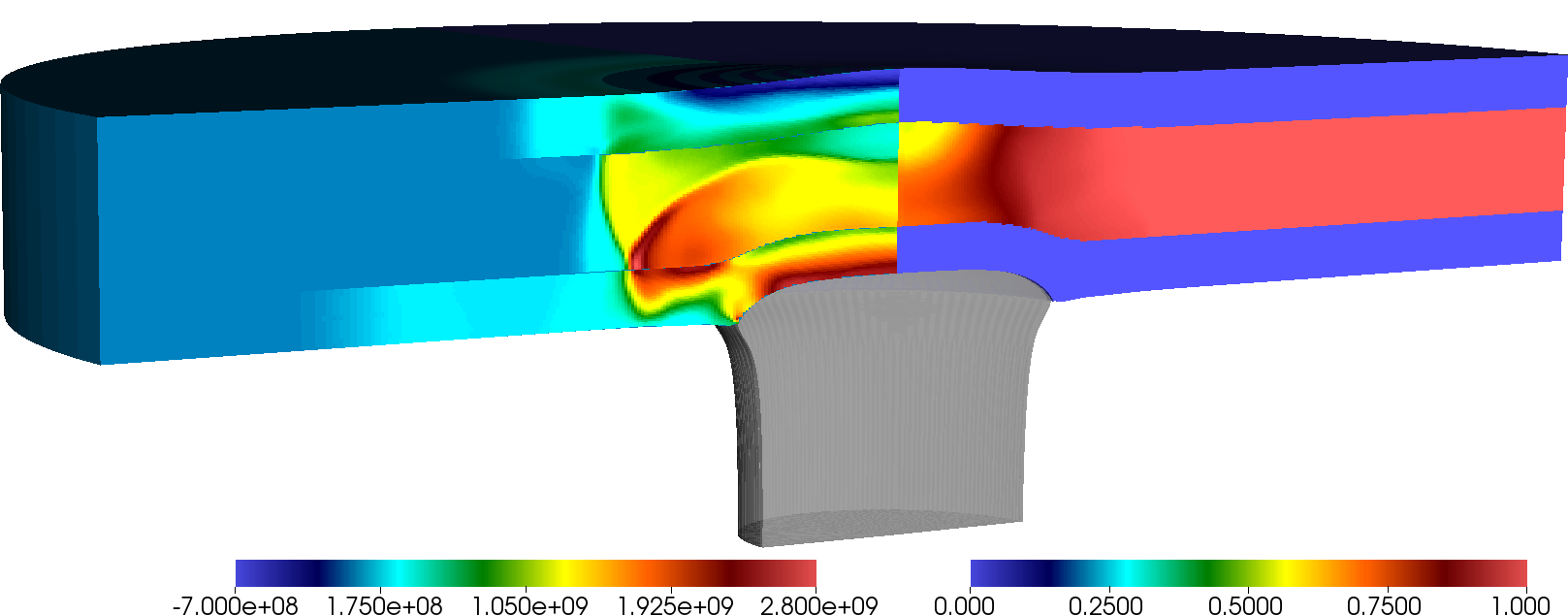}
        \end{subfigure}%
\quad
        \begin{subfigure}[t]{0.4\textwidth}
                \includegraphics[width=\textwidth]{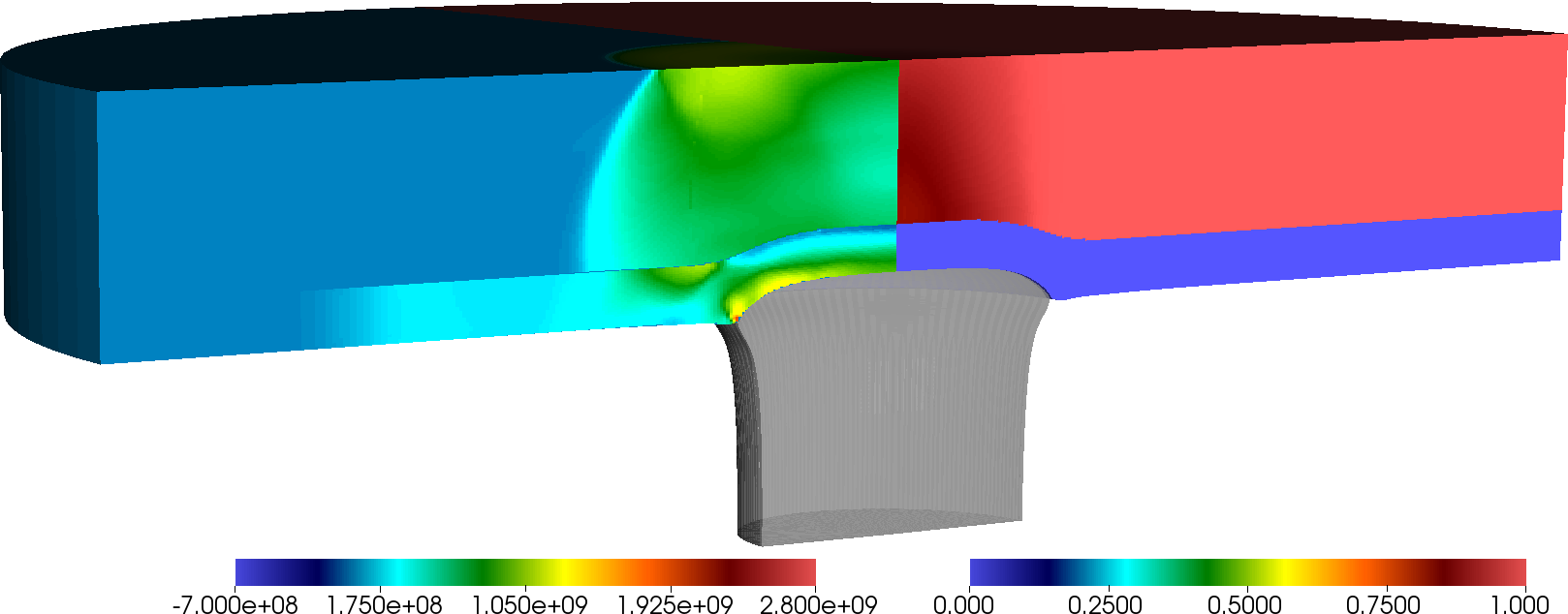}
        \end{subfigure}
\subcaption{$t=\SI{4.9}{\micro \second}$}
\end{minipage}
\caption{ The pressure distribution for all materials (left part of domain) and mass fraction of the reactants (right part of domain), at times $t=2.4$ and $\SI{4.9}{\micro \second}$, for the configuration of an explosive confined by one (right figures) or two (left figures) steel plates and impacted by steel projectile, as described in Sec.\ \ref{sandwiched_orig}. }
\label{sandwichC4} 
\end{figure}

In Fig.\ \ref{sandwichC4} (top), the mass fraction of the reactants is seen on the left and the pressure distribution on the right, at times $t=2.4$ and $\SI{4.9}{\micro \second}$. At $t=\SI{1.4}{\micro \second}$, minimal reaction, of the order of $\lambda=0.91$, is observed due to the shock wave generated at impact ($S_3$). The effect of the shock $S_5$ on the reaction is larger, leading to $\lambda=0.65$ at $t=\SI{2.4}{\micro \second}$ at the ignition site near the explosive/rear plate interface and to $\lambda=0.55$ at $t=\SI{4.9}{\micro \second}$ in the same site.

The same test is repeated with the rear plate removed and the explosive extended to have thickness of $\SI{10.82}{\mm}$ to {\color{black} demonstrate} the effect of the rear plate on the ignition of the explosive. In Fig.\ \ref{sandwichC4} (bottom), the mass fraction of the reactants is seen on the left and the pressure distribution on the right, at times $t=2.4$ and $\SI{4.9}{\micro \second}$.  At $t=\SI{1.4}{\micro \second}$, minimal reaction, of the order of $\lambda=0.91$, is observed due to the shock wave generated at impact ($S_3$). This is the same amount of reaction observed in the previous case as well, as up to this point, no waves have reached the rear plate. As mentioned earlier, if the rear plate is not present, only the wave $S_3$ affects the ignition process. At time  $t=\SI{2.4}{\micro \second}$, reaction of the order $\lambda=0.85$ is seen and at $t=\SI{4.9}{\micro \second}$ reaction of the order $\lambda=0.83$. The combination of the results of these two cases demonstrates the effect of the rear plate on accelerating the reaction. 

\subsection{Sandwich-plate impact with front air gap}
\label{Sec:AirGap}

The same test as in Sec.\ \ref{sandwiched_orig} is repeated with the inclusion of an air gap initially at atmospheric conditions (modelled as ideal gas with $\gamma=1.4$) of width $\SI{3.18}{\mm}$ between the front plate and the explosive. This demonstrates the full use of the MiNi16 model (air is phase 1, the explosive reactant is phase $\alpha$ and the explosive products is phase $\beta$; the last two form phase 2) coupled with the elastoplastic formulation. {\color{black} This problem is difficult to handle numerically, due to the big differences in the properties of the materials (reactants, products, air, solid) and the strong conditions involved.} In this scenario we {\color{black}demonstrate} how the presence of the air gap affects the ignition of the explosive so a direct comparison to the results of Sec.\ \ref{sandwiched_orig} is carried out. In both scenarios the impact generates a wave that travels in the front plate. In the air gap scenario, this reaches the air gap where a shock of the order of $p=0.5\SI{}{\mega \pascal}$ is transmitted in the air. After a short propagation in the air, at $t=1.5\SI{}{\micro \second}$ this wave reaches the explosive-air interface where it is transmitted in the explosive as a shock of the order of $p=0.5\SI{}{\mega \pascal}$. The strength of this shock is considerably lower than the strength of the wave generated upon impact on the no-air gap case ($p=2.1\SI{}{\giga \pascal}$). As a result, this wave does not generate any reaction at all in the explosive. At $t=2.3\SI{}{\micro \second}$ the deformed front plate has squeezed away the air gap and reaches the lower explosive face. Upon impact a new shock wave is generated in the explosive (and an opposite one in the front target plate) of the order of $p=2.5\SI{}{\giga \pascal}$. The strength of this wave is comparable to the strength of the wave generated upon impact on the no-air gap case. Comparing the air-gap and non-air-gap cases, at $1\SI{}{\micro \second}$ after the first wave impacts the explosive, a reaction of the order of $\lambda=0.91$ is observed in the non-air-gap case, whereas no reaction is seen in the air-gap scenario. At $2\SI{}{\micro \second}$ after the first wave impacts the explosive, a reaction of the order of $\lambda=0.65$ is observed in the non-air-gap case, and of the order of $\lambda=0.71$ in the air-gap scenario, demonstrating how the air gap hinders the reaction process. 

\begin{figure}[!t]
\begin{minipage}{\columnwidth}
\end{minipage}
\begin{minipage}{\columnwidth}
        \centering
        \begin{subfigure}[b]{0.42\textwidth}
                 \includegraphics[width=0.9\textwidth]{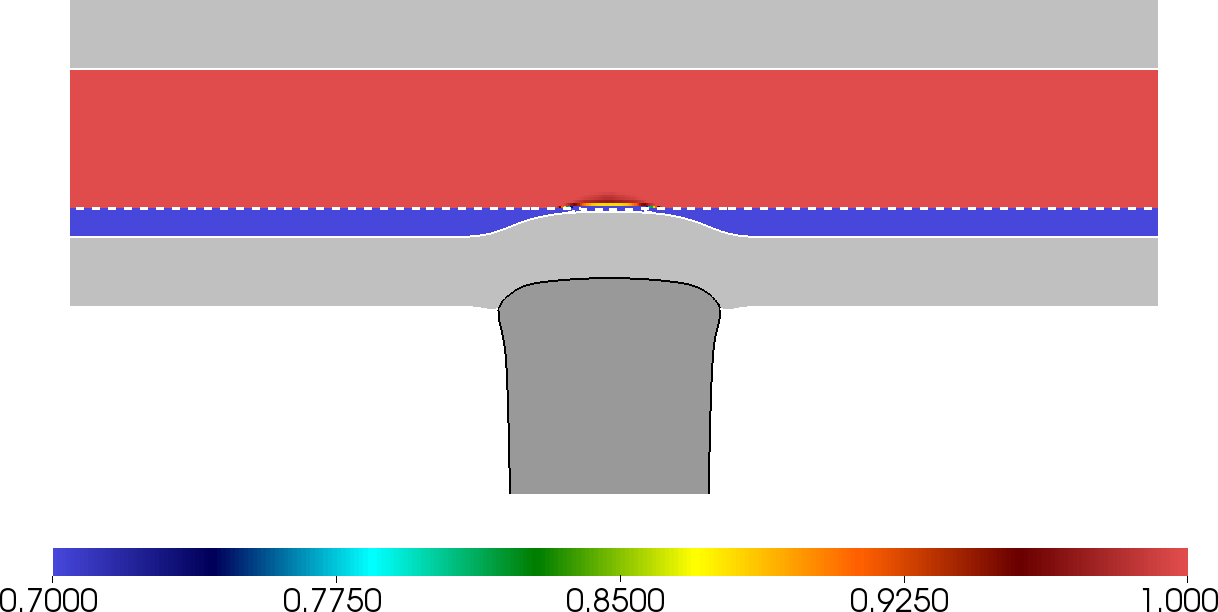}
       \end{subfigure}%
        \begin{subfigure}[b]{0.4\textwidth}
                \includegraphics[width=0.99\textwidth]{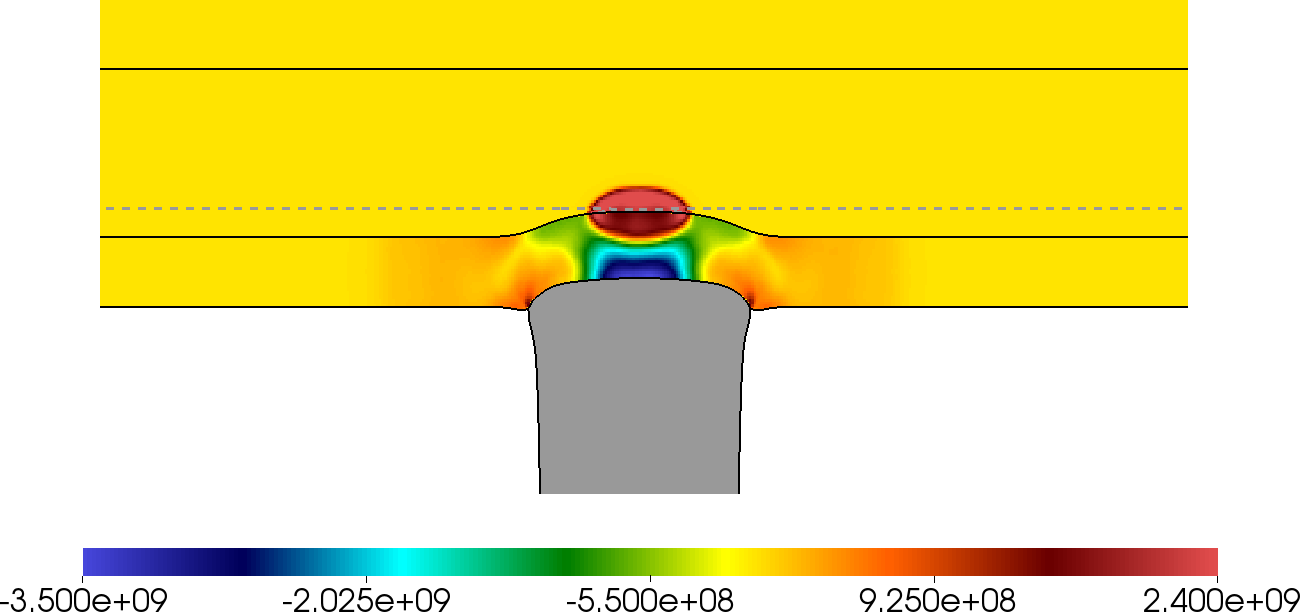}
         \end{subfigure}
        \centering
        \begin{subfigure}[b]{0.4\textwidth}
         \includegraphics[width=0.9\textwidth]{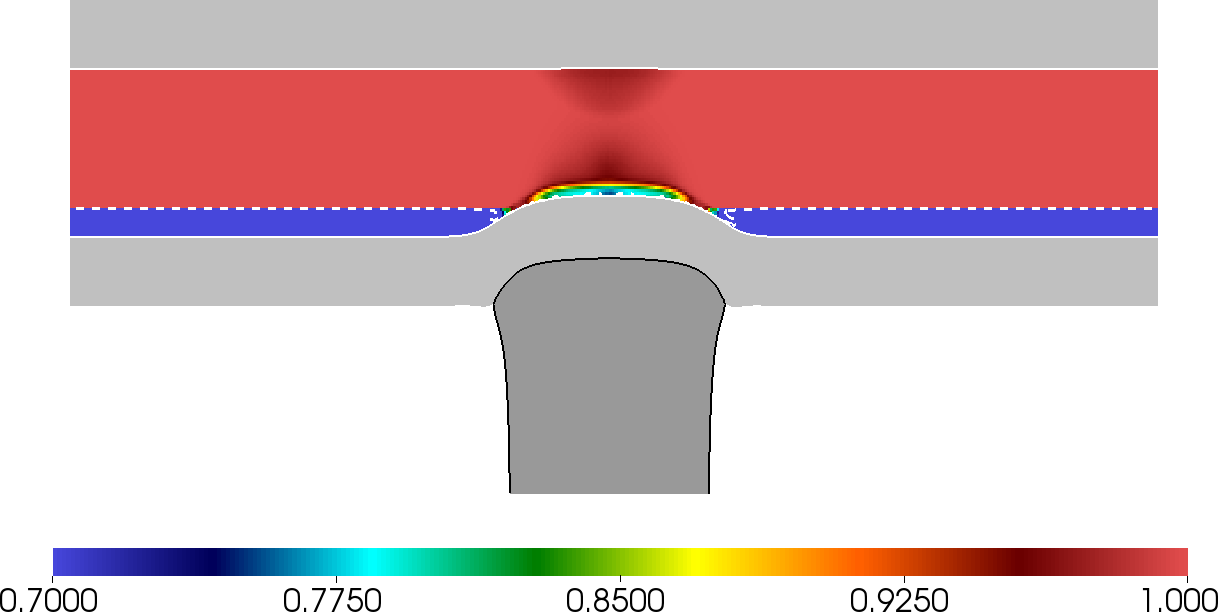} 
       \end{subfigure}%
        \begin{subfigure}[b]{0.4\textwidth}
           \includegraphics[width=\textwidth]{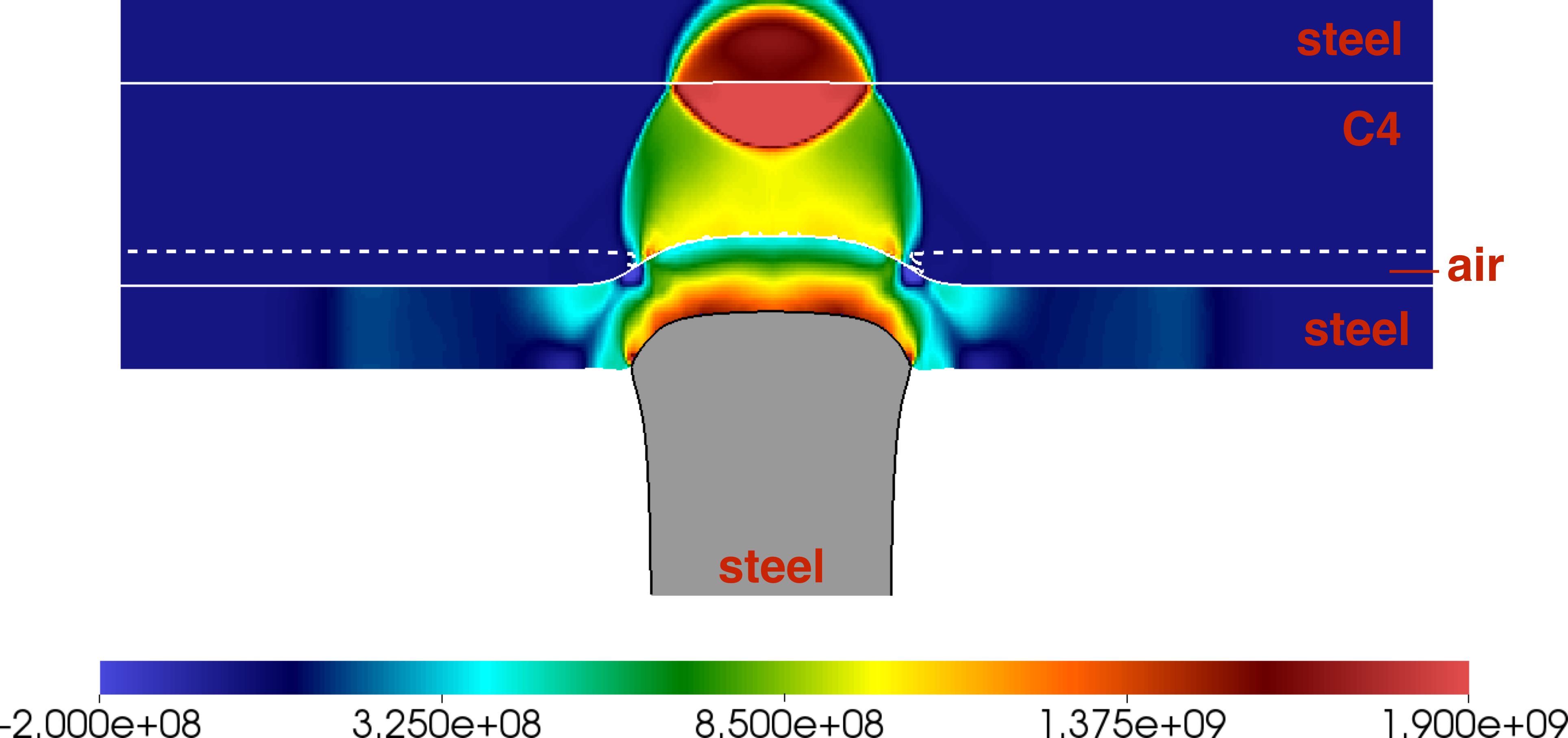} 
        \end{subfigure}
\end{minipage}
\caption{ The mass fraction of the reactants (left) and the pressure distribution {\color{black}for the solid plates, the explosive and air (right) are presented} at times $t=2.6$ and $\SI{4.6}{\micro \second}$, for the steel-confined explosive with an air gap, impacted by steel problem of Sec.\ \ref{Sec:AirGap}. {\color{black}Note that the figures have been stretched in the $y$-direction by a factor of 2, to allow for a clear view of the air gap.}}
\label{airgap} 
\end{figure}

 \subsection{Rod impact leading to detonation in a copper {\color{black}vessel}}
\label{Sec:ImpactcopperCan}

In this section, we present another showcase as a high-speed impact example involving {\color{black}multiphase reactive flow (explosive), elastoplastic structural (solid) response and inert air} response. Consider a copper can of outer diameter of $\SI{14}{\cm}$, inner diameter $\SI{10}{\cm}$ (resulting in a $\SI{2}{\cm}$ wall thickness) and height of $\SI{49.2}{\cm}$. The can is filled with a nitromethane charge of diameter $\SI{10}{\cm}$ and height of $\SI{45.2}{\cm}$ (i.e.\ no air gaps between the charge and the confiner). The can is impacted by a copper projectile travelling at $\SI{2000}{\m \per \second}$ with the same diameter as the explosive charge and a semi-infinite length, starting with length $\SI{6}{\cm}$ but extending also out of the domain. The configuration resides in air, of outer diameter $\SI{18}{\cm}$ and height $\SI{59.2}{\cm}$. All materials are initially at atmospheric conditions.

This test is similar to the confined explosion tests by Miller and Colella \cite{miller2002conservative}, Barton et al.\ \cite{barton2011conservative} and Schoch et al.\ \cite{schoch2013eulerian}. However, these tests describe the initiation of the explosive with a booster whereas here we consider directly the impact resulting to the initiation and detonation of the explosive. {\color{black}Moreover, in the aforementioned studies the explosive was either simpler and described with simpler EoS and/or the entire configuration resided in vacuum.}

{\color{black}In this work, we consider the explosive to be nitromethane, described by the Cochran-Chan EoS, which is of a Mie-Gr\"uneisen form given by Eq.\ \ref{MG} with reference curves:\begin{equation}
\label{CCeos1}
p_{\text{ref}}(\rho)=\mathcal{A}\Big{(}\frac{\rho_0}{\rho}{\Big{)}^{-\mathcal{E}_1}-\mathcal{B}\Big{(}\frac{\rho_0}{\rho}{\Big{)}^{-\mathcal{E}_2}}},\end{equation}
reference energy given by
\begin{equation}
\label{CCeos2}
e_{\text{ref}}(\rho)=\frac{-\mathcal{A}}{\rho_0(1-\mathcal{E}_1)}\Big{[}\Big{(}\frac{\rho_0}{\rho}\Big{)}^{1-\mathcal{E}_1}-1\Big{]} +\frac{\mathcal{B}}{\rho_0(1-\mathcal{E}_2)}\Big{[}\Big{(}\frac{\rho_0}{\rho}\Big{)}^{1-\mathcal{E}_2}-1\Big{]}
\end{equation}
Gr\"uneisen coefficient
$\Gamma(\rho)=\Gamma_0$
and parameters $\Gamma_0=1.19,\mathcal{A}=\SI{0.819}{\giga\pascal},\mathcal{B}=\SI{1.51}{\giga\pascal},\mathcal{E}_1=4.53,\mathcal{E}_2=1.42,\rho_0=\SI{1134}{\kilogram\per \meter \tothe{3}},c_v=\SI{1714}{\joule \per \kilogram \per \kelvin}$ and $Q=\SI{4.48}{\mega \joule \per \kilogram}$.  The chemical reaction follows a single-step, temperature dependent reaction rate law:
\begin{equation}
\label{Arrhenius}
K=\frac{d\lambda}{dt}=-\lambda Ce^{-T_A/T_{NM}},
\end{equation}
with $C=\SI{6.9e10}{\per \second}$ and activation temperature $T_A=\SI{11350}{\kelvin}$ \cite{paper2A} and nitromethane temperature recovered as $T_{NM}=T_2=\frac{p-p_{\text{ref}_2}(\rho)}{\rho_2\Gamma_2c_{v_2}}$. We neglect the explicit description of products so the explosives model reduces to augmented Euler.
}
The copper is described by the elastic Romenskii equation of state with parameters as given in Table \ref{eosTable} and perfect plasticity with yield of $\SI{70}{\mega \pascal}$. The entire configuration is residing in air, described as an ideal gas with $\gamma=1.4$, so we are able to visualise the transmission of the waves from the explosive, to the solid and finally to the gas. 
The simulation is performed with a base spatial resolution of $\Delta x = \Delta y = \SI{1}{\mm}$ and two levels of refinement each with refinement factor 2, resulting in an effective resolution of $\Delta x = \Delta y = \SI{0.25}{\mm}$.

\begin{figure}[!ht]
\centering
\begin{minipage}{\columnwidth}
        \begin{subfigure}[b]{0.49\textwidth}
                 \includegraphics[width=0.85\textwidth,center]{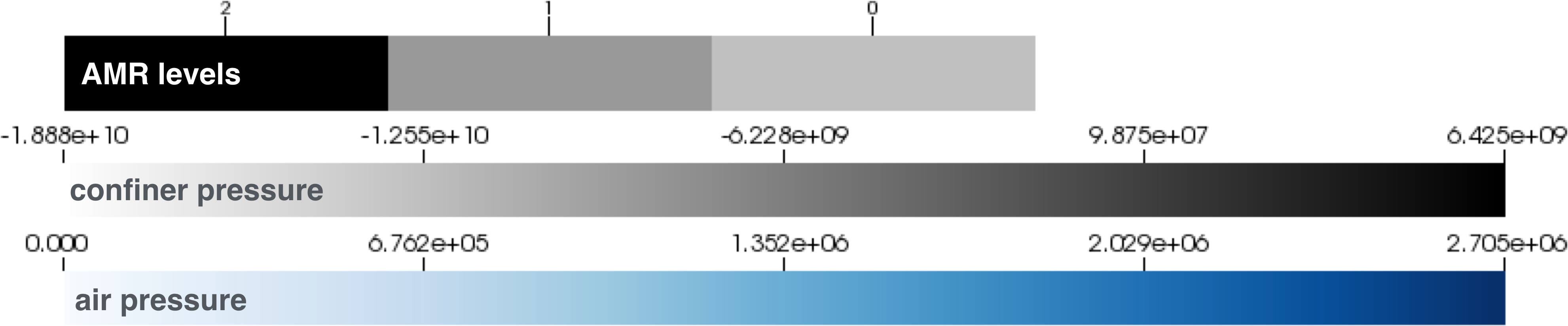}
       \end{subfigure}%
        \begin{subfigure}[b]{0.49\textwidth}
                 \includegraphics[width=0.85\textwidth,center]{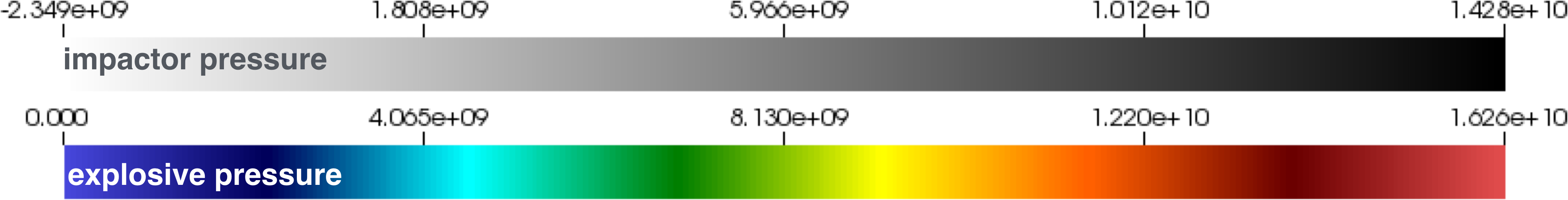}

         \end{subfigure}
\end{minipage}
\centering
\begin{minipage}{\columnwidth}
        \centering
        \begin{subfigure}[b]{0.49\textwidth}
                 \includegraphics[width=0.9\textwidth]{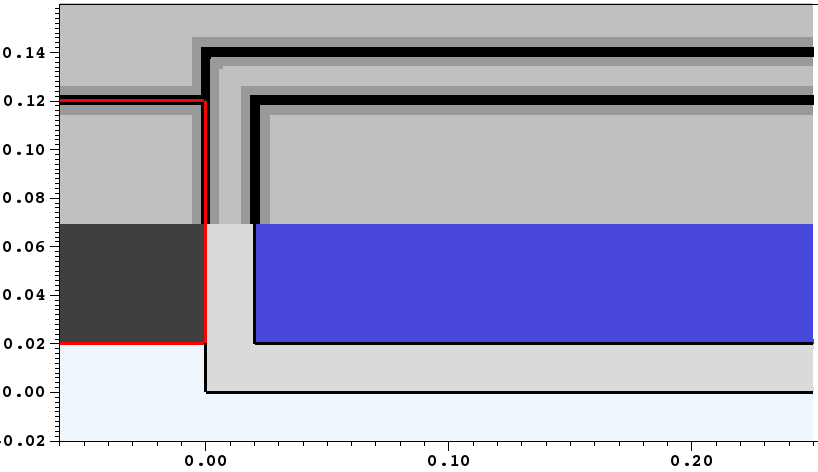}
                 \subcaption{$t=\SI{0}{\micro \second}$}
       \end{subfigure}%
        \begin{subfigure}[b]{0.49\textwidth}
                 \includegraphics[width=0.9\textwidth]{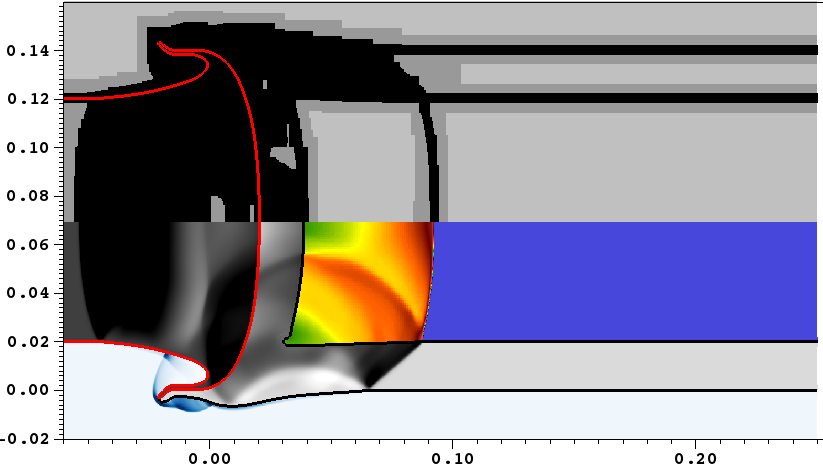}
                 \subcaption{$t=\SI{15}{\micro \second}$}
         \end{subfigure}
\end{minipage}
\centering
\begin{minipage}{\columnwidth}
        \centering
        \begin{subfigure}[b]{0.49\textwidth}
                 \includegraphics[width=0.9\textwidth]{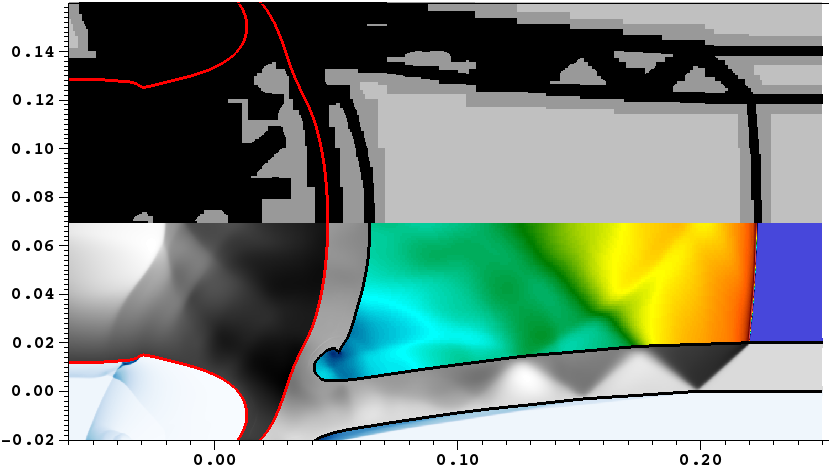}
                 \subcaption{$t=\SI{35}{\micro \second}$}
       \end{subfigure}%
        \begin{subfigure}[b]{0.49\textwidth}
                 \includegraphics[width=0.9\textwidth]{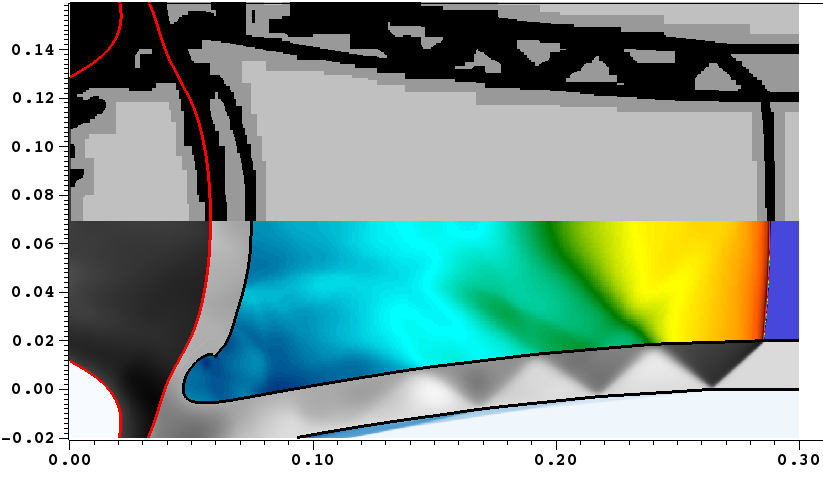}
                 \subcaption{$t=\SI{45}{\micro \second}$}
         \end{subfigure}
\end{minipage}
\centering
\begin{minipage}{0.9\columnwidth}
                \includegraphics[width=\textwidth]{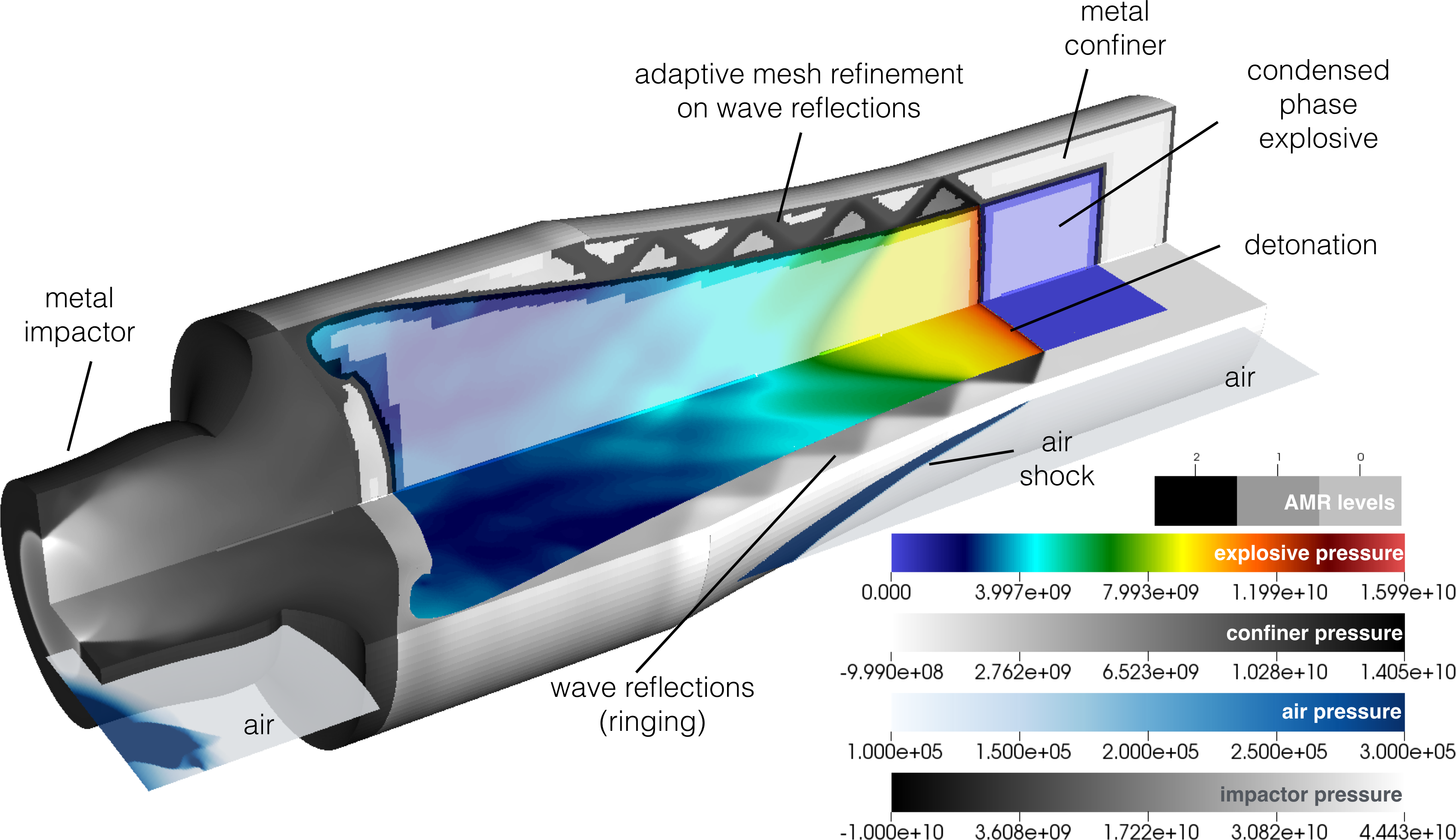}
                 \subcaption{$t=\SI{60}{\micro \second}$}
\end{minipage}
\caption{ The lower part of images (a)-(c) present the two-dimensional pressure distribution in the explosive, the solid, the projectile and the surrounding gas while the top part presents the AMR grid distribution in all materials, at times $t=0,15$ and $\SI{35}{\micro \second}$ of the showcase presented in Sec.\ \ref{Sec:ImpactcopperCan}. Im image (d) the same quantities are shown in a three-dimensional view for $\SI{60}{\micro \second}$.  }
\label{impactcopperCan} 
\end{figure}

{\color{black} Figure \ref{impactcopperCan} shows the pressure distribution in the explosive, the solid and the surround gas, as well as the AMR grids, at times $t=0,15,35$ and $\SI{60}{\micro \second}$. At the start of the simulation, the explosive is at ambient conditions, with pressure at $10^5\SI{}{\pascal}$. The impact sends a rightward-moving shock wave into the explosive and a leftward-moving wave in the projectile. Reflections of the waves in all materials also occur due to the confiner and the projectile not having the same width as the casing. The explosive is initiated and the reaction wave transits to a steady detonation. The detonation wave (red region in explosive) induces a shock wave in the solid (dark grey region in solid) which, upon reaching the copper-air boundary sends a shock wave in the air (dark blue region in air) and a release wave back into the solid. The repeated reflections at the material interfaces generate alternate regions of compression and tension in the solid and pressure waves in the explosive (often referred to as `ringing').  The effect of the detonation on the deformation of the can is seen by comparing the shape of the can in the early and late stages of the event.}
 
 \subsection{Rod impact leading to fuel ignition in a car fuel tank}
\label{Sec:ImpactcopperCan}

\begin{figure}[!t]
\begin{minipage}{\columnwidth}
        \centering
        \begin{subfigure}[b]{0.49\textwidth}
                 \includegraphics[width=0.9\textwidth]{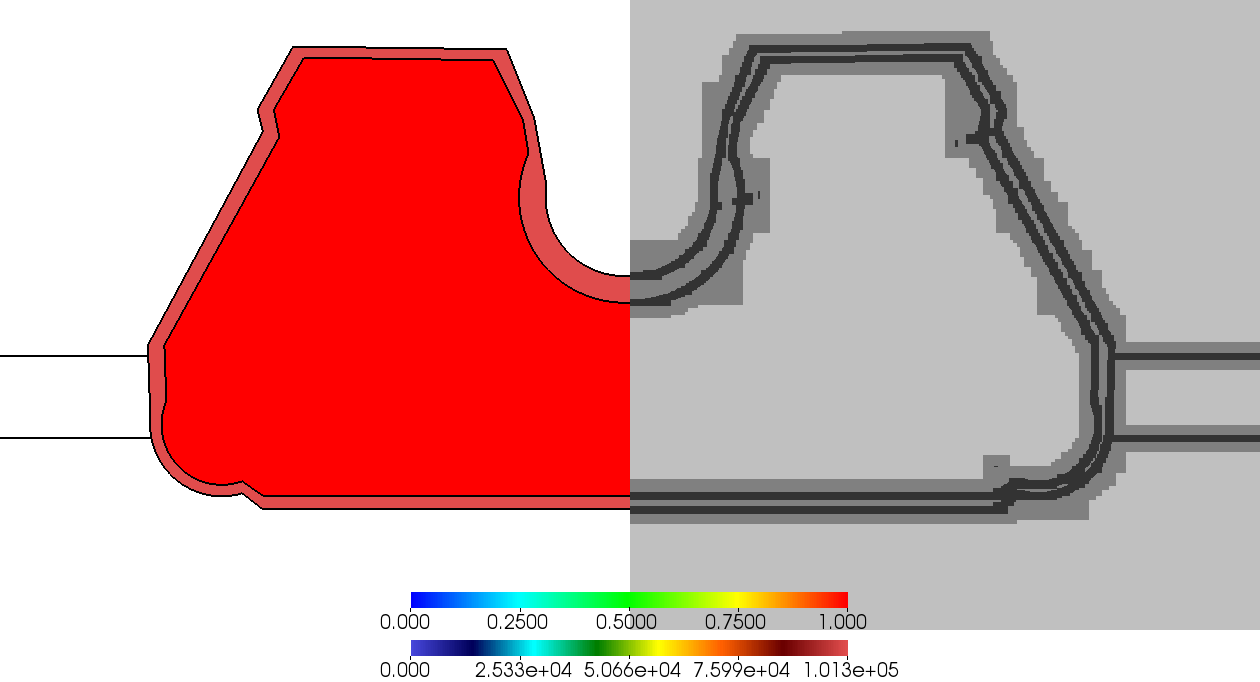}
        \subcaption{$t=\SI{0}{\second}$}
       \end{subfigure}%
        \begin{subfigure}[b]{0.49\textwidth}
                 \includegraphics[width=0.9\textwidth]{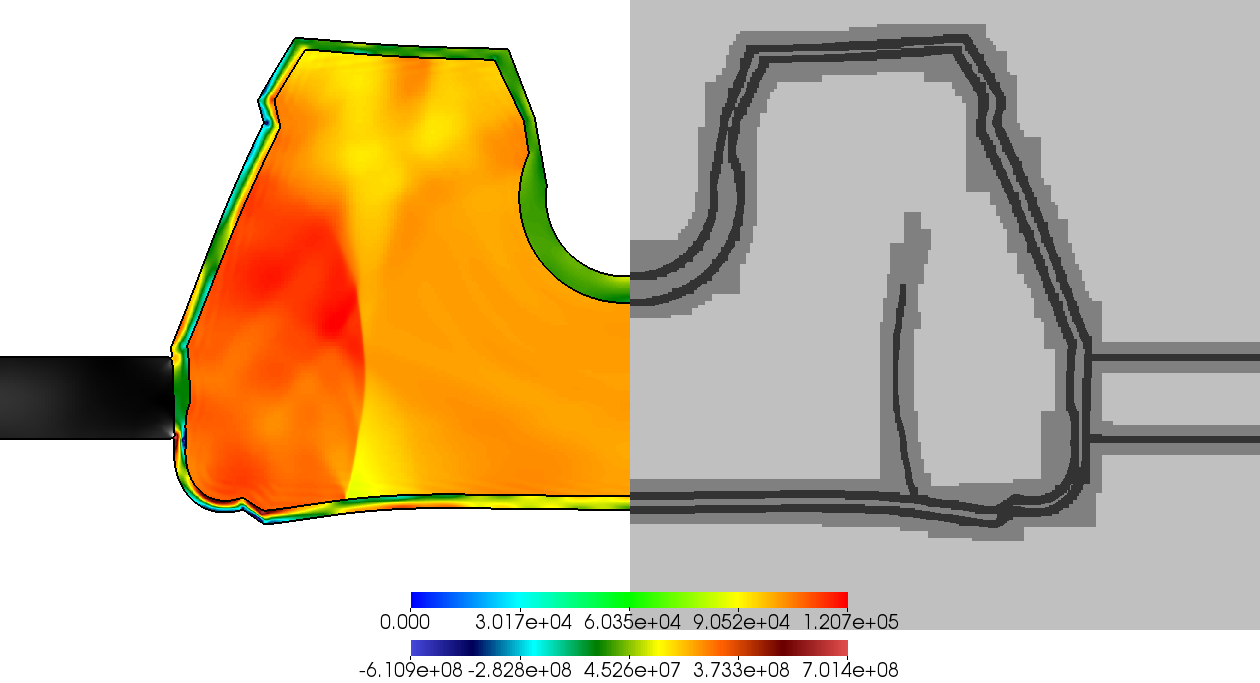}
        \subcaption{$t=\SI{0.0002}{\second}$}
         \end{subfigure}
\end{minipage}
\begin{minipage}{\columnwidth}
        \centering
        \begin{subfigure}[b]{0.49\textwidth}
                 \includegraphics[width=0.9\textwidth]{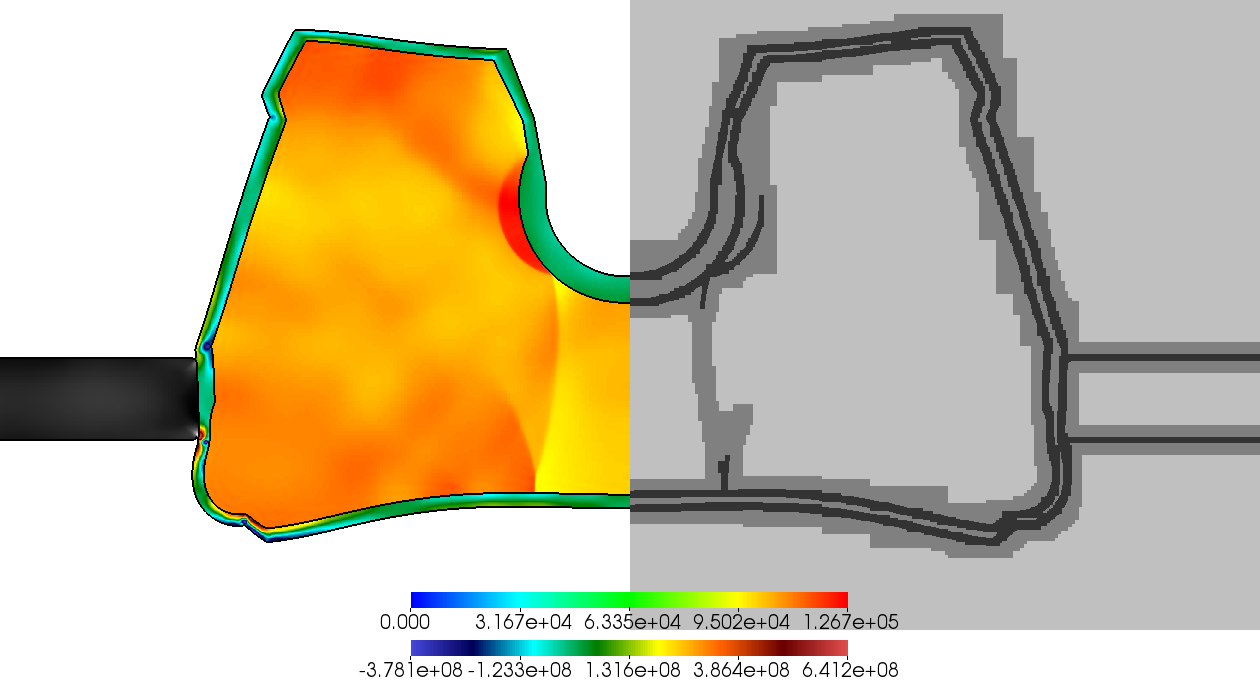}
        \subcaption{$t=\SI{0.0004}{\second}$}
       \end{subfigure}%
        \begin{subfigure}[b]{0.49\textwidth}
                 \includegraphics[width=0.9\textwidth]{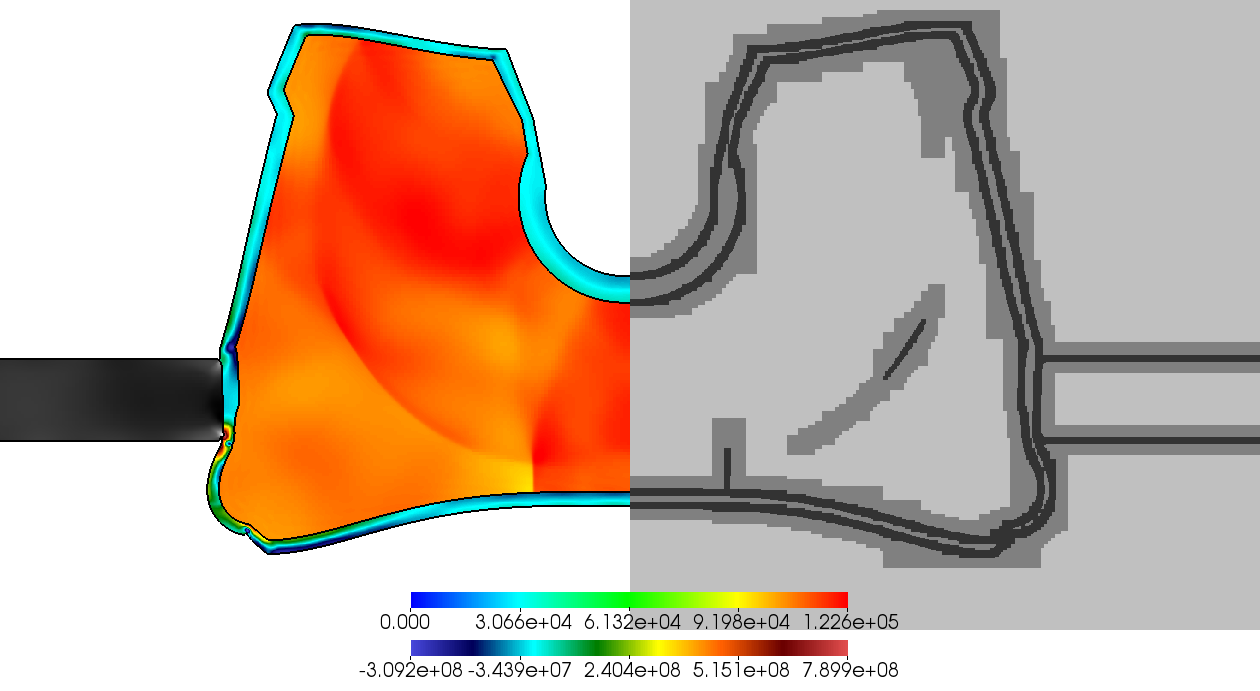}
        \subcaption{$t=\SI{0.0006}{\second}$}
        \end{subfigure}
\end{minipage}
\caption{The pressure distribution (left) and the distribution of AMR grids (right) for early stages of the fuel tank impact simulation. }
\label{fuelTankP} 
\end{figure}

In this section, we present a final showcase as a low-speed impact example involving {\color{black} reactive flow (fuel), elastoplastic structural (solid) response and inert air} response. Consider a fuel tank as presented in Fig.\ \ref{fuelTankP}a. The steel tank is filled with gaseous explosive and is impacted by a steel projectile travelling at $\SI{45}{\m \per \second}$, which can be considered to be the speed of a head-on car collision. The steel rod has diameter $\SI{3}{\cm}$ and a semi-infinite length, starting with length $\SI{5.5}{\cm}$  but extending also out of the domain. The configuration resides in air, in a domain with dimensions $\SI{23}{\cm}\times\SI{23}{\cm}$ with reflective right boundary condition and transmissive conditions at all other boundaries. The simulation is performed with a base spatial resolution of $\Delta x = \Delta y = \SI{1}{\cm}$ and one level of refinement with refinement factor 4, resulting in an effective resolution of $\Delta x = \Delta y = \SI{0.25}{\cm}$.

This use case aims to demonstrate the ability of the algorithm to handle low-speed impact scenarios involving multiple materials and combustion.  {\color{black}The fuel is modelled by an ideal gas EoS with $\gamma=1.4,c_v=\SI{718}{\joule \per \kilogram \per \kelvin}$ and $Q=\SI{0.497}{\mega \joule \per \kilogram} $ and its combustion by a single-step Arrhenius law as per Eq.\ \ref{Arrhenius} with $C=\SI{9.1e11}{\per \second}$ and activation temperature $T_A=\SI{7974.68}{\kelvin}$.} The fuel tank casing and the projectile are both taken to be {\color{black}mild steel, described} by the Hugoniot elastic equation of state with parameters:

\begin{equation}
(c_0 [\SI{}{\meter \per \second}], s, \rho_0 [\SI{}{\kilogram \per \meter \tothe{3}}], T_0 [\SI{}{\kelvin}] )=( 4569, 1.49, 7870, 298)
\end{equation}

 and perfect plasticity with yield of $\SI{137}{\mega \pascal}$. The entire configuration is residing in air, described as an ideal gas with $\gamma=1.4$. All materials are initially at atmospheric conditions. 

\begin{figure}[!t]
\begin{minipage}{\columnwidth}
        \centering
        \begin{subfigure}[b]{0.49\textwidth}
                 \includegraphics[width=0.9\textwidth]{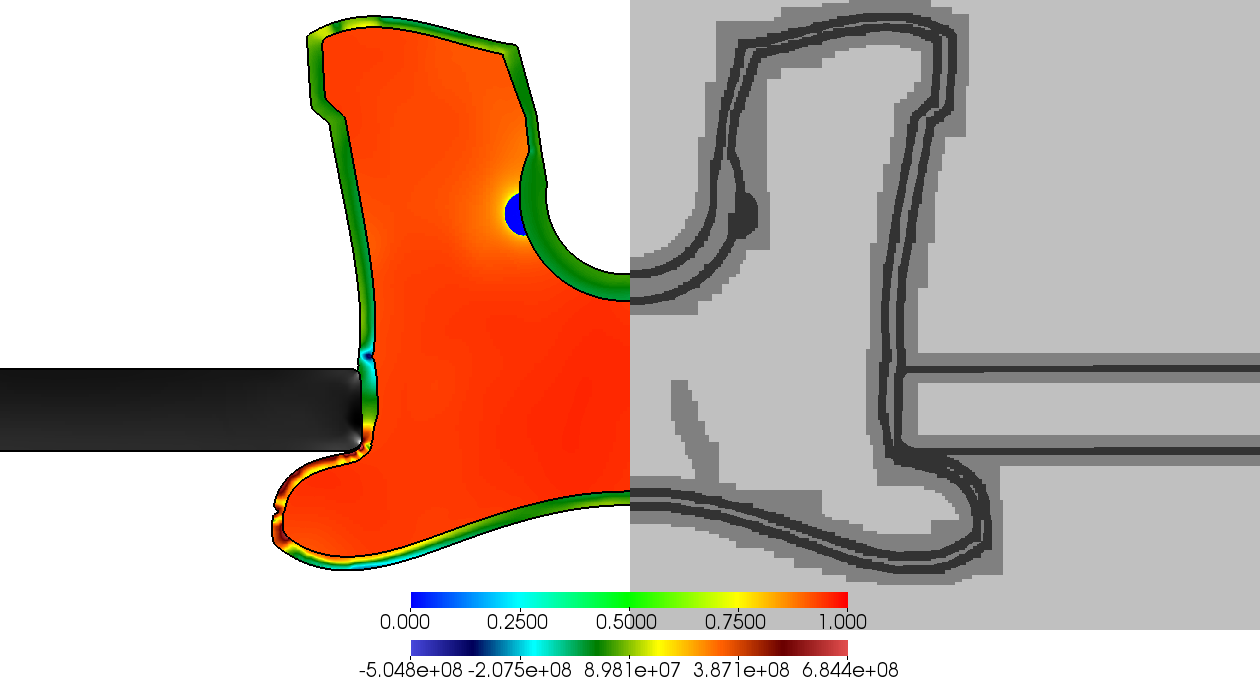}
        \subcaption{$t=\SI{0.00173}{\second}$}
       \end{subfigure}%
        \begin{subfigure}[b]{0.49\textwidth}
                 \includegraphics[width=0.9\textwidth]{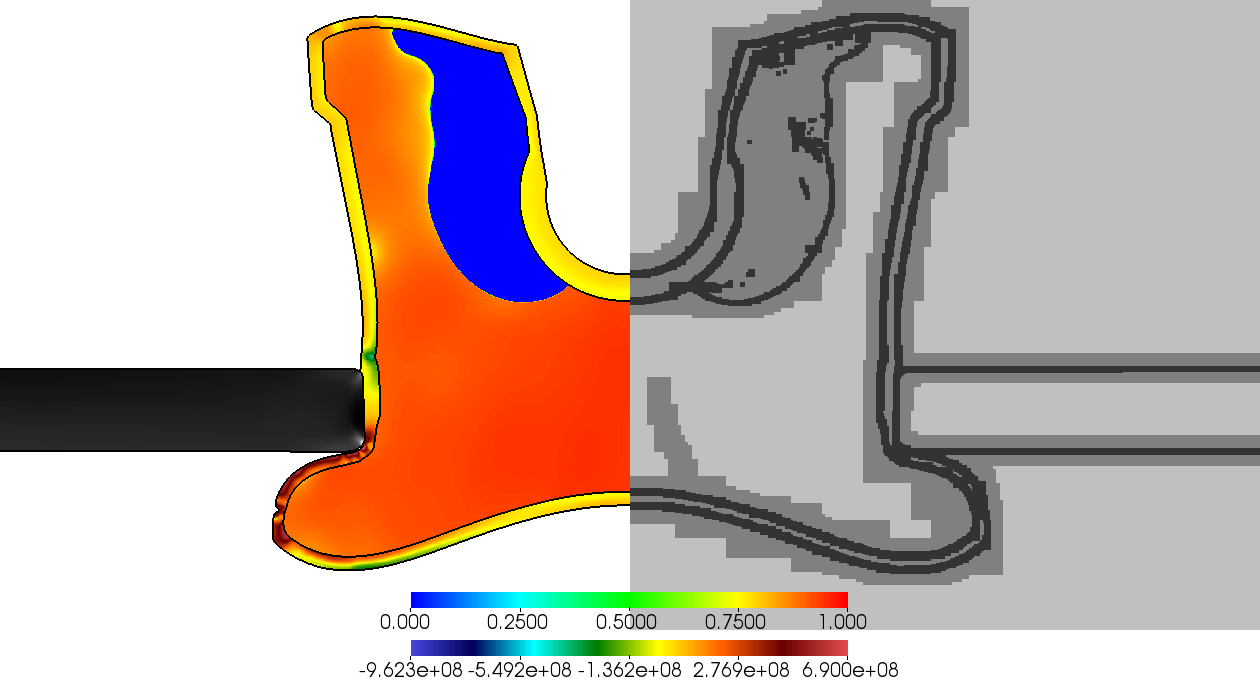}
                 \subcaption{$t=\SI{0.00175}{\second}$}
         \end{subfigure}
\end{minipage}
\begin{minipage}{\columnwidth}
        \centering
        \begin{subfigure}[b]{0.49\textwidth}
                 \includegraphics[width=0.9\textwidth]{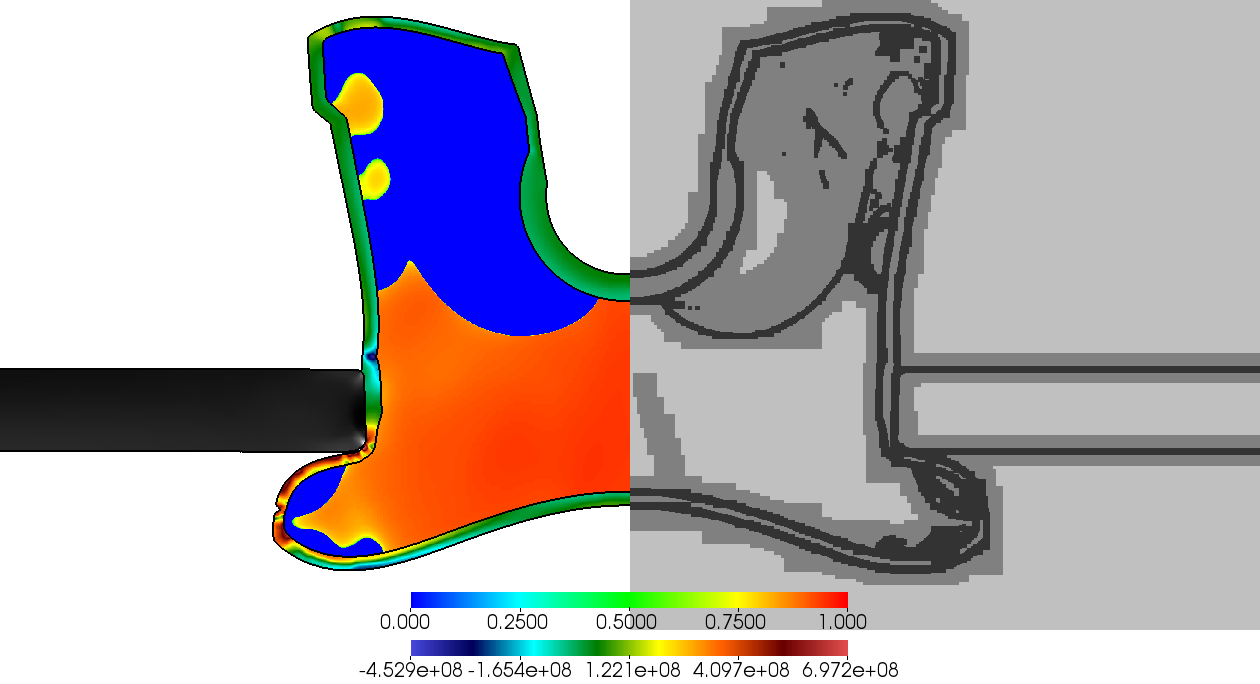}
        \subcaption{$t=\SI{0.00176}{\second}$}
       \end{subfigure}%
        \begin{subfigure}[b]{0.49\textwidth}
                 \includegraphics[width=0.9\textwidth]{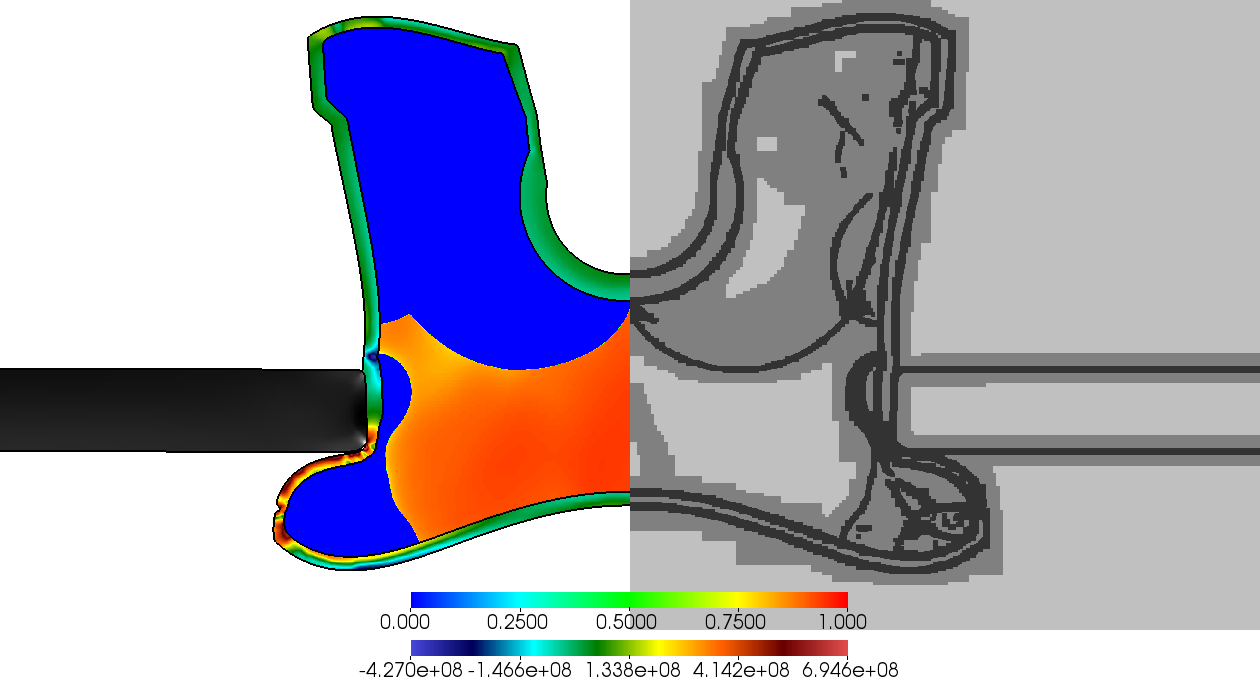}
        \subcaption{$t=\SI{0.00177}{\second}$}
         \end{subfigure}
\end{minipage}
\begin{minipage}{\columnwidth}
\centering
                 \includegraphics[width=0.7\textwidth]{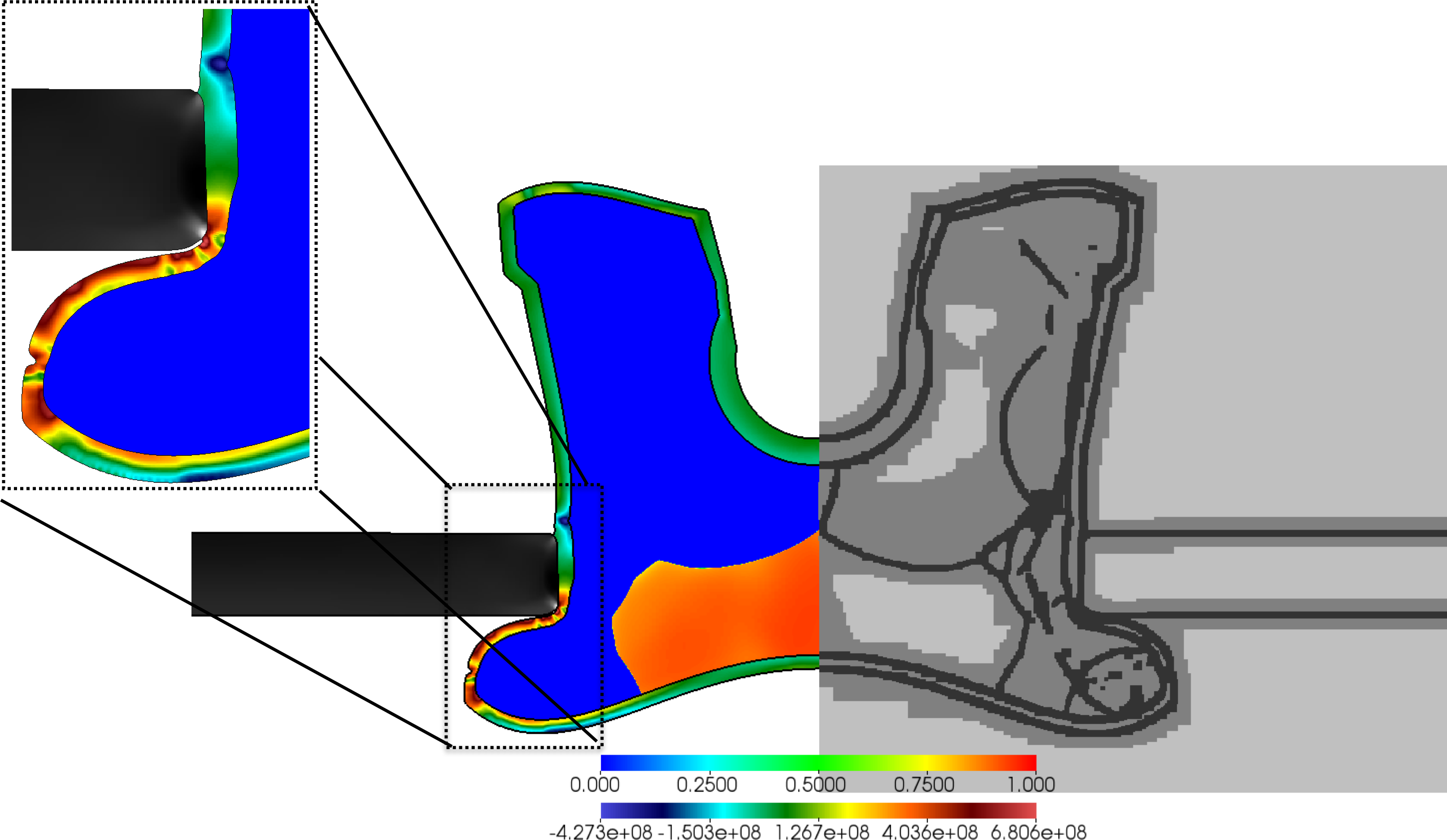}
        \subcaption{$t=\SI{0.00178}{\second}$}
\end{minipage}
\caption{The reaction progress variable (left) and the distribution of AMR grids (right) for late stages of the fuel tank impact simulation. The ignition of the fuel in a primary and secondary locations is seen, as well as the propagation and coalescence of the reaction fronts in the tank. A zoom on the tank shell is provided to demonstrate that the elastoplastic equations are solved both in the projectile and the tank shell simultaneous to the solution of the reactive flow equations in the fuel tank.}
\label{fuelTank} 
\end{figure}

 The left halves of the images in Fig.\ \ref{fuelTankP} show the pressure distribution in the fuel, in the casing and the impacting rod, while the right halves show the distribution of the AMR grids. Similarly, left halves of the images in Fig.\ \ref{fuelTankP} show the evolution of the reaction progress variable ($\lambda$) at late stages of the impact.  After several reflections of the pressure waves at the casing walls, initiation of the gas is observed in the blue region of Fig.\ \ref{fuelTank}a. The reaction region grows as depicted in  Figs.\ \ref{fuelTank}b-c. In  Fig.\ \ref{fuelTank}c a secondary initiation zone is seen to be generated at the leftmost (and symmetrically rightmost) bottom part of the tank. In Fig.\ \ref{fuelTank}f the distinct reaction fronts coalesce and move to deplete the remaining fuel in the tank. {\color{black}It should be noted that the elastoplastic equations are solved both in the projectile and the fuel tank shell. In  Fig.\ \ref{fuelTank}e we zoom in on the fuel tank shell to show waves that are travelling in this material. Being able to solve for the evolution of the fuel tank shell signifies that one can use this methodology to optimise the shape and material of the fuel tank shell for accident prevention. Similarly, as the impactor can be thought of as the car panel, our techniques can be used for optimising this car part as well.}
 

\section{Conclusions}

{\color{black}In this work we present a methodology for the numerical simulation of the two-way interaction of reactive flow and elastoplastic structural response for automotive and defence applications.} We present the coupling of a MiNi16 formulation and reduced models suitable for modelling {\color{black}gaseous fuels and condensed-phase explosives}, with fluid (compressible Euler) and elastoplastic solid formulations. The coupling utilises level set and Riemann ghost fluid methods to achieve communication between the materials described by different systems of equations. The fuel/explosive, solid and fluid models are all formulated as hyperbolic conservation laws and thus they can be solved by the same or similar numerical methods, facilitating the communication at material interfaces. The communication is achieved by solving mixed-material Riemann problems at the interfaces. To this end, we derive mixed Riemann solvers for each formulation pair considered in this work. These are based on the characteristic equations derived for each system pair, allowing for the computation of the star state in the `real' material, by taking into account the two different systems on either side of the interface and applying appropriate interface boundary conditions.

The elastoplastic solid model and the explosives model are in the first instance validated separately. This assesses the implementation of each model without the influence of the other formulation and the ghost fluid boundaries. The validation test chosen for the solid is at low enough speed to observe splitting of the elastic and plastic components of waves. The explosives model is validated for C4, as this is the explosive of interest in later applications. Pop-plots, as well as ZND and CJ conditions are demonstrated to match well with experiments.

The solid-explosive coupling is validated against several tests from the literature. Firstly, a one-dimensional mixed Riemann problem is utilised. We compare our approximate solutions for this test against exact solutions and other numerical solutions found in the literature. Then, we validate the coupling in cylindrical symmetry, in the context of sandwich-plate impact tests. The tests consider a non-reacting Detasheet material or a reacting C4 explosive residing between two steel plates. The sandwiched configuration is impacted by a steel projectile. In the inert scenario, we present and analyse the generated wave pattern and wave interaction. Pressure gauges are placed in Detasheet and pressure histories are compared with gauge results found in the literature, observing a very good match between the two. In the reacting scenario (ERA), we study how the generated wave pattern leads to the ignition of C4. We compare ignition times between the sandwich-plate configuration and a configuration without a rear steel plate, illustrating the effect of the plate in accelerating the ignition process.

{\color{black}We demonstrate the capability of the methodology to simulate high- and low-speed impact leading to the initiation of fuel/explosive in three use cases. The first use case, extends the ERA scenario by including an air gap between the front plate and the explosive and we demonstrate how the air gap hinders the initiation. The second use case is the impact of a copper can filled with reactive nitromethane. The impact initiates the nitromethane which transits to detonation. The detonation wave interacts with the confiner and shows wave-ringing in the confiner as reported in the literature. The deformation of the copper can is also highlighted, as well as the transmission of waves from the projectile to the explosive, to the confiner and finally to the surrounding air. The last use case is concerned with the accidental initiation of the fuel in the car fuel tank in crash scenarios. The speed of impact in this case is much lower than in the ERA scenario and it is demonstrated that the methodology and the code can handle this often difficult to handle scenario. The impact generates waves that reflect multiple times on the fuel tank boundaries until the fuel cannot sustain the excess energy and ignites. Multiple ignition sites are seen.  The multi-dimensional use cases illustrate the accurate implementation of mixed Riemann solvers for the different systems considered in this work, describing explicitly explosive, fluid and elastoplastic materials.} {\color{black} The methodology can also be used as part of the manufacturing process for optimising car (or other device) compartments in terms of shapes and materials.}

\section*{Acknowledgements}
This work was supported by Jaguar Land Rover and the UK-EPSRC grant EP/K014188/1 as part of the
jointly funded Programme for Simulation Innovation.


\bibliographystyle{elsarticle-num}
\bibliography{refs}

\end{document}